\documentclass[twocolumn]{aastex63}

\accepted{for publication in ApJ, June 10, 2022}

\shorttitle{Evolution of X-ray Activity}
\shortauthors{Getman, Feigelson, Garmire, Broos, Kuhn, Preibisch, Airapetian}

\graphicspath{{./}}

\begin{document}

\title{Evolution of X-ray Activity in $<25$ Myr Old Pre-Main Sequence Stars}

\correspondingauthor{Konstantin Getman}
\email{kug1@psu.edu}

\author[0000-0002-6137-8280]{Konstantin V. Getman}
\affiliation{Department of Astronomy \& Astrophysics \\
Pennsylvania State University \\ 
525 Davey Laboratory \\
University Park, PA 16802, USA}

\author[0000-0002-5077-6734]{Eric D. Feigelson}
\affiliation{Department of Astronomy \& Astrophysics \\
Pennsylvania State University \\ 
525 Davey Laboratory \\
University Park, PA 16802, USA}

\author[0000-0002-7371-5416]{Gordon P. Garmire}
\affiliation{Huntingdon Institute for X-ray Astronomy\\
LLC, 10677 Franks Road\\
Huntingdon, PA 16652, USA}

\author[0000-0002-7872-2025]{Patrick S. Broos}
\affiliation{Department of Astronomy \& Astrophysics \\
Pennsylvania State University \\ 
525 Davey Laboratory \\
University Park, PA 16802, USA}

\author[0000-0002-0631-7514]{Michael A. Kuhn}
\affiliation{Department of Astronomy\\
California Institute of Technology \\
Pasadena, CA 91125, USA}

\author[0000-0003-3130-7796]{Thomas Preibisch}
\affiliation{Universit\"{a}ts-Sternwarte M\"{u}nchen\\
Ludwig-Maximilians-Universit\"{a}t\\
Scheinerstr. 1, 81679, München, Germany}

\author[0000-0003-4452-0588]{Vladimir S. Airapetian}
\affiliation{American University\\ 4400 Massachusetts Avenue NW\\ Washington, DC 20016 USA}
\affiliation{NASA/GSFC/SEEC\\ Greenbelt, MD, 20771, USA}

\begin{abstract}
Measuring the evolution of X-ray emission from  pre-main sequence (PMS) stars gives insight into two issues: the response of magnetic dynamo processes to changes in interior structure and the effects of high-energy radiation on protoplanetary disks and primordial planetary atmospheres.  We present a sample of 6,003 stars with ages 7--25~Myr in ten nearby open clusters from {\it Chandra} X-ray and Gaia-EDR3 surveys. Combined with previous results in large samples of younger ($\lesssim 5$~Myr) stars in MYStIX and SFiNCs star forming regions, mass-stratified activity-age relations are derived for early phases of stellar evolution. X-ray luminosity ($L_X$) is constant during the first few Myr, possibly due to the presence of extended X-ray coronas insensitive to temporal changes in stellar size. $L_X$ then decays during the 7--25~Myr period, more rapidly as stellar mass increases.  This decay is interpreted as decreasing efficiency of the $\alpha^2$ dynamo as radiative cores grow and a solar-type $\alpha \Omega$ dynamo emerges. For more massive 3.5--7~M$_{\odot}$ fully radiative stars, the X-ray emission plummets indicating lack of an effective magnetic dynamo. The findings provide improved measurements of high energy radiation effects on circumstellar material, first the protoplanetary disk and then the atmospheres of young planets. The observed X-ray luminosities can be so high that an inner Earth-mass rocky, unmagnetized planet around a solar-mass PMS star might lose its primary and secondary atmospheres within a few-several million years. PMS X-ray emission may thus have a significant impact on evolution of early planetary atmospheres and the conditions promoting the rise of habitability.  
\end{abstract}

\section{Introduction} \label{sec:intro2}

Pre-main-sequence (PMS) stars are fully or partially convective, rapidly rotating young stars that exhibit enhanced magnetic activity, which indicates the presence of powerful magnetic dynamos. Evidence includes cool starspots covering considerable fractions of the surface, high surface magnetic fields, and magnetic reconnection flares producing $\sim$1,000--10,000 times more X-ray emission than the contemporary Sun \citep{Bouvier93, Donati97, Feigelson1999, Gudel2004, Preibisch05, Feigelson10, Gregory2010, Stelzer2017}.

\begin{figure*}[ht!]
\epsscale{1.15}
\plotone{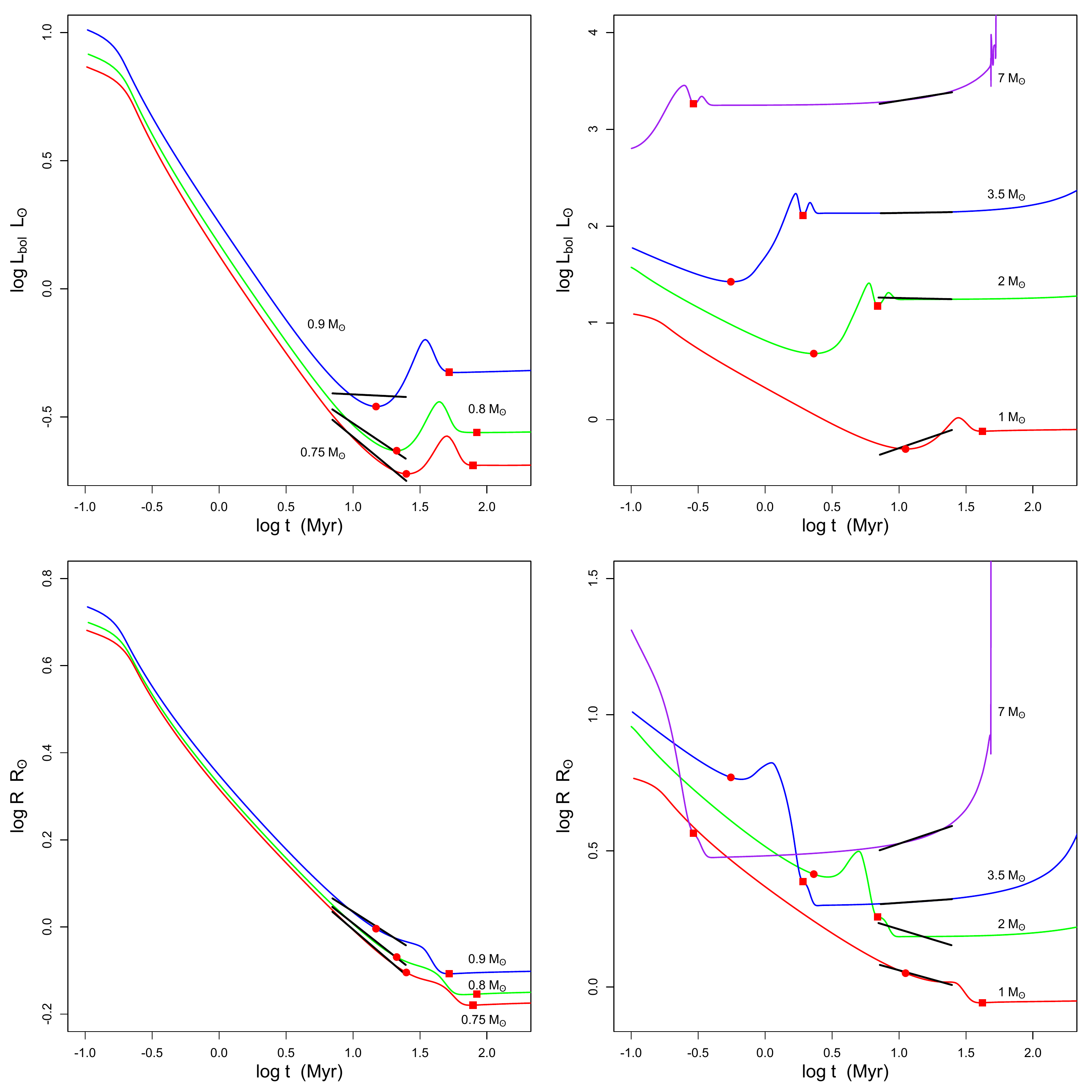}
\caption{Temporal evolution of stellar bolometric luminosity and radius for 7 stellar masses as predicted by the PARSEC 1.2S evolutionary models. The corresponding Hayashi-Henyey and Henyey-ZAMS evolutionary boundaries are depicted by the small circle and square symbols, respectively. The black solid lines indicate linear regression fits to the model data within the $t=$~7--25~Myr age span of the open clusters of interest in the current study. Corresponding inferred slopes $b$ for the $L_{bol} \propto t^{b}$ and $R_{\star} \propto t^{b}$ relations within this age range are listed in Table~\ref{tab:temporal_evolution_table}.  \label{fig:lbol_r_vs_time}}
\end{figure*}

But, except for very low mass stars that remain fully convective, the interior of stars change dramatically before they arrive at the Zero Age Main Sequence (ZAMS). This evolution was calculated in the prescient study of \citet{Iben65} and is shown in Figure~\ref{fig:lbol_r_vs_time} using modern PARSEC 1.2S evolutionary models \citep{Bressan12, Chen14}.  Stars around a solar mass first contract with falling bolometric luminosities along Hayashi tracks for $10-15$~Myr years.  We will call this the early-PMS (e-PMS) phase.  The star then enters the late-PMS (l-PMS) phase when the surface temperature rapidly increases, the outer layer contraction slows, and the luminosity rises somewhat.  A core radiative zone, already emerging during the e-PMS phase, now occupies most of the volume and mass of the interior. 

This l-PMS phase, sometimes called the Henyey track, lasts a few million years for solar-mass stars.  At the end of the l-PMS phase, core convection might emerge briefly due to $^{12}$C burning, after which the star settles on the Zero Age Main Sequence (ZAMS) with continuous hydrogen burning.  The duration of these phases is strongly mass-dependent with higher mass stars rapidly leaving the e-PMS Hayashi track and migrating across the Hertzsprung-Russell diagram in the l-PMS phase in $\lesssim 1$~Myr (right panels of Figure~\ref{fig:lbol_r_vs_time}). 

The l-PMS phase is relatively poorly studied because large and reliable star samples are difficult to obtain.  Many star clusters have dispersed, the infrared-excess protoplanetary disks have disappeared, and the stars have drifted away from their natal molecular clouds which themselves may have dissipated \citep{Kuhn19, Alexander2014, Richert18, Feigelson96}. There are no conveniently nearby l-PMS samples such as the e-PMS Orion Nebula cluster at distance $d \simeq 0.4$~kpc or lower mass groups Taurus-Auriga or Ophiuchus around $d \simeq 0.15$~kpc. The nearest l-PMS samples $-$ such as the $\beta$ Pic and TW Hya moving groups and the Sco-Cen Association $-$ are spread broadly across the sky, inhibiting the study of  large samples of l-PMS stars facilitated by compact clusters. 

Consequently, there is limited empirical evidence for the effects of the interior changes on  surface magnetic activity. It is widely believed that fully convective e-PMS (and late-M main sequence) stars generate magnetic fields through a distributed turbulent $\alpha^2$-type magnetic dynamo rather than solar-type tachoclinal $\alpha \Omega$-type dynamos \citep{Durney1993, Yadav2015, Cohen2017, Warnecke2020}.  It is unclear whether $\alpha^2$ dynamos will grow fields up to equipartition strength or whether they will saturate at weaker levels.  The morphology of the fields emerging onto the surface of e-PMS stars phase may be distinctive; \citet{Cohen2017} predict they are concentrated at high-latitudes; this is seen in some e-PMS stars \citep{Donati07} but not others \citep{Argiroffi2017}.  

In the X-ray band where magnetic activity is readily investigated, one sample of l-PMS stars has been studied: $\sim 400$ stars in the 13~Myr old cluster NGC~869 = h Per at a distance of $d \simeq 2.5$~kpc. Here \citet{Argiroffi2016} find a dependence of X-ray luminosity on rotation for 1--2~M$_{\odot}$, suggesting that activity in  stars is now powered by a $\alpha\Omega$-type dynamo.  

In the present study, we investigate the X-ray emission  from the e-PMS through the l-PMS phases in more detail with a larger and more diverse sample: 6,003 young stars in 10 open clusters at distances $0.3 \lesssim d \lesssim 2.5$~kpc with ages ranging from 7 to 25~Myr (Table~\ref{tab:cluster_props}).  This sample is combined with our earlier MYStIX (Massive Young Star-Forming Complex Study in Infrared and X-Ray) and SFiNCs (Star Formation In Nearby Clouds) surveys of 40,041 e-PMS stars (ages 0.5--5~Myr) in 42 star forming regions at distances $0.3 \lesssim d \lesssim 3$~kpc \citep{Feigelson13, Getman2019}. Large samples are needed to distinguish dependencies on mass, age and rotation in the presence of scatter produced by flare variability.  Here all of these samples are observed with the same instrument and analyzed in a homogeneous fashion, capture stars with different masses (including the rare intermediate mass stars) at different stages of their PMS evolution.  

Our characterization of mass-stratified X-ray activity-age relations in early stellar evolution from $0.5$ to $25$~Myr has one astronomical goal and two astrophysical goals:
\begin{enumerate}

    \item It fills a `missing link' in the activity-rotation-age relations for solar-type stars \citep{Skumanich72}.  For the PMS phases, it is important to add stellar mass as a variable due to strong dependence of interior structure on mass (Figure~\ref{fig:lbol_r_vs_time}).  Past empirical measurements of the mass-stratified evolution of X-ray activity have little information between the e-PMS and ZAMS phases \citep{PreibischFeigelson2005, Gregory16}. Our cluster sample is large enough that, with careful statistical analysis, we can provide distribution functions of X-ray activity measures rather than just first moment (e.g. median value) summaries. Our much larger cluster sample should offer more accurate estimates of mass-stratified X-ray activity as a function of age.
    
    \item Our current study, with a larger cluster sample and a range of ages, can verify the evidence of \citet{Argiroffi2016} that the transition from $\alpha^2$ to $\alpha\Omega$ dynamos in the stellar interior has observational consequences in stellar X-ray activity, as outlined above.  
    
    \item The high fluence X-ray radiation of PMS stars will penetrate deep into nearby molecular environments. For the e-PMS phase, it is well-established that the X-rays ionize the protoplanetary disk and have a major role in its dissipation \citep{Alexander2014}.  For the l-PMS phase when disk gas is mostly gone, the X-rays may have significant effects on the primordial atmospheres of close-in planets. \citet{Lammer2003} was the first to show that stellar X-ray and ultraviolet irradiation will dominate thermal evaporation in atmospheric loss, although their original calculation has been revised in later studies (reviewed by \citet{Owen19}).  Our empirical findings can assist the analytical studies like the one by \citet{Johnstone2021} who use mass-stratified changes in the X-ray activity as inputs for  their calculations on early planetary atmospheres.
    
\end{enumerate}

The paper is organized as follows. Past studies on the X-ray evolution in PMS stars are reviewed in \S \ref{sec:intro_to_past_studies}. The sample of ten 7--25~Myr-old open clusters and associated {\it Chandra} X-ray data and reduction procedures are described in \S \ref{sec:targets_chandra}. Stellar membership assignments for the open clusters are based on a combination of {\it Chandra} and {\it Gaia} properties (\S \ref{sec_membership}). Properties of the open clusters and their stellar members are derived in \S\S \ref{sec:clus_mem_properties}-\ref{sec:Xray_luminosities}. The younger e-PMS MYStIX and SFiNCs stars supplementing our l-PMS sample are described in \S \ref{sec:mystix_sfincs}. Results on X-ray-mass-age relations are presented in \S\S \ref{sec:xrays_vs_mass}-\ref{sec:xray_activity_slopes} and Appendix sections \ref{sec:appendix_temporal_fx_lxlbol_pms}-\ref{sec:temporal_highest_masses}. Comparison with past results on the X-ray evolution in PMS stars is presented in \S \ref{sec:prev_literature}. Astrophysical implications for dynamos and surface activity are discussed in \S\ref{sec:dynamo_activity}, and implications for ionization of disks and planetary atmospheres are discussed in \S\ref{sec:disks_atmos}.  Companion studies will analyze the l-PMS X-ray activity dependence on stellar rotation, and will investigate the presence of mega- and super-flares in l-PMS stars that are common in e-PMS MYStIX/SFiNCs stars \citep{Getman2021}.  

\section{Past Studies on X-ray Evolution in PMS Stars} \label{sec:intro_to_past_studies}
\citet{PreibischFeigelson2005} introduce two different approaches to study the age-dependence of the stellar X-ray luminosity ($L_X$). One (approach A) deals with $L_X$ changes over a short time range, first 10~Myr of stellar evolution, based on the single X-ray dataset of stars in Orion Nebula \citep[COUP;][]{Getman05} with the assumption that the related isochronal age spread of $\sim 10$~Myr in the Orion Nebula is real. The stellar sample is limited to lightly-absorbed Orion stars that are likely members of the Orion Nebula Cluster (ONC). In contrast, the second approach (B) relies on median stellar X-ray luminosities and ages for several stellar clusters, spanning a wide age range of a few Gyr. These include very young $t<10$~Myr clusters ONC, NGC 2264, and Chameleon; older $t=$~100--600~Myr open clusters Pleiades and Hyades; and field stars. One of the major findings of this study is a double-slope decay of X-ray activity in 0.5--1.2~M$_{\odot}$ stars: mild decay $L_X \propto t^{-0.3}$ over the first 10~Myr of stellar evolution (based on approach A) followed by much faster decay of $L_X \propto t^{-0.75}$ over the large age range of a few Gyr (approach B). But the authors note that ``Comparison [of ONC] with the NGC 2264 and Chamaeleon I clusters suggests there may be no decay at all during the PMS phase [i.e., based on approach B]''.

Using approach A, \citet{Gregory16} extend the X-ray activity evolution studies towards five young clusters, ONC, NGC 2264, IC 348, NGC 2362, and NGC 6530 (the latter is the primary ionizing cluster of Lagoon Nebula). The X-ray datasets for all but one cluster were taken from our MYStIX project \citep{Feigelson13}. In the stellar mass range of \citet{PreibischFeigelson2005} (0.1--2~M$_{\odot}$), the inferred slopes $b$ in the $L_X \propto t^{b}$ relation appear similar within errors between the two studies. But Gregory et al. report slightly steeper decay for more massive 2--3~M$_{\odot}$ stars. Overall, Gregory et al. find that the PMS stars on Henyey tracks undergo faster X-ray activity changes due to the development of radiative cores.

\begin{deluxetable*}{lrrrrrrrc}
\tabletypesize{\footnotesize}
\tablecaption{Sample of 10 Late-PMS Open Clusters\label{tab:cluster_props}}
\tablewidth{0pt}
\tablehead{
\colhead{Region} & \colhead{$l$} &
\colhead{$b$} & \colhead{$N_{mem}$} &  \colhead{$D$} &
\colhead{$\sigma_{D}$} & \colhead{$A_{V}$} & \colhead{Age} & \colhead{$M_{lim}$}\\
\colhead{} & \colhead{deg} &  \colhead{deg} & & \colhead{pc} & \colhead{pc} & \colhead{mag} & \colhead{Myr} & \colhead{M$_{\odot}$}\\
\colhead{(1)} & \colhead{(2)} & \colhead{(3)} & \colhead{(4)} & \colhead{(5)} & \colhead{(6)} & \colhead{(7)} & \colhead{(8)} & \colhead{(9)}
}
\startdata
NGC1502 & 143.6742 & 7.6567 & 256 (231) & 1078 & 13 & 2.1 & 7 & 0.7\\
NGC2169 & 195.6189 & -2.9264 & 104 (90) & 959 & 6 & 0.7 & 7 & 0.4\\
IC2395 & 266.6473 & -3.6082 & 246 (179) & 717 & 4 & 0.5 & 9 & 0.4\\
NGC3293 & 285.8533 & 0.0726 & 944 (743) & 2552 & 34 & 0.9 & 11 & 0.7\\
NGC869\tablenotemark{a} & 134.6311 & -3.7401 & 1370 (1096) & 2451 & 22 & 1.7 & 13 & 1.0\\
NGC884\tablenotemark{a} & 135.0176 & -3.5958 & 1294 (895) & 2475 & 23 & 1.7 & 13 & 1.0\\
NGC4755 & 303.2035 & 2.5031 & 1219 (1003) & 2101 & 14 & 1.2 & 15 & 0.7\\
NGC1960 & 174.5388 & 1.0702 & 406 (358) & 1192 & 13 & 0.8 & 22 & 0.7\\
NGC3766 & 294.1170 & -0.0231 & 1968 (1291) & 2074 & 15 & 0.8 & 22 & 0.7\\
NGC2232 & 214.4996 & -7.4062 & 117 (117) & 319 & 1 & 0.2 & 25 & 0.4\\
\enddata
\tablecomments{Column 1: Cluster name. Columns 2-3: Galactic coordinates for the cluster center in degrees. Column 4: Total number of cluster members identified across entire {\it Chandra} fields (\S\S \ref{sec_membership} and \ref{sec:clus_mem_properties}) and presented in Table \ref{tab:stellar_props_open_clusters}; totaling 7,924 X-ray and non-X-ray stars. The values enclosed in the parentheses indicate stellar members, located in the central parts of the clusters (except for the most sensitive {\it Chandra} mosaic of the nearest NGC~2232 cluster) and employed in the analyses of the X-ray-mass-age relations (\S\S \ref{sec:xrays_vs_mass}-\ref{sec:xray_activity_slopes} and Appendix \ref{sec:temporal_highest_masses}); totaling 6,003 X-ray and non-X-ray stars. Columns 5-6: Median cluster distance from the Sun and its 68\% bootstrap error, derived in \S \ref{sec_membership}. Columns 7-8: Cluster average extinction in visual band and cluster age, derived in \S \ref{sec:clus_mem_properties}. Column 9: Mass completeness limits for our cluster member samples, derived in \S \ref{sec:clus_mem_properties}.} 
\tablenotetext{a}{NGC 869 = h Per;  NGC 884 = $\chi$ Per}
\end{deluxetable*}

But the approach A that rests on individual stellar ages may be questionable. Debates have waged for decades over interpretation of observed age spreads in rich clusters, such as ONC \citep[e.g.,][]{Reggiani2011}. The difficulties of interpretation are due to various observational and theoretical reasons such as photometric variability, multiplicity, accretion history, binarity, extinction uncertainty, veiling from accretion, scattering and absorption by disks, stellar interiors model uncertainty, and distance uncertainty \citep[][and references therein]{Preibisch2012,Getman14a}. Due to possibly unaccounted magnetic effects, old generation PMS evolutionary models may provide inconsistent ages between intermediate- and low-mass stars, a phenomenon referred as ``radius inflation''  \citep[][and references therein]{Richert18}. For instance, the HRD diagrams in Figure 3 of \citet{Gregory16} show clear isochronal age biases between the intermediate- and low-mass stellar members of NGC 2264 and IC 348. While recent studies do provide strong evidence for the presence of astrophysical age spreads in rich nearby star forming regions and individual clusters, these generally do not exceed 1~Myr per 1 parsec scale \citep{Reggiani2011, Getman14a, Getman2014b, Beccari2017, Getman2018a}, which is much smaller than for instance, $\Delta t \gtrsim 10$~Myr per 2 parsec scale of ONC assumed in their method A by \citet{PreibischFeigelson2005,Gregory16}.

\citet{Tu2015,Johnstone2021} consider temporal mass-stratified changes in the X-ray activity of young and older stars using a different, semi-analytical-empirical, approach. The analytical part includes a modification of the angular momentum evolution model by \citet{Gallet2015}, which takes into consideration three major physical processes, such as star-disk interaction, momentum loss due to stellar wind, and redistribution of angular momentum in stellar interior. With the knowledge of the angular momentum, fractional X-ray luminosity ($R_X = L_X/L_{bol}$) is then predicted assuming (based on past empirical results) two regimes (``saturation'' and ``non-saturation'') described as two different power-law dependencies of $R_X$ on Rossby number (ratio of rotation period to convective turnover time). Various model parameters are adjusted based on the comparison of the model outcomes with the stellar X-ray and rotation empirical data for clusters of different ages. For a wide age range of up to a few Gyr, \citet{Tu2015} and \citet{Johnstone2021} calculate and provide X-ray luminosity temporal tracks for stars in mass ranges near 1~M$_{\odot}$ and 0.1--1.2~M$_{\odot}$, respectively. Within the 0.5--25~Myr age range (targeted by our current paper), \citet{Tu2015,Johnstone2021} use empirical data sets for only a handful of clusters: X-ray data for $\sim 2$~Myr old Taurus cloud stars and 13~Myr old h Per cluster; stellar rotation data for $\sim 2$~Myr old Lagoon Nebula stars and 13~Myr old h Per cluster.

Unlike in \citet{Tu2015,Johnstone2021}, the investigation of the X-ray evolution here is based solely on empirical data. We improve the previous studies of \citet{PreibischFeigelson2005, Gregory16, Tu2015,Johnstone2021} by expanding significantly the X-ray stellar samples, from a few clusters per study to a dozen nearby, very young 0.5--5~Myr stellar clusters/groups from our previously published {\it Chandra} MYStIX/SFiNCs projects \citep[][and see \S \ref{sec:mystix_sfincs} in the current paper]{Feigelson13,Getman17} combined with the new {\it Chandra} X-ray data for 10 nearby, older 7--25~Myr open clusters (\S \ref{sec:targets_chandra}). Due to the aforementioned problems with age spreads we further apply approach B rather than approach A to our data.

\begin{deluxetable*}{ccccrrccr}
\tablecaption{X-ray Sources Towards 10 Open Clusters \label{tab:xray_photometry}}
\tablewidth{0pt}
\tablehead{
\colhead{Region} & \colhead{CXOU J} & \colhead{R.A.} &
\colhead{Decl.} & $C_{net}$& \colhead{$\sigma_{net}$} & \colhead{$PFlux$} & \colhead{$ME$} &
\colhead{Group} \\
\colhead{} & \colhead{} &  \colhead{deg} &
\colhead{deg} & \colhead{cnts} & \colhead{cnts} & \colhead{ph cm$^{-2}$ s$^{-1}$} & \colhead{keV} &
\colhead{}\\
\colhead{(1)} & \colhead{(2)} & \colhead{(3)} & \colhead{(4)} & \colhead{(5)} &
\colhead{(6)} & \colhead{(7)} & \colhead{(8)} & \colhead{(9)} 
}
\startdata
NGC1502 & 040628.63+621745.1 & 61.619304 & 62.295876 & 24.3 & 6.7 & -5.640 & 3.8 & 10~~\\
NGC1502 & 040632.61+621612.3 & 61.635902 & 62.270094 & 10.4 & 5.3 & -6.012 & 4.9 & 10~~\\
NGC1502 & 040635.11+622057.9 & 61.646325 & 62.349421 & 8.0 & 3.5 & -5.677 & 2.1 & 10~~\\
NGC1502 & 040638.68+621951.4 & 61.661188 & 62.330956 & 7.9 & 4.3 & -6.153 & 2.9 & 10~~\\
NGC1502 & 040639.70+621928.8 & 61.665445 & 62.324694 & 15.8 & 5.2 & -5.863 & 1.6 & 5~~\\
NGC1502 & 040643.05+622040.0 & 61.679393 & 62.344472 & 2181.3 & 47.1 & -3.719 & 1.6 & 7~~\\
NGC1502 & 040643.15+622025.4 & 61.679799 & 62.340410 & 33.5 & 7.1 & -5.535 & 1.7 & 5~~\\
\enddata
\tablecomments{This table is available in its entirety (14,222 X-ray sources) in machine-readable form in the online journal. These {\it Chandra}-X-ray source positions and photometric quantities are provided by the {\it ACIS Extract} package. Column 1: Cluster name. Column 2: IAU designation. Columns 3-4: Right ascension and declination (in decimal degrees) for epoch J2000.0. The X-ray photometric quantities listed in Columns 5-8 are calculated in the $(0.5-8)$~keV band. Columns 5-6: Net counts and average of the upper and lower 1-$\sigma$ errors. Column 7: $\log$ of apparent photometric flux as the ratio of the net counts to the mean Auxiliary Response File value (product of the local effective area and quantum efficiency) and exposure time. Column 8: Background-corrected median photon energy. Column 9: Source class: Groups 5-10 (\S\ref{sec_membership}).}
\end{deluxetable*}

\section{Target Sample And {\it Chandra} Data Reduction} \label{sec:targets_chandra}

Our sample comprises ten rich, nearby, 7--25~Myr-old open clusters listed in Table~\ref{tab:cluster_props} in order of increasing age. They were selected from the literature based on age, proximity and richness.  The table lists refined numbers of cluster members and cluster distances, extinctions, and ages based on the analyses presented later in \S\S \ref{sec_membership} and \ref{sec:clus_mem_properties}. 

Table \ref{tab:log_chandra_observations} in Appendix \ref{sec:appendix_chandra_log_table} provides information about 37 observations of these clusters made by the {\it Chandra} X-ray Observatory. Eight clusters were specifically targeted by {\it Chandra} during observation Cycles 20-21 as part of this large GO/GTO program (PIs, Getman and Garmire). Data for two clusters, NGC 3293 and NGC 869, were obtained from the {\it Chandra} archive; previous results for these clusters were reported by \citet{Preibisch2017}, \citet{Argiroffi2016}, and \citet{Townsley2019}. 

All observations were performed using {\it Chandra}'s Advanced CCD Imaging Spectrometer imaging array \citep[ACIS-I;][]{Garmire03}. Numerous $17\arcmin \times 17\arcmin$ images from four contiguous CCD chips were aimed at the centers of these clusters. All but three observations were taken in Very Faint Timed Exposure mode; ObsIDs 9912, 9913, and 12021 were taken in Faint mode.

\begin{figure*}[ht!]
\epsscale{1.15}
\plotone{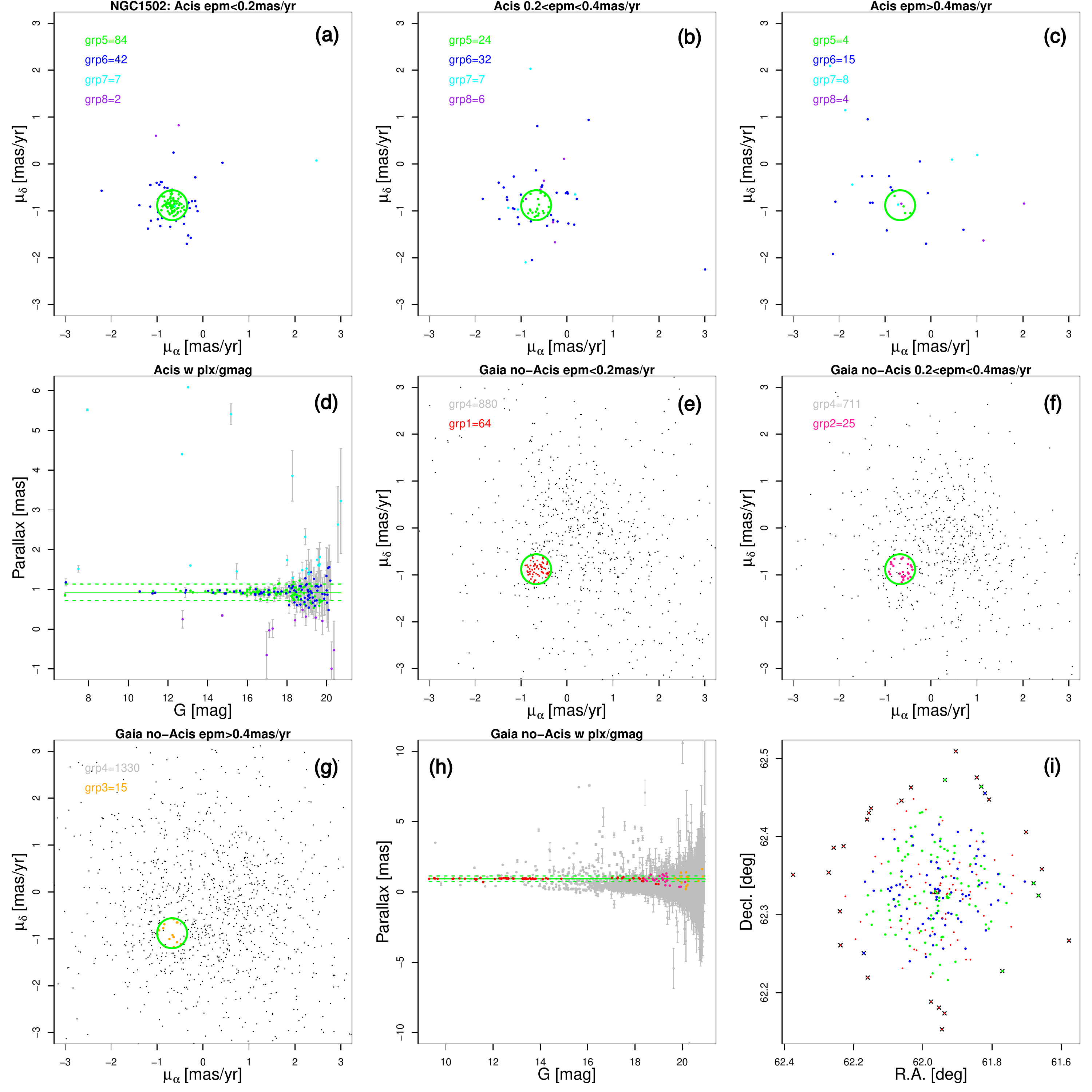}
\caption{{\it Gaia} astrometry for X-ray and non-X-ray sources towards NGC 1502. Source classes discussed in the text are color coded as: red (Group 1), pink (Group 2), orange (Group 3), grey (Group 4), green (Group 5), blue (Group 6), cyan (Group 7), and purple (Group 8). Figure legends give the number of stars in each group.  Panels a, b, c:  Proper motion diagrams for X-ray sources with increasing astrometric uncertainties: $<0.2$, 0.2--0.4 and $>0.4$~mas~yr$^{-1}$. The green circle defining the initial cluster membership is discussed in the text. Panel d: {\it Gaia} parallax as a function of magnitude for the X-ray sources. The solid and dashed lines show the median and $2 \times MADN$ spread for Group 5 sources with accurate astrometry. Panels e, f, g, h: Similar to panels a-d but for the non-X-ray sources. Panel i: Spatial distribution of Group 1, 2, 5, and 6 cluster member candidates. The $\times$ symbols mark sources, located at the edges of the {\it Chandra} ACIS-I fields, that are excluded from our mass stratified activity-age analyses. The complete figure set for all the ten open clusters is available in the full electronic version of the paper. \vspace{0.7in} \label{fig:astrometry_ngc1502}}
\end{figure*}

\begin{figure*}[ht!]
\epsscale{1.15}
\plotone{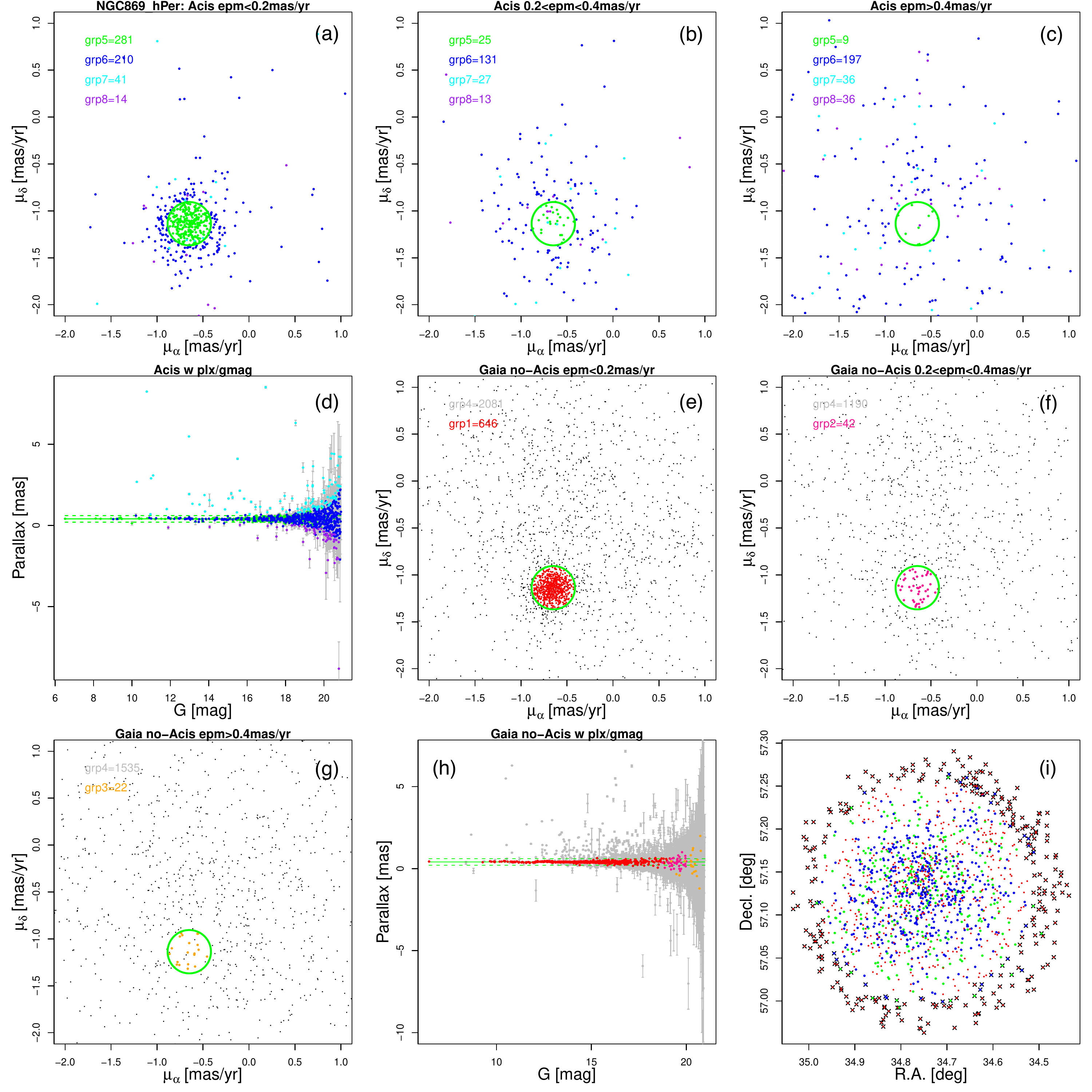}
\caption{Gaia astrometry of X-ray and non-X-ray sources towards NGC 869 = h Per. See Figure \ref{fig:astrometry_ngc1502} caption for explanation.  \label{fig:astrometry_hper}}
\end{figure*}

\begin{figure*}[ht!]
\epsscale{1.15}
\plotone{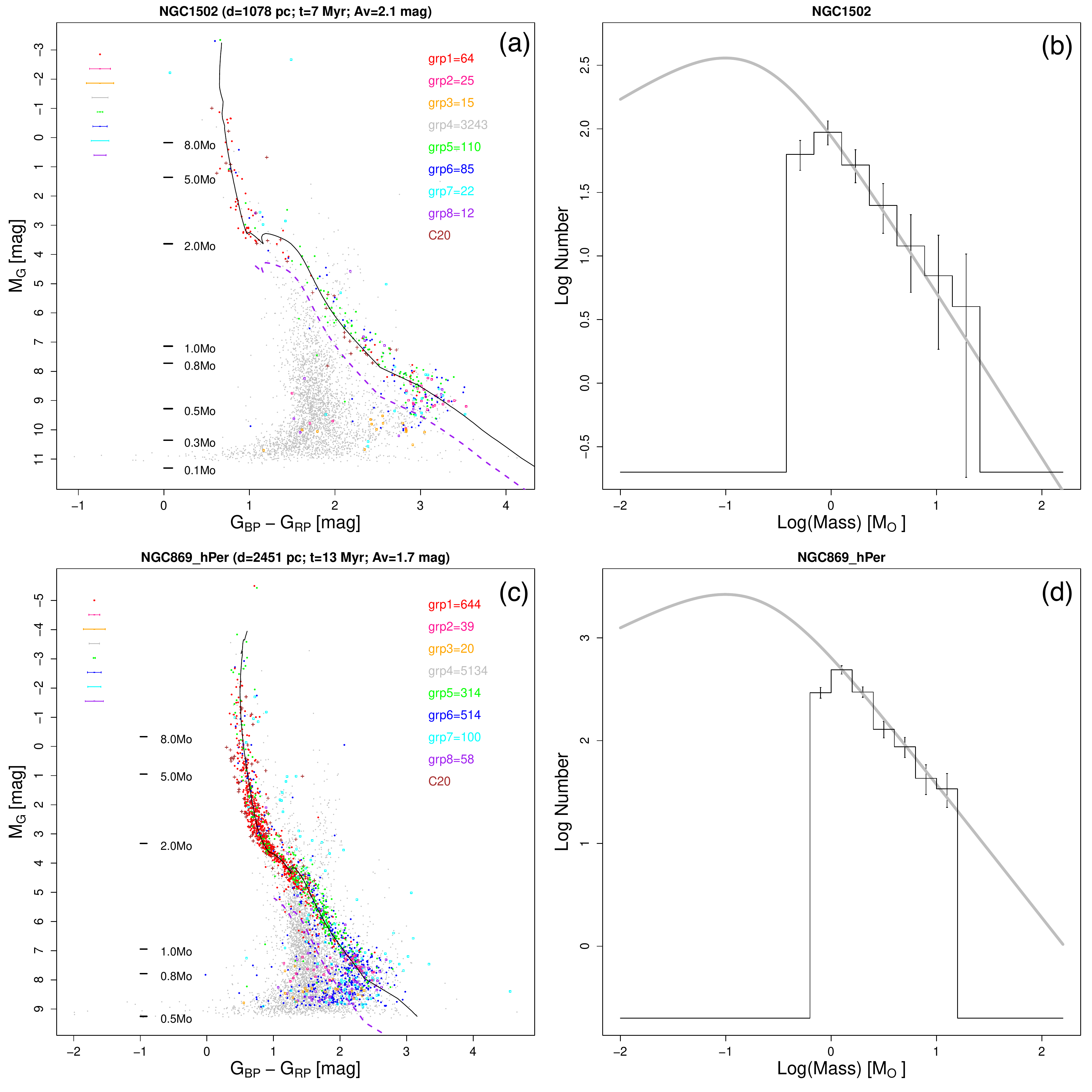}
\caption{{\it Gaia} color-magnitude diagrams and mass functions for cluster member stars in NGC 1502 (upper panels) and NGC 869 (lower panels). Panels a and c: CMDs for X-ray and non-X-ray classes.  Groups 1 through 8 are color coded as in Figure~\ref{fig:astrometry_ngc1502} and brown $+$ symbols are distant cluster members from C20.  Typical {\it Gaia} EDR3 photometric uncertainties are depicted in the left-upper corners of the figure panels. Legends provide the number of stars in each group.  The black curve represents the best fit PARSEC 1.2S evolutionary models for the Group 1, 2, 5, and 6 stars. Corresponding stellar masses are indicated by black markers. X-ray and non-X-ray stars (even from Groups 1, 2, 5 and 6) that are located to the left of the purple dashed lines are omitted from the member lists.  Panels b and d: Histogram of the mass function of member stars (Groups 1, 2, 5, and 6 stars located to the right of the purple dashed lines on the CMD panels). Histogram error bars are approximations to 95\% confidence intervals of a Poissonian distribution \citep{Gehrels86}. The gray curve shows the best-fit theoretical IMF from \citet{Maschberger2013}. The complete figure set for all the ten open clusters is available in the electronic version of the paper. \vspace*{0.5in} \label{fig:cmd_imf_ngc1502_hper}}
\end{figure*}

\begin{deluxetable*}{cccccccccccccc}
\tabletypesize{\footnotesize}
\tablecaption{Gaia-EDR3 Properties Of Non-X-ray And X-ray Sources Towards 10 Open Clusters\label{tab:gaia_photometry}}
\tablewidth{0pt}
\tablehead{
\colhead{Region} & \colhead{R.A.} &
\colhead{Decl.} & \colhead{Group} & \colhead{$\omega$} & \colhead{$\sigma_{\omega}$} & \colhead{$\mu_{\alpha\star}$} &
\colhead{$\sigma_{\mu_{\alpha\star}}$} & \colhead{$\mu_{\delta}$} &
\colhead{$\sigma_{\mu_{\delta}}$} & \colhead{$G$} & \colhead{$\sigma_{G}$} & \colhead{$B-R$} & \colhead{$\sigma_{BR}$}  \\
\colhead{} & \colhead{deg} &  \colhead{deg} & \colhead{} & \colhead{mas} & \colhead{mas} & \colhead{} & \colhead{} & \colhead{} & \colhead{} & \colhead{mag} & \colhead{mag} & \colhead{mag} & \colhead{mag}\\
\colhead{(1)} & \colhead{(2)} & \colhead{(3)} & \colhead{(4)} & \colhead{(5)} & \colhead{(6)} & \colhead{(7)} & \colhead{(8)} & \colhead{(9)} & \colhead{(10)} & \colhead{(11)} & \colhead{(12)} & \colhead{(13)} & \colhead{(14)} 
}
\startdata
NGC1502 & 61.970594 & 62.327418 & 1 & 0.955 & 0.018 & -0.870 & 0.021 & -0.791 & 0.018 & 12.864 & 0.003 & 0.909 & 0.005\\
NGC1502 & 61.960166 & 62.329657 & 1 & 0.954 & 0.014 & -0.790 & 0.017 & -0.693 & 0.015 & 12.359 & 0.003 & 0.838 & 0.005\\
NGC1502 & 61.988023 & 62.308490 & 1 & 0.997 & 0.119 & -0.430 & 0.139 & -0.725 & 0.122 & 17.922 & 0.007 & 2.506 & 0.054\\
NGC1502 & 61.931138 & 62.326292 & 1 & 0.937 & 0.015 & -0.660 & 0.018 & -0.842 & 0.015 & 10.285 & 0.003 & 0.795 & 0.005\\
NGC1502 & 61.928628 & 62.327341 & 1 & 0.955 & 0.014 & -0.720 & 0.017 & -0.946 & 0.015 & 9.545 & 0.003 & 0.739 & 0.005\\
\enddata
\tablecomments{This table is available in its entirety in the machine-readable form in the online journal. It presents 10,534 stars in Groups 1, 2, 3, 5, 6, 7, and 8, of which 7,924 are cluster members from Groups 1, 2, 5, and 6 (see the membership list in Table \ref{tab:stellar_props_open_clusters}). Column 1: Cluster name. Columns 2-3: Gaia right ascension and declination (in decimal degrees) for epoch J2000.0. Column 4: Group assignment. Columns 5-10: Gaia parallax and proper motions and their uncertainties. Proper motions are in mas yr$^{-1}$.  Columns 11-14: Gaia magnitude and color and their uncertainties.}
\end{deluxetable*}

X-ray data analysis closely followed procedures from our earlier projects studying young stellar populations \citep[e.g.,][]{Kuhn2013a, Getman17, Townsley2019}, and is briefly summarized here. The suite of {\it ACIS Extract} and related tools described by \citet{Broos10,Broos2012} provides very sensitive and reliable identification of faint X-ray sources in {\it Chandra} images. A deep catalog of candidate sources is obtained from bumps in a smoothed map of the field based on maximum likelihood deconvolution using the known spatially variable telescope point-spread function. Local background levels are iteratively calculated for each candidate source. This candidate source identification procedure is performed on the merged image from multiple exposures (Table \ref{tab:log_chandra_observations}). 

A global astrometric correction to the pointing direction is made by removing any offset between bright X-ray sources and Gaia-DR2 counterparts. This allows sub-arcsecond positional accuracy for on-axis sources. Photons are extracted for each candidate source in small regions scaled to the local point-spread function, typically containing 90\% of the expected photons. Local background rates are subtracted to give net source counts. The local exposure time is obtained from the merged exposure map. 

A variety of X-ray properties are then calculated from the extracted events: net count rate corrected for local exposure time and point-spread function tails; source locations with errors depending on off-axis angle and net count rate; probability of source existence based on Poisson distributions for the source and background photon rates; hypothesis tests for variability in the photon arrival times using a Kolmogorov–Smirnov test; apparent photomeric flux; and median energy of the net counts.

For NGC 3293 and NGC 869, the same reduction procedures were performed by \citet{Preibisch2017, Townsley2019}, and the final X-ray data products were shared with us by these authors.  

A total of 14,222 candidate X-ray sources were identified across the ten open clusters. Table \ref{tab:xray_photometry} presents their locations and X-ray photometric quantities.  These X-ray sources consist of PMS cluster members of interest here, extragalactic contaminants and Galactic field contaminants. Our classification of the sources is described in the next section.

\section{Cluster Membership} 
\label{sec_membership}

We identify members of these young clusters by combining results from the {\it Chandra} source catalog in Table~\ref{tab:xray_photometry} and results from the public Early Data Release 3 (EDR3) of the {\it Gaia} mission \citep{GaiaMission2016, GaiaEDR32021}. {\it Gaia} EDR3 offers an all-sky, deep astrometric and photometric source catalog in the visual band. All cluster members we identify have {\it Gaia} counterparts but only a fraction are detected with {\it Chandra}; thus both X-ray and non-X-ray stars can be members.  Membership is constrained to lie within the {\it Chandra} fields listed in Table~\ref{tab:log_chandra_observations} of Appendix \ref{sec:appendix_chandra_log_table}. Since these are lightly absorbed clusters without nebula background emission (unlike MYStIX and SFiNCs clusters embedded in star forming regions), a cluster census down to some limiting stellar mass can be obtained from a magnitude limited survey. Thus if contaminants can be effectively removed, the {\it Gaia}-based sample will be complete above this limiting mass.

The steps of cluster member identification are outlined here and illustrated for two clusters in Figures \ref{fig:astrometry_ngc1502}-\ref{fig:cmd_imf_ngc1502_hper}.  Similar figures for all ten clusters are available in the online figure sets. 

First, Gaia-EDR3 \citep{GaiaEDR32021} and {\it Chandra} catalog source positions (Table~\ref{tab:xray_photometry}) are cross-correlated within a constant search radius of $1\arcsec$. Most {\it Chandra} sources are localized to better than $1\arcsec$; a very small fraction of X-ray sources may miss their {\it Gaia counterpart} due to the large point spread function at the outer regions of the ACIS-I field. Out of 14,222 X-ray sources $5,027$ have {\it Gaia} counterparts.

Second, since the X-ray emission of cool members of such young open clusters is expected to be elevated $\sim 100-1000$ times above typical main sequence levels \citep{PreibischFeigelson2005}, the X-ray sources with {\it Gaia} counterparts represent a reliable subsample of cluster members.  We constrain this initial sample to stars with accurate Gaia proper motions, $\sigma_{\mu_{\alpha}} < 0.2$~mas/yr. A circle in the proper motion diagram is then drawn where these X-ray emitting stars are most concentrated as shown in Figures \ref{fig:astrometry_ngc1502}a and \ref{fig:astrometry_hper}a.  The circle, chosen to have radius $R_{cl,pm}$ as 1.3 normalized median absolute deviations (MADNs), includes $60-90$\% of such X-ray stars. This is a conservative choice as many {\it Chandra-Gaia} cluster members with less accurate proper motions lie outside the circle (panels b and c in Figures~\ref{fig:astrometry_ngc1502}-\ref{fig:astrometry_hper}).  But a small circle reduces contamination of non-cluster members in the {\it Gaia} survey.  

Third, we estimate the distance to each cluster.  The parallax of the cluster is chosen to be the median parallax of this initial member sample, and the distance is the inverse of this median. None of the individual stellar parallaxes are negative so the median of parallaxes or inverse parallaxes (stellar distances) produce the same results. The uncertainty of the distance is obtained from the 68\% quantiles of a bootstrap resampling of parallaxes.  These distances and uncertainties are listed in Table~\ref{tab:cluster_props}.   

These distances are highly reliable.  \citet{Getman2019} show that for different parts of the Orion A cloud, such simple distance estimates are consistent within 1\% with the probabilistic distance estimates by \citet{Bailer-Jones2018}. Cluster distances inferred from weighted median and mean parallaxes \citep{Kuhn19} lie within the 95\% confidence intervals of our values. Stars with {\it Gaia} re-normalized unit weight errors (RUWE) in excess of 1.4 (a threshold considered as a possible indicator of poor Gaia astrometric solution) represent only $\lesssim 6$\% in these stellar samples and do not affect our cluster distance estimates. 

In addition to the uncertainty of the median parallax, we define a spread of parallax values about the median to be two normalized median absolute deviations of these values, $PlxSpread = 2 \times MADN$.  This spread is used below for identification of additional members, and is shown as a band of dashed lines in panels d and h of Figures \ref{fig:astrometry_ngc1502}-\ref{fig:astrometry_hper}.
 
We now classify all {\it Chandra-Gaia} and non-{\it Chandra-Gaia} sources located within the {\it Chandra} fields into the following 10 groups.  These are visually shown using different color symbols in Figures~\ref{fig:astrometry_ngc1502}-\ref{fig:astrometry_hper}.  Group assignments for X-ray sources are listed in Table~\ref{tab:xray_photometry}. 

Groups 1 (red), 2 (pink), and 3 (orange): These are non-{\it Chandra Gaia} sources with proper motion uncertainties of $\sigma_{\mu_{\alpha}} < 0.2$~mas/yr, $0.2 < \sigma_{\mu_{\alpha}} < 0.4$~mas/yr, and  $\sigma_{\mu_{\alpha}} > 0.4$~mas/yr, respectively whose proper motion values lie inside the $R_{cl,pm}$ circle and their parallax confidence intervals intercept the $PlxSpread$ band (panels e-h). After inspection of their photometric properties (e.g., Figure \ref{fig:cmd_imf_ngc1502_hper}), we find that the Group 1 stars are typically higher mass ($>1-2$~M$_{\odot}$) cluster members and many Group 2 stars are lower mass ($<1$~M$_{\odot}$) cluster members.  Group 3 stars include both low mass ($<0.5-0.8$~M$_{\odot}$) cluster member candidates and foreground and background contaminants. 

Group 4 (gray): This contains the remaining non-{\it Chandra Gaia} stars within the {\it Chandra} fields (panels e-h). These are mostly field stars unrelated to the open clusters. 

Groups 5 (green) and 6 (blue): These are {\it Chandra-Gaia} stars with their parallax confidence intervals intercepting the $PlxSpread$ band and their proper motion positions lying inside (Group 5) and outside (Group 6) the $R_{cl,pm}$ circle, shown in panels a-d. {\it Gaia} photometry indicates that the Group 5 objects are mainly cluster members with masses $>0.5$~M$_{\odot}$ and the Group 6 includes both $<2$~M$_{\odot}$ cluster members and stellar contaminants.

Groups 7 (cyan) and 8 (purple): {\it Chandra-Gaia} stars with their lower (upper) error-bar on parallax being above (below) the $PlxSpread$ band (panels h and d) for Group 7 (Group 8). The Group 7 is predominantly foreground stars and the Group 8 is predominantly background stars.

Groups 9 and 10. Group 9 comprises {\it Chandra-Gaia} sources without astrometry information. Some may be very low-mass cluster members and some may be contaminants. Group 10 incorporates {\it Chandra} sources without {\it Gaia} counterparts. The vast majority of these are extragalactic contaminants or spurious (noise) X-ray sources.  These groups are not shown in the figures.

Table \ref{tab:gaia_photometry} presents {\it Gaia} astrometric and photometric properties for non-X-ray (groups 1, 2, and 3) and X-ray (groups 5, 6, 7, and 8) sources towards the ten open clusters. Among those, Groups 1, 2, 5, and 6 contain the vast majority of cluster member candidates. 

\begin{deluxetable*}{cccccccccccc}
\tabletypesize{\footnotesize}
\tablecaption{Stellar Properties Of Non-X-ray And X-ray Members Of 10 Open Clusters\label{tab:stellar_props_open_clusters}}
\tablewidth{0pt}
\tablehead{
\colhead{Region} & \colhead{R.A.} &
\colhead{Decl.} & \colhead{Group} &  \colhead{$\log(T_{eff})$} &
\colhead{$M$} & \colhead{$\log(L_{bol})$} &
\colhead{$\log(L_{X})$} & \colhead{$\log(L_{X,up})$} & \colhead{F1} & \colhead{Ph} & \colhead{F2}\\
\colhead{} & \colhead{deg} &  \colhead{deg} & \colhead{} & \colhead{K} & \colhead{$M_{\odot}$} & \colhead{$L_{\odot}$} & \colhead{erg s$^{-1}$} & \colhead{erg s$^{-1}$} & \colhead{} & \colhead{HRD} & \colhead{}\\
\colhead{(1)} & \colhead{(2)} & \colhead{(3)} & \colhead{(4)} & \colhead{(5)} & \colhead{(6)} & \colhead{(7)} & \colhead{(8)} & \colhead{(9)} & \colhead{(10)} & \colhead{(11)} & \colhead{(12)}
}
\startdata
NGC1502 & 62.257371 & 62.385897 & 2 & 3.54 & 0.55 & -0.82 & \nodata & 29.54 & 1 & 1 & 0\\
NGC1502 & 61.944486 & 62.153143 & 1 & 4.04 & 2.40 & 1.55 & \nodata & 29.76 & 1 & 5 & 0\\
NGC1502 & 61.577576 & 62.266908 & 2 & 3.57 & 0.71 & -0.62 & \nodata & 29.80 & 1 & 1 & 0\\
NGC1502 & 61.904523 & 62.509605 & 1 & 3.55 & 0.63 & -0.72 & \nodata & 29.77 & 1 & 1 & 0\\
NGC1502 & 62.374648 & 62.351225 & 2 & 3.54 & 0.54 & -0.83 & \nodata & 29.73 & 1 & 1 & 0\\
NGC1502 & 61.665573 & 62.324773 & 5 & 3.69 & 1.25 & 0.02 & 29.91 & \nodata & 1 & 2 & 0\\
NGC1502 & 61.679861 & 62.340361 & 5 & 3.72 & 1.40 & 0.21 & 30.24 & \nodata & 1 & 2 & 0\\
NGC1502 & 61.758327 & 62.315094 & 5 & 3.66 & 1.05 & -0.20 & 29.80 & \nodata & 0 & 2 & 0\\
NGC1502 & 61.764918 & 62.335727 & 6 & 3.54 & 0.54 & -0.83 & 29.42 & \nodata & 0 & 1 & 0\\
NGC1502 & 61.769816 & 62.227697 & 5 & 3.75 & 1.55 & 0.45 & 30.49 & \nodata & 1 & 2 & 0\\
\enddata
\tablecomments{This table is available in its entirety (7,924 cluster members) in the machine-readable form in the online journal. Only sources from Groups 1, 2, 5, and 6 with available mass estimates are presented. And only lower-mass stars with colors redder than the purple dashed lines in Figure \ref{fig:cmd_imf_ngc1502_hper}a,c are included. Out of the total 7,924 cluster members, 6,003 are located in the central parts of the clusters (Column 10 flag $=0$), and are included in the science analysis of the X-ray-mass-age relations (\S\S \ref{sec:xrays_vs_mass}-\ref{sec:xray_activity_slopes} and Appendix \ref{sec:temporal_highest_masses}). Column 1: Cluster name. Columns 2-3: Gaia right ascension and declination (in decimal degrees) for epoch J2000.0. Column 4: Source class: groups 1, 2 (non-X-ray stars), and groups 5, 6 (X-ray stars). Columns 5-7: Stellar effective temperature, mass, and bolometric luminosity derived from the Gaia color-magnitude diagrams (Figure \ref{fig:cmd_imf_ngc1502_hper}a,c). Columns 8-9: X-ray luminosity (for X-ray members; groups 5 and 6) and upper limits to X-ray luminosity (for non-X-ray members; groups 1 and 2). Column 10: A flag indicating whether the star is located inside ($=0$) or outside ($=1$) the spatial circle around the cluster center depicted in Figures \ref{fig:astrometry_ngc1502}i and \ref{fig:astrometry_hper}i. Column 11: The evolutionary status of the star on the Hertzsprung–Russell diagram based on the predictions of the PARSEC 1.2S evolutionary models: $=1$ - Hayashi track; $=2$ or $=3$ - Henyey track; $\geq 4$ - ZAMS and beyond. Column 12: A flag indicating sources with inconsistency ($=1$) between the {\it Gaia} $G$-band, and BP and RP photometry. There are 595 stars with F2$=1$; of those 485 are located within the central cluster regions, i.e., have F1$=0$.}
\end{deluxetable*}

Due to the degradation of the {\it Chandra} ACIS-I point source sensitivity with increasing distance from the detector center the fraction of X-ray detections of cluster members decreases considerably in the outer regions of the detector. For this reason, the cluster member candidates lying outside of a $R=7.5\arcmin$ circle around the cluster centers ($\times$ symbols in Figures \ref{fig:astrometry_ngc1502}i and \ref{fig:astrometry_hper}i) for all but one (NGC~2232) clusters will be ignored from the analyses of mass-stratified activity-age relations presented later in \S\S \ref{sec:xrays_vs_mass}-\ref{sec:xray_activity_slopes} and Appendix \ref{sec:temporal_highest_masses}. The outer members will be retained in the case of NGC~2232 because this nearest region offers the best point source X-ray sensitivity among our open clusters (see Column 9 in Table \ref{tab:cluster_props}). 

In a final step to reduce contamination by field stars, we require that cluster members lie close to the PMS and main sequence isochrones on the {\it Gaia} color-magnitude diagram (Figure \ref{fig:cmd_imf_ngc1502_hper}).  We fit Group 1, 2, 5, and 6 stars with the PARSEC 1.2S evolutionary models \citep{Bressan12, Chen14} at the same time omitting faint objects with blue colors as possible stellar contaminants. These are objects located to the left of the dashed purple curves depicted in panels a and c of Figure \ref{fig:cmd_imf_ngc1502_hper}. These purple curves are chosen to follow the best-fit isochrones shifted towards bluer colors by the typical confidence interval for the colors of faint sources. 

The result of these selection criteria is a sample of 7,924 young stars (from Groups 1, 2, 5, and 6) in the ten clusters with ages 7--25~Myr. These stars are listed in Table~\ref{tab:stellar_props_open_clusters}, whose content is described below. Omitting cluster members lying outside the spatial circles in Figures \ref{fig:astrometry_ngc1502}i and \ref{fig:astrometry_hper}i gives a sub-sample of 6,003 cluster members further employed in the analyses of mass-stratified activity-age relations (\S\S \ref{sec:xrays_vs_mass}-\ref{sec:xray_activity_slopes} and Appendix \ref{sec:temporal_highest_masses}).

\section{Cluster and member properties} 
\label{sec:clus_mem_properties}

Cluster absorption and ages are estimated as follows.  The extinction coefficients from \citet{Luhman2020} are used to redden theoretical PARSEC 1.2S isochrones on the {\it Gaia} color-magnitude diagram for a range of absorptions and ages. These curves are fitted to the {\it Gaia} EDR3 photometry data of the cluster members using minimum chi-squared estimation weighted with the $G_{BP} - G_{RP}$ errors. The CMD positions of the brightest stars, typically with $M>2$~M$_{\odot}$, are more sensitive to the changes in the cluster extinctions while the positions of the lower-mass stars are more sensitive to the changes in the cluster ages.

The inferred cluster distances (\S \ref{sec_membership}), average extinctions in visual band, cluster ages, and numbers of the cluster members $-$  Group 1, 2, 5, and 6 sources located to the right of the dashed purple lines on the CMD diagrams (Figure \ref{fig:cmd_imf_ngc1502_hper}a,c) $-$ are summarized in Table \ref{tab:cluster_props}. Galactic coordinates for the cluster centers are also provided.

Effective temperatures, bolometric luminosities, and masses for individual cluster members are obtained through a comparison of the observed $G$-band absolute magnitudes with theoretical predictions for the best-fit reddened PARSEC 1.2S isochrone. Table \ref{tab:stellar_props_open_clusters} lists these {\it Gaia}-based stellar properties together with flags indicating source's proximity to the cluster centers.  An additional flag indicates the evolutionary phase of each star:  e-PMS Hayashi track, l-PMS  Henyey track, or ZAMS.  The boundaries between these phases are shown as large symbols in Figure~\ref{fig:lbol_r_vs_time}.

According to \citet{Riello2021}, the ${BP}$-band flux of some faint Gaia-EDR3 objects ($G_{BP} > (20.3-20.9)$~mag) may be overestimated. Because of this effect our cluster member lists may miss true very low mass young stars with their CMD positions shifted to the locus of background stars (Figure~\ref{fig:cmd_imf_ngc1502_hper}). Among the 7,924 open cluster members (Table \ref{tab:stellar_props_open_clusters}), 13\% and 5\% have $G_{BP}$ magnitudes in excess of 20.3 and 20.9, respectively. Ignoring these stars from the CMD fitting does not change the inferred cluster ages. The mass and bolometric luminosity estimates of such stars are not affected since these are based on the $G$-band magnitudes only. 

Following \citet{Riello2021}, we also checked our open cluster members with high EDR3 blend fraction ($\beta > 0.1$) for possible contamination from nearby sources. Among 7,924 cluster members, there are only 68 cases of ``target'' stars with $\beta > 0.1$ lying near (within 1.05$\arcsec$) a potential ``blending'' Gaia-EDR3 source, where $F_{G,target}/F_{G,blending} < 4$. Vast majority of such stars are members of two distant and rich clusters, NGC~4755 and NGC~3766. Ignoring such stars from the analysis of X-ray evolution (\S \ref{sec:xray_activity_slopes}) does not affect any science results.

\citet{GaiaDR22018, Riello2021} suggest that high excess in the sum of the Gaia BP-band and RP-band fluxes relative to the G-band flux may indicate that star’s photometry is susceptible to the effects of source crowding and/or nebulosity. We find that 7\% of our 7,924 open cluster members have their corrected BP and RP flux excesses as $|C^{\star}| > 5 \sigma_{C^{\star}}$ \citep{Riello2021,Anders2022}, indicating inconsistency between the G-band, and BP and RP photometry. In Table~\ref{tab:stellar_props_open_clusters}, such stars can be identified through the selection $F2=1$. We verified that these stars affect neither the CMD fitting nor X-ray evolution results (\S \ref{sec:xray_activity_slopes}). 

See Appendix C of \citet{Getman2021} and \S \ref{sec:mystix_sfincs} for details on applying PARSEC 1.2S models to derive MYStIX/SFiNCs e-PMS stellar properties. 

Once individual masses are estimated, they can be considered as an ensemble to examine the Initial Mass Function (IMF).  This can give confidence that our cluster memberships are complete above a limiting mass. IMF histograms for NGC~1502 and NGC~869 are shown in panels b and d of Figure \ref{fig:cmd_imf_ngc1502_hper}.  They are fitted to \citet{Maschberger2013} generalized log-log formulation of the IMF.  Our membership analysis delivers cluster member mass distributions consistent with the power law tail of Maschberger's IMF. To obtain mass completeness limits, starting from low masses of 0.2~M$_{\odot}$, mass cutoffs are iteratively increased until the Anderson-Darling goodness-of-fit test shows statistically acceptable fits ($p > 0.05$) between the unbinned c.d.f.'s for the data and Maschberger's model. The mass completeness limit for each cluster is listed in Table~\ref{tab:cluster_props}; these are generally around $0.4-0.7$~M$_\odot$ but are higher for two distant and more absorbed clusters, NGC~869 and NGC~884.  Our cluster memberships represent roughly 20\% of the total intrinsic cluster population and 80\% of the cluster mass.

Finally, we note that these clusters were subject to previous membership analysis with a multi-stage statistical techniques applied to the {\it Gaia} DR2 \citep{GaiaDR22018} catalog without considering any {\it Chandra} sources \citep[][hereafter C20] {CantatGaudinAnders2020,Cantat-Gaudin2020}.  All but a handful of C20 members located inside the {\it Chandra} ACIS-I fields have counterparts belonging to our Groups 1, 2, 5, and 6. Those without counterparts (none or a few C20 members per cluster) generally have astrometry discrepancies between the {\it Gaia} DR2 and EDR3 catalogs\footnote{See, for instance, a DR2-EDR3 parallax discrepancy in a NGC 1502 at $(\alpha,\delta) = (61.937884, 62.317771)$.}. Due to the improved accuracy of the Gaia-EDR3 data and availability of {\it Chandra}-X-ray data, our member lists provide numerous new low-mass cluster members that were not located by C20.  Our catalogs increase the cluster census within the {\it Chandra} fields by factors $1.6$ (NGC 2232), $(2-3)$ (NGC 1502, NGC 2169, NGC 1960, IC 2395), and  $(3-6)$ for most distant clusters (NGC 3293, NGC 884, NGC 3766, NGC 4755, NGC 869). C20 members which reside outside the {\it Chandra} fields, ranging from a dozen to a hundred stars per cluster, are added to the Gaia CMDs in Figure \ref{fig:cmd_imf_ngc1502_hper}a,c (brown $+$). Their color-magnitude positions are in complete agreement with those of our Groups 1, 2, 5, and 6.

\section{X-ray Luminosities and Upper Limits}
\label{sec:Xray_luminosities}

For younger members of MYStIX/SFiNCs star forming regions in the e-PMS phase (\S \ref{sec:mystix_sfincs}), intrinsic X-ray luminosities were derived using the non-parametric method XPHOT \citep{Getman10} based on a concept that is similar to the long-standing use of color–magnitude diagrams in optical and infrared astronomy. However, XPHOT requires the presence of both soft ($<2$~keV) and hard ($>2$~keV) X-ray photons. The members of the ten l-PMS open clusters considered here are generally lightly absorbed and older than the MYStIX/SFiNCs stars with lower flare activity. They thus often lack hard X-ray photons. The XPHOT method could be applied to roughly half of the cluster members.  

\begin{deluxetable*}{cccccccccc}
\tabletypesize{\small}
\tablecaption{X-ray Spectral Fits Of Stacked Data  \label{tab:xspec_fits}}
\tablewidth{0pt}
\tablehead{
\colhead{Sample} & \colhead{$ME$~range} &
\colhead{$NC$} & \colhead{$\chi^{2}_\nu$} & \colhead{dof} & \colhead{$N_{H}$} & \colhead{$kT_2$} & \colhead{$EM_{1}$} & \colhead{$EM_2$} &  \colhead{Flux}\\
\colhead{} & \colhead{} &
\colhead{} & \colhead{} & \colhead{} & \colhead{$10^{22}$} & \colhead{} & \colhead{$10^{52}$} & \colhead{$10^{52}$} &  \colhead{$10^{-15}$}\\
\colhead{} & \colhead{(keV)} & \colhead{(cnts)} & \colhead{} & \colhead{} & \colhead{(cm$^{-2}$)} & \colhead{(keV)} &  \colhead{(cm$^{-3}$)} & \colhead{(cm$^{-3}$)} & \colhead{(erg~s$^{-1}$~cm$^{-2}$)}\\
\colhead{(1)} & \colhead{(2)} & \colhead{(3)} & \colhead{(4)} & \colhead{(5)} & \colhead{(6)} &
\colhead{(7)} & \colhead{(8)} & \colhead{(9)} &
\colhead{(10)}
}
\startdata
NGC1502\_S1 & $[0.9-1.5]$ & 1278 & 1.5 & 57 & 0.42 & $1.6\pm0.2$ & $ 0.12 \pm 0.06 $ & $ 10.79 \pm 2.55 $ & 6.88\\
NGC1502\_S2 & $[1.5-1.7]$ & 1283 & 0.7 & 56 & 0.42 & $2.8\pm0.8$ & $ 3.18 \pm 3.14  $ & $ 10.43 \pm 5.90 $ & 9.49\\
NGC1502\_S3 & $[1.7-3.4]$ & 1263 & 1.1 & 55 & 0.42 & $4.3\pm0.6$ & $ 0.04 \pm 0.03 $ & $ 11.96 \pm 1.61 $ & 8.61\\
NGC2169\_S1 & $[1.0-1.5]$ & 874  & 0.9 & 34 & 0.14 & $2.0\pm0.5$ & $ 7.57 \pm 6.15 $ & $ 7.90 \pm 3.28 $ & 7.65\\
NGC2169\_S2 & $[1.5-1.6]$ & 944  & 1.2 & 37 & 0.14 & $6.0\pm3.1$ & $ 27.22 \pm 10.79 $ & $ 7.60 \pm 3.43 $ & 18.65\\
NGC2169\_S3 & $[1.6-1.8]$ & 633  & 1.4 & 23 & 0.14 & $3.2\pm0.5$ & $ 0.01 \pm 0.00 $ & $ 15.12 \pm 2.86 $ & 11.78\\
IC2395\_S1 & $[1.0-1.5]$ & 1468 & 0.9  & 61 & 0.10 & $2.3\pm0.5$ & $ 1.02 \pm 0.62 $ & $ 4.55 \pm 1.80 $ & 8.19\\
IC2395\_S2 & $[1.5-1.6]$ & 1512 & 0.9  & 65 & 0.10 & $2.9\pm0.6$ & $ 11.91 \pm 11.90 $ & $ 20.33 \pm 4.68 $ & 37.13\\
IC2395\_S3 & $[1.6-4.2]$ & 1414 & 1.0  & 65 & 0.10 & $6.2\pm2.6$ & $ 6.35 \pm 3.06 $ & $ 3.91 \pm 2.42 $ & 11.73\\
NGC3293\_S1 & $[0.7-1.3]$ & 1375 & 1.4 & 53 & 0.18 & $5.0\pm1.5$ & $ 17.63 \pm 3.80 $ & $ 0.18 \pm 0.06 $ & 2.08\\
NGC3293\_S2 & $[1.3-1.5]$ & 1381 & 1.0 & 59 & 0.18 & $6.6\pm3.4$ & $ 10.80 \pm 8.74 $ & $ 6.24 \pm 2.24 $ & 2.83\\
NGC3293\_S3 & $[1.5-5.3]$ & 1354 & 1.0 & 64 & 0.18 & $6.6\pm2.0$ & $ 1.96 \pm 1.86 $ & $ 11.71 \pm 4.59 $ & 2.43\\
NGC869\_S1 & $[0.8-1.3]$ & 6726 &  1.0 & 195 & 0.34 & $2.5\pm0.5$ & $ 17.81 \pm 2.66 $ & $ 3.27 \pm 1.54 $ & 1.63\\
NGC869\_S2 & $[1.3-1.5]$ & 6713 &  1.1 & 195 & 0.34 & $3.1\pm0.4$ & $ 8.68 \pm 4.29 $ & $ 12.56 \pm 2.83 $ & 2.23\\
NGC869\_S3 & $[1.5-5.6]$ & 6626 &  1.2 & 228 & 0.34 & $6.8\pm1.4$ & $ 2.70 \pm 2.55 $ & $ 11.16 \pm 1.42 $ & 2.08\\
NGC884\_S1 & $[0.9-1.6]$ & 2102 &  1.2 & 105 & 0.34 & $2.7\pm0.6$ & $ 7.36 \pm 6.89 $ & $ 2.05 \pm 1.89 $ & 1.69\\
NGC884\_S2 & $[1.6-1.8]$ & 2100 &  1.0 & 109 & 0.34 & $2.9\pm0.4$ & $ 0.20 \pm 0.10 $ & $ 19.88 \pm 5.46 $ & 2.59\\
NGC884\_S3 & $[1.8-5.8]$ & 2090 &  1.1 & 125 & 0.34 & $7.1\pm2.3$ & $ 0.02 \pm 0.01 $ & $ 5.13 \pm 4.43 $ & 1.99\\
NGC4755\_S1 & $[0.6-1.5]$ & 2688 & 1.3 & 111 & 0.24 & $3.7\pm1.7$ & $ 10.40 \pm 3.68 $ & $ 1.86 \pm 1.09 $ & 1.73\\
NGC4755\_S2 & $[1.5-1.7]$ & 2675 & 0.8 & 115 & 0.24 & $3.7\pm0.7$ & $ 4.76 \pm 4.66 $ & $ 12.46 \pm 2.32 $ & 2.91\\
NGC4755\_S3 & $[1.7-6.6]$ & 2660 & 1.2 & 148 & 0.24 & $17.3\pm8.1$ & $ 4.18 \pm 2.90 $ & $ 7.38 \pm 1.61 $ & 2.33\\
NGC1960\_S1 & $[0.8-1.4]$ & 1101 & 1.1 & 47 & 0.16 & $2.4\pm1.1$ & $ 6.05 \pm 2.96 $ & $ 1.47 \pm 1.19 $ & 3.02\\
NGC1960\_S2 & $[1.4-1.7]$ & 1088 & 1.1 & 44 & 0.16 & $2.9\pm0.7$ & $ 1.18 \pm 1.16 $ & $ 6.80 \pm 2.13 $ & 3.99\\
NGC1960\_S3 & $[1.7-5.8]$ & 1030 & 1.3 & 47 & 0.16 & $5.6\pm3.2$ & $ 1.39 \pm 0.66 $ & $ 4.94 \pm 2.36 $ & 4.56\\
NGC3766\_S1 & $[0.6-1.4]$ & 3218 & 1.2 & 124 & 0.16 & $3.6\pm0.7$ & $ 14.28 \pm 1.65 $ & $ 0.32 \pm 0.08 $ & 1.84\\
NGC3766\_S2 & $[1.4-1.6]$ & 3208 & 1.0 & 140 & 0.16 & $3.9\pm0.9$ & $ 12.55 \pm 4.20 $ & $ 5.91 \pm 2.23 $ & 2.37\\
NGC3766\_S3 & $[1.6-6.5]$ & 3123 & 1.1 & 182 & 0.16 & $10.3\pm2.5$ & $ 0.01 \pm 0.01 $ & $ 7.15 \pm 0.63 $ & 1.82\\
NGC2232\_S1 & $[0.9-1.4]$ & 302 & 1.8 & 9 & 0.04 & $2.8\pm4.9$ & $ 3.55 \pm 3.54 $ & $ 1.02 \pm 0.64 $ & 27.27\\
NGC2232\_S2 & $[1.4-1.5]$ & 293 & 1.4 & 9 & 0.04 & $2.0\pm0.3$ & $ 0.30 \pm 0.20 $ & $ 4.58 \pm 3.39 $ & 86.96\\
NGC2232\_S3 & $[1.5-2.1]$ & 211 & 0.9 & 10 & 0.04 & $2.6\pm0.4$ & $ 0.33 \pm 0.33 $ & $ 4.18 \pm 0.58 $ & 38.57\\
\enddata
\tablecomments{Column 1: Cluster name and spectral stratum id. Columns 2: Range of X-ray source median energies. Column 3: Total number of net (background-corrected) X-ray counts in a spectrum. Columns 4-5: Reduced $\chi^{2}$ for the overall spectral fit and degrees of freedom. Column 6: Fixed value of cluster's X-ray column density.  Column 7: Inferred temperature of the hot plasma component and its 1~$\sigma$ error. Columns 8-9: Inferred emission measures and their 1~$\sigma$ errors for each plasma component. Column 10: Inferred absorption corrected incident X-ray flux in the 0.5--8~keV band.}
\end{deluxetable*}

For all X-ray detected cluster members (regardless the presence/absence of XPHOT estimates), we obtain factors converting the apparent X-ray photometric fluxes given in Table \ref{tab:xray_photometry} to intrinsic X-ray luminosities by assuming the stars have similar spectra. Individual X-ray spectra are stacked in three energy bands and the stacked spectra are fitted with optically thin thermal plasma models. Stacked luminosities are then obtained using Gaia-based cluster distances (Table \ref{tab:cluster_props}). Individual stellar X-ray luminosities are then estimated from their contributed fraction of the stacked spectrum.  Only young stars with X-ray net counts $<500$ counts belonging to the Groups 5 and 6 (Table \ref{tab:xray_photometry}) are included in the spectral stacking. 

\begin{figure*}[ht!]
\epsscale{1.15}
\plotone{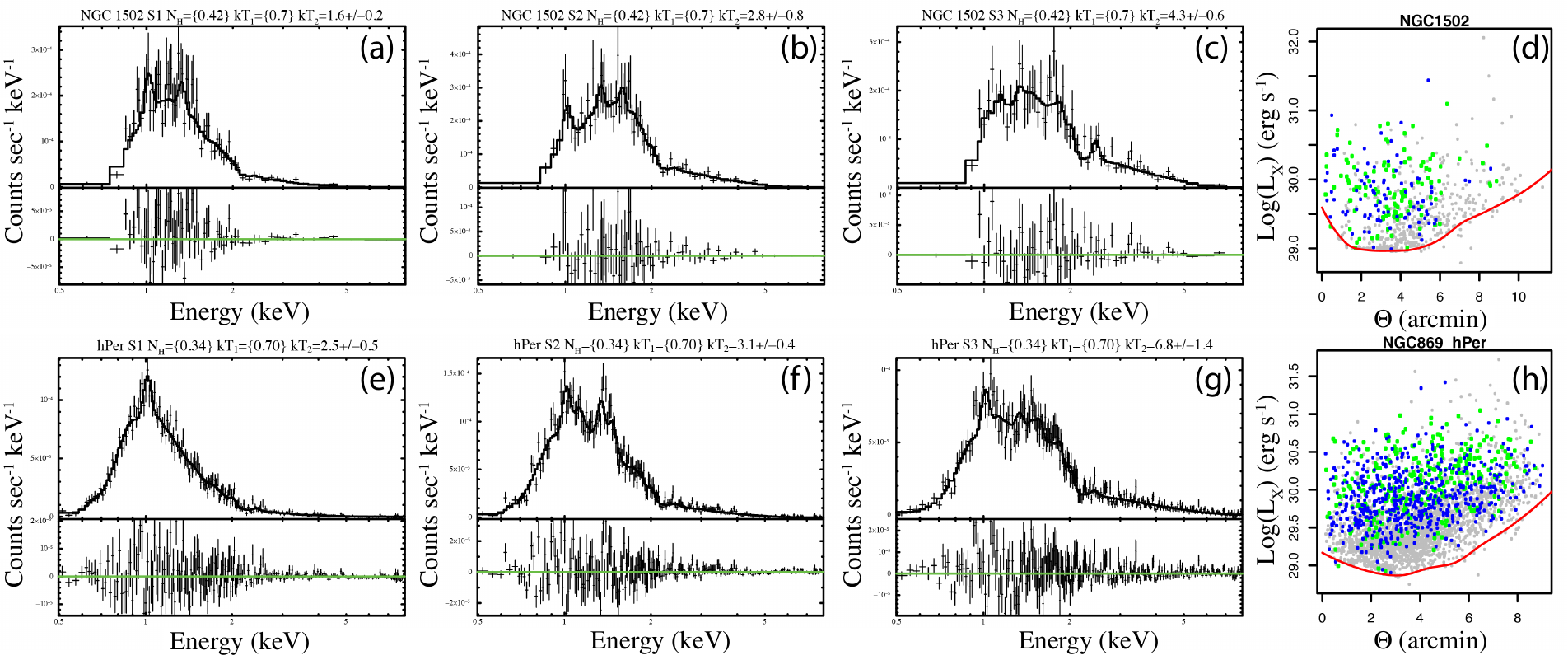}
\caption{Fitting plasma models to X-ray spectra to derive  X-ray luminosities for the NGC 1502 and NGC 869 clusters. Panels a, b, and c: Spectral fitting of the stacked X-ray spectra of Group 5 and 6 cluster members that have soft, intermediate, and hard hardness ratios. Panels e, f, and g: Similar for NGC 869 members. The hot plasma components have typical temperatures around 2~keV, 3~keV and 4-7~keV, respectively, in addition to a cool plasma component with $0.7$~keV.  Panels d and h: Cluster member X-ray luminosities as a function of angular distance from the cluster center.  The local regression fit to the lower envelope is used to estimate X-ray upper limits to undetected cluster members. Group 5 members are shown in green, Group 6 in blue, and Groups 7, 8, 9, and 10 (mostly extragalactic contaminants) in gray.  The complete figure set for the ten open clusters is available in the  electronic version of the paper. \label{fig:xray_spectra}} 
\end{figure*}

The sources' X-ray median energy (Table \ref{tab:xray_photometry}) is employed to separate the stars into three energy bands. The choice of the three energy bands varies from cluster to cluster with the goal to accumulate similar numbers of X-ray counts per merged spectrum. Individual X-ray spectra are merged and grouped according to the procedure described in the {\it ACIS Extract} User Guide\footnote{ \url{http://personal.psu.edu/psb6/TARA/ae_users_guide.pdf}.}. 

Spectral fitting is performed with the {\it XSPEC} package \citep{Arnaud1996}. Grouped spectra were fitted using the $\chi^2$ statistic with two-temperature VAPEC plasma emission models \citep{Smith2001} subject to TBABS absorption \citep{Wilms2000}. We assumed the following typical coronal elemental abundances for PMS or extremely active stars \citep{Gudel2007b}. These are defined as relative to the solar photospheric abundances from \citet[][as default abundance table in XSPEC]{AndersGrevesse1989}: C = 0.45, N =
0.788, O = 0.426, Ne = 0.832, Mg = 0.263, Al = 0.5, Si = 0.309, S = 0.417, Ar = 0.55, Ca = 0.195, Fe = 0.195, Ni = 0.195. Occasionally, elemental abundances were varied to fit prominent spectral lines. 

Based on the spectral analyses of the {\it Chandra} and {\it EUVE} data for very young ($t \sim (1-3)$~Myr) ONC and much older ($t > (0.06-1)$~Gyr) nearby stars, \citet{Preibisch05,Sanz-Forcada2003} suggest the presence of a ``fundamental'' coronal structure with plasma temperature of around $8-10$~MK. Assuming that such a ``fundamental'', high-density, low-temperature coronal structure exists in PMS stars as a common feature of coronally active stars, we fix the cool temperature component at a similar value of 0.7~keV ($8$~MK).

The X-ray column density was fixed at the average cluster extinction value (Table \ref{tab:cluster_props}) assuming a gas-to-dust ratio from \citet{Zhu2017}.

Table~\ref{tab:xspec_fits} lists the spectral fit results. For each stacked spectrum, it gives the range of source's X-ray median energy, total number of X-ray net counts, quality of the overall fit, column density, temperature of the hot plasma component, emission measures, and incident absorption corrected X-ray flux for the overall stacked spectrum. This flux is derived using the XSPEC command {\it flux} assuming zero absorption. 

The total intrinsic luminosity for the overall stacked spectrum ($L_{X,tot}$) is the product of such flux and $4 \pi D^2$, where $D$ is the distance to the cluster from the Sun.  The sum of the apparent photomeric fluxes ($PFlux$; Table~\ref{tab:xray_photometry}) for all the sources, whose data are included in the stacked spectrum, gives the total apparent flux for the overall stacked spectrum ($PFlux_{tot}$). The X-ray luminosity of an individual source ($L_{X,i}$) included in the stacked spectrum is then estimated as the product of its apparent photometric flux from Table~\ref{tab:xray_photometry} and the conversion factor $L_{X,tot}/PFlux_{tot}$.

Figure \ref{fig:xray_spectra} illustrates the fitted X-ray spectra for NGC 1502 and NGC 869. The resulting inferred stellar X-ray luminosities are listed in Table \ref{tab:stellar_props_open_clusters}. No biases are found upon comparison of the X-ray luminosities obtained from the spectral stacking and XPHOT procedures.

The ratio of the ``hot'' to ``cool'' emission measures does not depend on cluster age, but indicates the dominance of the hot plasma in the majority of the S2 (medium hardness) and S3 (high hardness) spectra likely due to the presence of larger flares (Table~\ref{tab:xspec_fits}).

Upper limits to the X-ray luminosities of the non-X-ray members (Groups 1 and 2) are obtained from the X-ray faint source luminosity sensitivity for each cluster. The effect of the reduced {\it Chandra} point source sensitivity with increasing off-axis angle \citep{Feigelson2002} is taken into consideration. Panels d and h in Figure \ref{fig:xray_spectra} show detected X-ray luminosities as a function of off-axis angle for Group 5 and 6 cluster members, as well as X-ray sources unrelated to the clusters (Groups 7-10). The latter X-ray luminosities were obtained as above with the incorrect assumption that the sources lie in the star clusters. The red curves show likelihood-based local quadratic regression fits to the lower envelope data of the X-ray luminosity distribution generated using the {\it locfit.robust} function from the R CRAN {\rm locfit} package \citep{Loader99,Loader20}. Using these curves, upper limits to the X-ray luminosity based on the source's angular distances from the cluster centers are assigned to non-X-ray cluster members (Groups 1 and 2). The inferred upper limits are listed in Table \ref{tab:stellar_props_open_clusters}.

\section{MYStIX/SFiNCs Stars} 
\label{sec:mystix_sfincs}

The interpretation of the 7--25~Myr clusters (with many PMS stars) examined here greatly benefits from comparison to previously reported X-ray studies of 0.5--5~Myr stellar clusters (with mostly e-PMS stars) in the MYStIX and SFiNCs projects.  The MYStIX \citep{Feigelson13} {\it Chandra} survey covers 20 regions dominated by multiple O-type stars at typical distances $1.5 \leq d \leq 2.5$~kpc, while the SFiNCs \citep{Getman17} survey covers 22 regions dominated by single O- or multiple B-type stars at typical distances  $0.3 \leq d \leq 0.8$~kpc. The original catalogs of over 40,000 MYStIX and SFiNCs probable cluster members in the 42 star forming regions are provided in \citet{Broos13, Getman17}. 

Numerous MYStIX/SFiNCs-based papers on methodology and topics of cluster formation and stellar activity have been published. Related to the current project, \citet{Getman14a,Getman18b,Richert18,Getman2021} calculate stellar masses and cluster ages using various theoretical evolutionary models. The MYStIX/SFiNCs cluster ages range between 0.5~Myr and 5~Myr. For the current study we adopt stellar masses and cluster ages that are based on the PARSEC 1.2S evolutionary models \citep[Appendix C in][]{Getman2021}. 

Stellar extinctions, effective temperatures, masses, and bolometric luminosities were derived using near-IR color-magnitude diagrams.  However,  MYStIX/SFiNCs cluster memberships do not include {\it Gaia}-selected stars due to the absorption and nebulosity associated with these embedded e-PMS populations. This can lead to systematic biases such as an excess of high-$L_X$ stars in the X-ray luminosity function in MYStIX/SFiNCs populations.  This bias should be less pronounced in our older clusters where X-ray upper limits are treated.   

Table \ref{tab:mystix_sfincs_table} gives inferred stellar and X-ray properties for the MYStIX and SFiNCs samples. Out of 40,041 MYStIX/SFiNCs stellar members, mass estimates are available for 26,681 most massive members. Out of these 26,681 most massive members, X-ray luminosity estimates are available for 16,011 most massive and X-ray luminous members.

\begin{deluxetable*}{cccccccccccc}
\tabletypesize{\footnotesize}
\tablecaption{Stellar Properties of MYStIX and SFiNCs Members\label{tab:mystix_sfincs_table}}
\tablewidth{0pt}
\tablehead{
\colhead{Region} & \colhead{Source} & \colhead{R.A.} &
\colhead{Decl.} & \colhead{$ME$} & \colhead{$A_{V}$} &
\colhead{$t$} & \colhead{$\log(T_{eff})$} & \colhead{$M$} & \colhead{$\log(L_{bol})$} &
\colhead{$\log(L_{X})$} & \colhead{Evol}  \\
\colhead{} & \colhead{} &\colhead{deg} &  \colhead{deg} & \colhead{ke)} & \colhead{} & \colhead{Myr} & \colhead{K} & \colhead{$M_{\odot}$} & \colhead{$L_{\odot}$} & \colhead{erg s$^{-1}$} & \colhead{phase}\\
\colhead{(1)} & \colhead{(2)} & \colhead{(3)} & \colhead{(4)} & \colhead{(5)} & \colhead{(6)} & \colhead{(7)} & \colhead{(8)} & \colhead{(9)} & \colhead{(10)} & \colhead{(11)} & \colhead{(12)}
}
\startdata
be59 & 000033.87+672446.2 & 0.141150 & 67.412846 & 2.6 & 9.3 & 0.9 & 3.66 & 1.0 & 0.35 & 31.1 & 1\\
be59 & 000036.43+672658.5 & 0.151798 & 67.449596 & 1.9 & 5.3 & 1.7 & 3.70 & 1.7 & 0.56 & 31.0 & 2\\
be59 & 000045.20+672805.8 & 0.188345 & 67.468297 & 1.7 & 5.1 & 1.7 & 3.69 & 1.4 & 0.43 & \nodata & 1\\
be59 & 000046.19+672358.2 & 0.192477 & 67.399503 & 1.6 & \nodata & \nodata & \nodata & \nodata & \nodata & 30.4 & \nodata\\
be59 & 000050.10+672721.4 & 0.208781 & 67.455954 & 1.8 & 4.1 & 2.0 & 3.59 & 0.7 & -0.27 & 30.1 & 1\\
\enddata
\tablecomments{This table is available in its entirety (40,041 MYStIX/SFiNCs members) in the machine-readable form in the online journal.  Column 1: Star forming region. Column 2: Source's IAU designation. This identifier is used in the MYStIX and SFiNCs member tables of \citet{Broos13} and \citet{Getman17}. Columns 3-4: Right ascension and declination (in decimal degrees) for epoch J2000.0. Column 5: X-ray median energy in the $(0.5-8)$~keV band. Column 6. Visual extinction. Column 7. Stellar age estimated from the $Age_{JX}$ chronometer \citep{Getman14a}. Columns 8-10. Stellar effective temperature, mass, and bolometric luminosity derived from $JHK$ color-magnitude diagrams. Column 11: X-ray luminosity in the $(0.5-8)$~keV band. Column 12: The evolutionary status of the star on the Hertzsprung–Russell diagram based on PARSEC 1.2S model tracks (Figure~\ref{fig:lbol_r_vs_time}): 1 = Hayashi track; 2 or 3 = Henyey track; $\geq 4$ = ZAMS.}
\end{deluxetable*}

\section{PMS X-ray Emission Dependence on Mass}  \label{sec:xrays_vs_mass}

Main sequence stellar activity is primarily a function of rotation \citep{Skumanich1972}.  But e-PMS activity depends on bulk properties like mass, surface area and volume \citep{Preibisch05, PreibischFeigelson2005, Telleschi07}.  The evolution of interior structure during the l-PMS phase shown in Figure~\ref{fig:lbol_r_vs_time} is also strongly dependent on mass.  For these reasons, it is essential that our study of the evolution of X-ray emission consider mass strata separately. We treat both the e-PMS-dominated population from the MYStIX/SFiNCs surveys (\S\ref{sec:mystix_sfincs}) and the older l-PMS stars in ten 7--25~Myr open clusters presented here. 

\begin{figure*}[ht!]
\epsscale{1.15}
\plotone{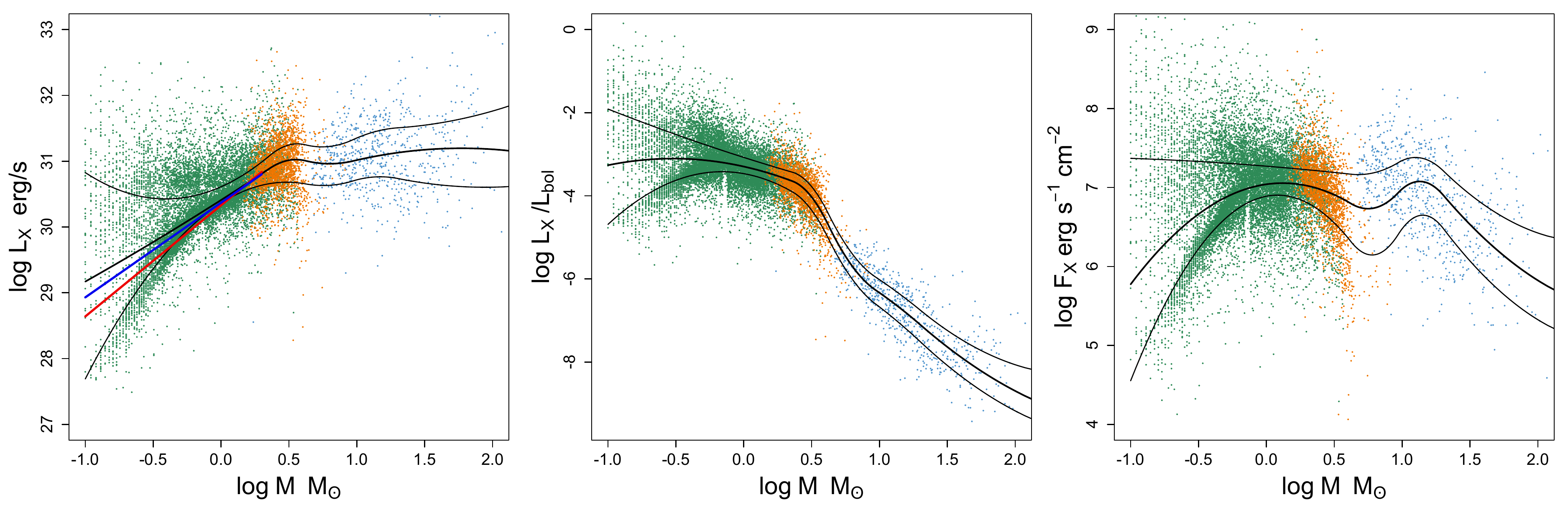}
\caption{X-ray emission quantities as functions of stellar mass for $16,011$ stellar members of the MYStIX/SFiNCs regions. The stars are color-coded according to their evolutionary status in the Hertzsprung-Russell diagram: stars on the Hayashi track (teal), Henyey track (orange), and ZAMS or beyond (blue). The black curves are spline fits to the 25\%, 50\% and 75\% quartiles of the X-ray distribution. In the first panel, the straight lines are regression fits from \citet[COUP;][]{Preibisch05} (blue) and \citet[XEST;][]{Telleschi07} (red) surveys}.  \label{fig:xray_vs_mass_mystixsfincs}
\end{figure*}

Figures \ref{fig:xray_vs_mass_mystixsfincs} shows three measures of X-ray emission as a function of stellar mass for the 16,011 MYStIX/SFiNCs stellar members with available mass and X-ray luminosity estimates.  These measures are the $(0.5-8)$~keV-band X-ray luminosity $L_X$ useful for measuring ionizing fluence on the environment, $L_X/L_{bol}$ giving the ratio of magnetic activity to bolometric stellar output, and $F_X$ giving the ratio of the X-ray luminosity to the stellar surface area.

The points are displayed in three colors representing three phases of early stellar evolution demarcated in Figure~\ref{fig:lbol_r_vs_time}: the e-PMS Hayashi track for low mass stars with M$\lesssim$2~M$_\odot$, the l-PMS Henyey track for masses $2 \lesssim M \lesssim 4$~M$_\odot$, and the ZAMS for masses $4 \lesssim M \lesssim 100$~M$_\odot$.  Widely used linear fits to the $L_X-M$ relation for e-PMS stars are shown from the Orion COUP \citep[blue;][]{Preibisch05} and Taurus XEST \citep[red;][]{Telleschi07} surveys. But the $L_X-M$ pattern is complicated and is better modeled by the black curves showing B-spline regression fits to the $25$\%, $50$\% (median), and $75$\% quartiles of the X-ray measure distribution.  These curves are generated using R CRAN package {\it cobs} \citep{Ng2007,Ng2020}.

\begin{figure*}[ht!]
\epsscale{1.05}
\plotone{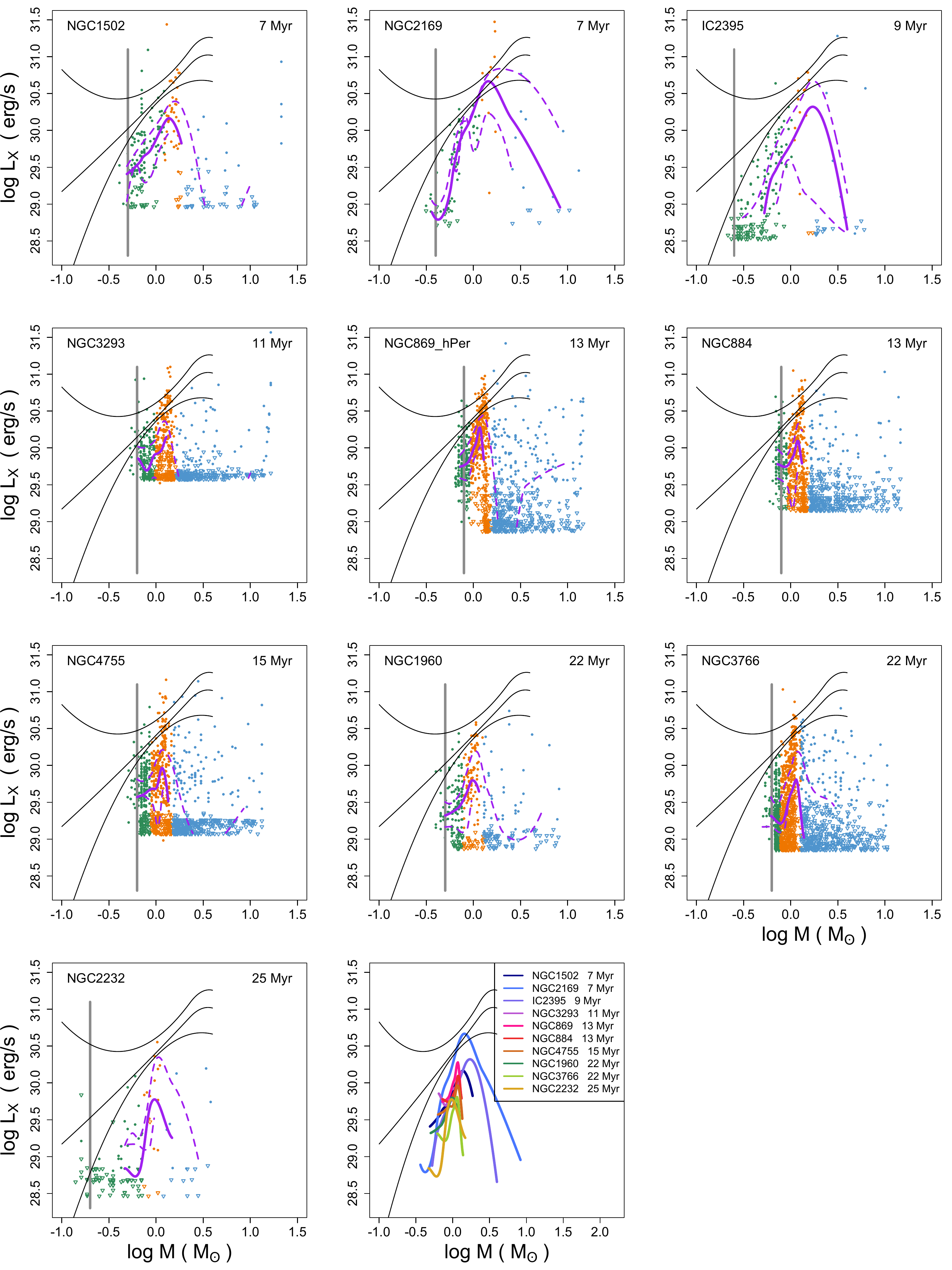}
\caption{X-ray luminosities as a function of stellar mass for members of the ten open clusters in Table~\ref{tab:cluster_props}. The vertical gray line shows the mass completeness limit for each cluster.  The purple curves show smoothed  Kaplan-Meier quartiles of the luminosity distributions (see text).  The black quartile curves from the MYStIX/SFiNCs surveys are reproduced from the left panel of  Figure \ref{fig:xray_vs_mass_mystixsfincs}. The last panel collects the median curves from the other panels to facilitate comparison among the clusters. \label{fig:lx_vs_mass}}
\end{figure*}

Figure~\ref{fig:lx_vs_mass} shows $L_X-M$ distributions for the ten 7--25~Myr clusters treated here. This is one of the principal observational results of this study. These plots include both ACIS (points) and non-ACIS (triangles indicating upper limits)\footnote{
Upper limits are not available for the MYStIX/SFiNC sample as reliable {\it Gaia} memberships cannot be obtained due to absorption and nebulosity. Based on the X-ray luminosity function (XLF) and IMF analyses \citep{Kuhn15a,Getman2021}, the X-ray samples from different MYStIX/SFiNCs regions are approximately complete above different mass thresholds.}.  Approximate mass completeness limits obtained from the cluster IMF distributions (Figure \ref{fig:cmd_imf_ngc1502_hper}) are marked by vertical grey lines, and smoothed quartiles of the MYStIX/SFiNCs $L_X-M$ distribution from Figure~\ref{fig:xray_vs_mass_mystixsfincs} are shown as black curves. 

It is difficult to visually compare the cluster $L_X-M$ distributions given the wide range of luminosities and different fractions of non-detections.  To assist in this comparison we calculate smooth curves representing the median values (solid purple curves), and the 25\% and 75\% quartiles (dashed purple curves) of the $L_X$ values as a function of mass. These curves are calculated as follows.  Adaptive stellar mass bins are introduced to accumulate $N_{tot}/10$ stars per bin, where $N_{tot}$ is the total number of stars per cluster included in this analysis. To treat both X-ray detections and upper limits, we use the nonparametric Kaplan-Meier estimator \citep[KM;][]{KaplanMeier58} for the $L_X$ distributions in each stellar mass bin.  The KM calculations are made with R CRAN package {\it survival} \citep{Therneau20}. Local quadratic fits using R's {\it loess}  \citep{Cleveland1992} are then calculated to provide smoothed representations of the $25$\%$-50$\%$-75$\% KM quartiles.

For mass bins with high fractions of X-ray nondetections, the KM 25\% and 50\% (and sometimes even the 75\% level) quartiles may be unavailable. The median curves are superposed in the last panel to help compare the cluster $L_X-M$ patterns.  
 
Three important results emerge from both the MYStIX/SFiNCs and the older cluster $L_X-M$ relationships.

{\bf 1.} The e-PMS $L_X-M$ relation in the MYStIX/SFiNCs surveys is not well-described by a power law. Dozens of studies have adopted the $L_X \propto M^{1.7}$ relation from the Taurus sample of \citet{Telleschi07} as a valid parameterization of how magnetic activity increases with stellar mass for stars on the Hayashi track.  But the left panel of Figure~\ref{fig:xray_vs_mass_mystixsfincs} shows this is an inadequate description of a more complex distribution. 

The power law is a reasonable fit to the median $L_X$ in the  0.3--2~M$_{\odot}$ range of our MYStIX/SFiNCs. But this masks a $\sim 30$-fold range in $L_X$ at any mass, and a bifurcation between a minority of stars around $\log L_X \simeq 30.5-31.0$ erg~s$^{-1}$ and a majority of stars around $29.5-30.5$~erg~s$^{-1}$. The upper fork can be attributed to the capture of individual super- and mega-flares in the {\it Chandra} exposures \citep{Getman2021}.  This bimodality is more prominent for masses below 0.5~M$_\odot$ where the $L_X-M$ relation for most stars steepens considerably. Recall that the MYStIX/SFiNCs samples are missing most of the low mass stellar population of their clusters, and the true $L_X$ distributions will have more low $L_X$ values than shown in Figure~\ref{fig:xray_vs_mass_mystixsfincs}. The bifurcation and steepening in the $L_X$ dependence on mass, inconsistent with a simple power law relation, is also seen in the dependence of $L_X/L_{bol}$ and $F_X$ on mass.

These wide and non-Gaussian spreads can not reasonably be attributed to inaccuracies in X-ray measurements, or mass, or usage of different mass-scales based on different evolutionary models, i.e., PARSEC~1.2S (here) versus \citet{Siess00} (as in \citet{Preibisch05,Telleschi07}). Visual inspection of Figure~3 in \citet{Preibisch05} indicates that the $L_X-M$ relationship showed a wide spread and diverged from a single power law at lower masses ($<0.4$~M$_{\odot}$). The latest census of Taurus e-PMS stars based on the {\it Gaia} catalog indicates that the XEST Taurus sample used by \citet{Telleschi07} was deficient in $0.2-0.6$~M$_\odot$ stars \citep{Luhman2018}. This may be responsible for their overestimation of the $L_X-M$ relation at low masses.

{\bf 2.} The median X-ray luminosities of l-PMS stars in our open clusters are weaker than median luminosities from e-PMS stars at all masses. This is seen in the comparison of smooth distributions shown in the last panel of Figure~\ref{fig:lx_vs_mass}.  This drop in X-ray luminosity continues through the l-PMS phase. The difference between the median $L_X$ values of the open clusters and MYStIX/SFiNCs clusters increases from a factor of $\simeq$3 drop for $t \simeq 7$~Myr to a factor of $\simeq$6 at $t \simeq 25$~Myr.

{\bf 3.} At any given age, intermediate mass stars on the l-PMS Henyey track have higher X-ray luminosities than either lower mass Hayashi track or higher mass ZAMS stars. This is seen in Figure~\ref{fig:lx_vs_mass} where the highest luminosities are orange points.  

The last panel's smooth distributions show narrow peaks in X-ray luminosities at intermediate masses compared to other masses. In particular, higher-mass stars in the range $M \gtrsim 1-3$~M$_{\odot}$ have  X-ray luminosities plummeting with increasing stellar mass. These X-ray luminosity decreases take place for stars that are completing their changes on the Henyey track or already reside on the ZAMS. 

The combination of the above effects causes a distinctive drift of the $\log(L_X)-\log(M)$ turnover point with age. Again this is best seen in the smooth median $L_X-M$ curves in the last panel of Figure~\ref{fig:lx_vs_mass}.

In summary, the historically reported strong correlation of X-ray luminosity with mass for e-PMS Hayashi track stars with $M \lesssim 3$~M$_{\odot}$ is confirmed, but a simple power law relationship does not apply.  Most stars exhibit a relation steeper than $L_X \propto M^{1.8}$ while other stars are dominated by powerful flares and show a shallower relation. The median X-ray luminosities decline gradually as stars begin passage along the l-PMS Henyey track and then plummet rapidly as they approach the ZAMS.  We find definite systematic decreases in surface magnetic activity as the interior changes from fully convective to mostly radiative energy transport. Overall, it is clear that the X-ray activity of the stars diminishes with increasing age and the Henyey/ZAMS stars experience the most drastic X-ray activity changes. We quantify these changes in \S \ref{sec:xray_activity_slopes}.

Finally, we note that our discussion of l-PMS activity evolution has been limited to X-ray luminosities with no mention of $L_X/L_{bol}$ and $F_X$. Plots of these quantities against mass for the 7--25~Myr open clusters show complex structure changes (such plots are omitted from this paper). Often the changes in X-ray emission are overwhelmed by bolometric luminosity and stellar radius changes (Figure~\ref{fig:lbol_r_vs_time}). $L_X/L_{bol}$ and $F_X$ have long proved useful in plots against stellar rotational measures on the main sequence in discussions of saturation and tachoclinal dynamo processes \citep{Gudel2004}.  But they are not useful for the l-PMS phase when luminosities and radii are rapidly changing.

\section{PMS X-ray Emission Dependence on Age}  \label{sec:xray_activity_slopes}

\begin{figure*}[ht!]
\epsscale{1.15}
\plotone{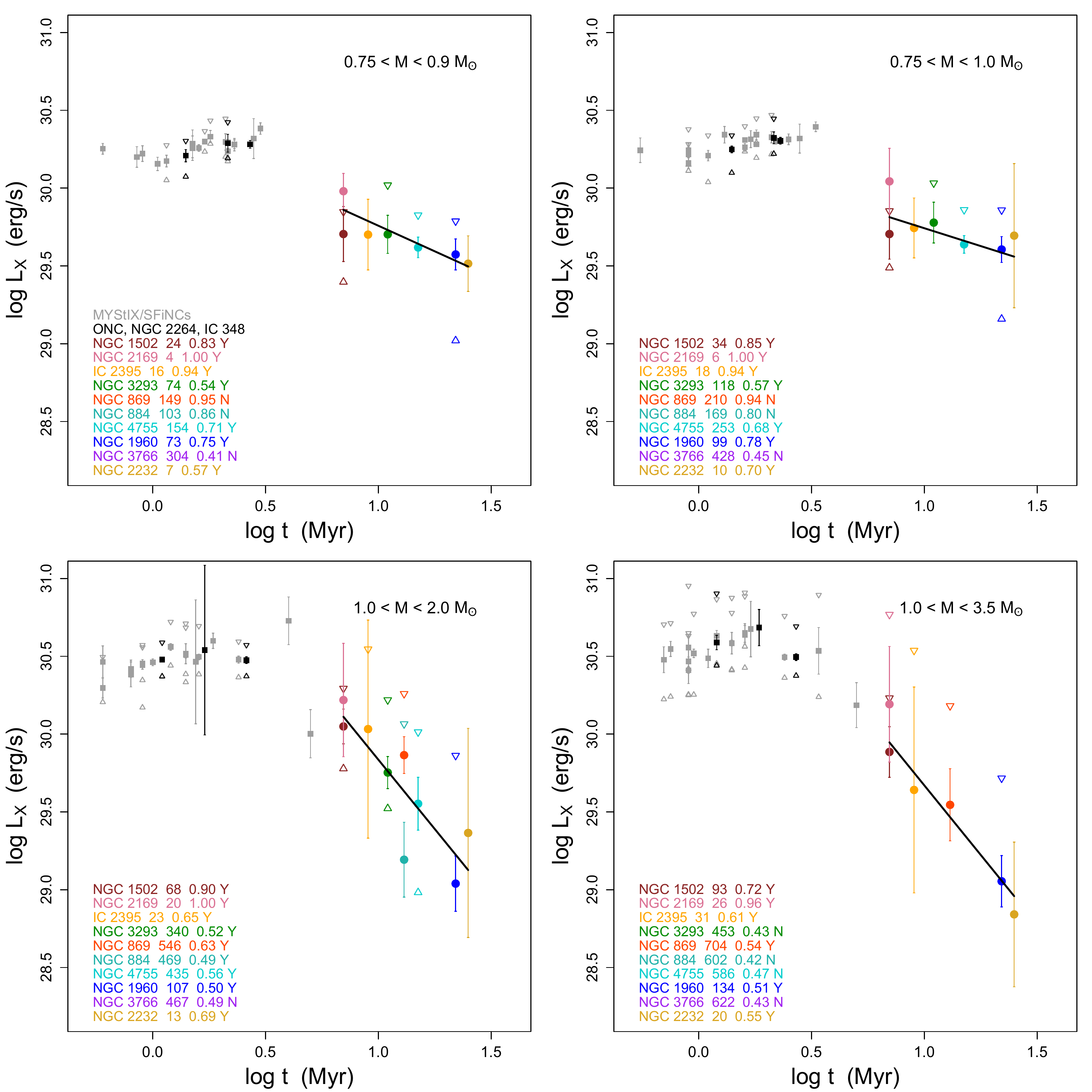}
\caption{Temporal evolution of X-ray luminosity for four mass strata. Corresponding mass bins are given in the figure legends. e-PMS stellar samples from nearby MYStIX/SFiNCs regions are shown as grey points with error bars representing 68\% confidence intervals on median (solid squares). 25\% and 75\% quartiles are shown for samples with $>20$ stars (open triangles). Stars from the lightly-absorbed e-PMS sub-populations are marked by black points. The l-PMS open clusters are depicted by colored points with 68\% error bars on the 50\% quartiles (solid circles) of the KM estimators. 25\% and 75\% quartiles are  shown only for samples with over 20 stars (open triangles); 25\%-quartile KM estimators are often not available due to high numbers of X-ray upper limits. The black lines indicate linear regression fits to the median X-ray luminosities of the open clusters. The regression fit results are listed in Table~\ref{tab:temporal_evolution_table}. The legends also list the numbers of stars in the open clusters, whose stellar masses lie within the mass range of interest, and the fractions of X-ray detections. The flag "Y" indicates that the mass completeness limit (Table~\ref{tab:cluster_props}) is below the mass range of interest, the 50\% quartile of the KM estimator is available, and the cluster' median $L_X$ is plotted and involved in the regression calculation. \label{fig:lx_vs_time}}
\end{figure*}

We now address how the X-ray emission of  ensembles of stars evolves from the fully convective e-PMS phase through the l-PMS changes and settling on the ZAMS. Here we stratify stellar masses into a few bins.  This is challenging because, due to differing distances and {\it Chandra} observation sensitivities, X-ray detections are complete to different limiting masses in a sample of PMS clusters.  The addition of {\it Gaia}-only members does not fully alleviate this difficulty as the {\it Gaia} survey also has a distance-dependent sampling effect.  The results of this analysis presented in Figure \ref{fig:lx_vs_time} show the  mass-stratified evolution of the X-ray luminosity for the stellar groups from  MYStIX/SFiNCs regions and our 7--25~Myr open clusters.

For the e-PMS phase, the analysis is limited to the nearby MYStIX/SFiNCs stellar samples with better sensitivities towards low-mass stars and completeness limits down to $0.5-0.75$~M$_{\odot}$.\footnote{
The included MYStIX/SFiNCs regions are (in order of increasing distance from the Sun out to $d \simeq 1$~kpc; see Table~1 in \citet{Getman2021}): IC 348, OMC~2-3, ONC Flanking Fields, Orion Nebula, NGC~2068, W~40, LkH$_{\alpha}$~101, NGC~2264, IC~5146, Cep~B, RCW~36, NGC~7160, Be~59, NGC~2362, and Lagoon Nebula.} 
Each region is shown as a gray point in Figure \ref{fig:lx_vs_time}.  We also consider lightly-absorbed (X-ray median energy $<2$~keV) sub-populations from the deep {\it Chandra} surveys of Orion Nebula (COUP), MYStIX-NGC2264, and SFiNCs-IC348, which exhibit even better completeness limits down to $0.1-0.4$~M$_{\odot}$ (black points). 

For the l-PMS phase, the analysis is limited to mass strata in each cluster where the cluster's IMF completeness limit is below the lower boundary of the mass stratum and the KM estimator is able to produce a 50\% quartile value with uncertainty. This avoids low mass samples that are dominated by X-ray upper limits.

Historically, a power law relation $L_X \propto t^{b}$ has been used to characterize the decay of magnetic activity.  From the Orion Nebula Cluster COUP study, \citet{PreibischFeigelson2005} report, for instance, $b \simeq -0.32$ in the 0.4--1~M$_{\odot}$ range over the age range $0.1-10$~Myr, and a steeper decline $b \simeq -0.75$ when older ZAMS stars are included. In Figure~\ref{fig:lx_vs_time}, we find that the power law model does provide good fits to the evolution of X-ray luminosities during the l-PMS phase. Results from weighted least squares regression, using the R {\it lm} function \citep{Sheather09}, for $L_X \propto t^b$ in the age range 7--25~Myr are listed in Table~\ref{tab:temporal_evolution_table}.

\begin{deluxetable*}{ccrrccc}
\tabletypesize{\small}
\tablecaption{Evolution of $L_X$, $F_X$, $L_X/L_{bol}$, $L_{bol}$, and $R_{\star}$ in 7--25~Myr-old Stars \label{tab:temporal_evolution_table}}
\tablewidth{0pt}
\tablehead{
\colhead{Mass Stratum} & \colhead{Quantity} &
\colhead{$N_{clusters}$} & \colhead{$N_{stars}$} & \colhead{$\log(a)$} & \colhead{$b$} & \colhead{$p$-val}\\
\colhead{$M_{\odot}$} & \colhead{} &
\colhead{} & \colhead{} & \colhead{} & \colhead{} & \colhead{}\\
\colhead{(1)} & \colhead{(2)} & \colhead{(3)} & \colhead{(4)} & \colhead{(5)} & \colhead{(6)} &
\colhead{(7)}
}
\startdata
$0.75-0.9$ & $L_X$	& 7~~~~~~ & 352 &	$30.42 \pm 0.19$ & $-0.66 \pm 0.17$ & 0.01\\
$0.75-0.95$ & $L_X$	& 7~~~~~~ & 443 &	$30.41 \pm 0.18$ & $-0.63 \pm 0.15$ & 0.01\\
$0.75-1$ & $L_X$	& 7~~~~~~ & 538 &	$30.20 \pm 0.21$ & $-0.46 \pm 0.18$ & 0.05\\
$1-2$ & $L_X$	& 9~~~~~~ & 2021 &	$31.62 \pm 0.44$ & $-1.78 \pm 0.41$ & 0.00\\
$1-3.5$ & $L_X$	& 6~~~~~~ & 1008 &	$31.46 \pm 0.21$ & $-1.79 \pm 0.19$ & 0.00\\
$1-3.5$ & Q85\% $L_X$	& 10~~~~~~ & 3271 &	$31.03 \pm 0.35$ & $-0.67 \pm 0.30$ & 0.06\\
$3.5-5$ & Q85\% $L_X$	& 5~~~~~~ & 224 &	$32.48 \pm 0.46$ & $-2.25 \pm 0.42$ & 0.01\\
$3.5-7$ & Q85\% $L_X$	& 7~~~~~~ & 374 &	$34.06 \pm 0.84$ & $-3.65 \pm 0.88$ & 0.01\\
$0.75-0.9$ & $L_X/L_{bol}$	& 7~~~~~~ & 352 &	$-2.85 \pm 0.19$ & $-0.44 \pm 0.17$ & 0.05\\
$0.75-0.95$ & $L_X/L_{bol}$	& 7~~~~~~ & 443 &	$-2.81 \pm 0.21$ & $-0.49 \pm 0.18$ & 0.04\\
$0.75-1$ & $L_X/L_{bol}$	& 8~~~~~~ & 966 &	$-2.73 \pm 0.44$ & $-0.62 \pm 0.39$ & 0.16\\
$1-2$ & $L_X/L_{bol}$	& 10~~~~~~ & 2488    &	$-0.87 \pm 0.38$ & $-3.14 \pm 0.39$ & 0.00\\
$1-3.5$ & $L_X/L_{bol}$	& 7~~~~~~ & 1630    &	$0.16 \pm 0.53$ & $-4.55 \pm 0.48$ & 0.00\\
$1-3.5$ & Q85\% $L_X/L_{bol}$	& 10~~~~~~ & 3271 &	$-2.65 \pm 0.25$ & $-0.71 \pm 0.23$ & 0.01\\
$3.5-5$ & Q85\% $L_X/L_{bol}$	& 5~~~~~~ & 224 &	$-3.14 \pm 0.50$ & $-2.65 \pm 0.45$ & 0.01\\
$3.5-7$ & Q85\% $L_X/L_{bol}$	& 7~~~~~~ & 374 &	$-4.36 \pm 1.17$ & $-1.76 \pm 1.04$ & 0.15\\
$0.75-0.9$ & $F_X$	& 7~~~~~~ & 352 &	$7.18 \pm 0.19$ & $-0.22 \pm 0.17$ & 0.24\\
$0.75-0.95$ & $F_X$	& 7~~~~~~ & 443 &	$7.23 \pm 0.19$ & $-0.27 \pm 0.17$ & 0.18\\
$0.75-1$ & $F_X$	& 7~~~~~~ & 538 &	$6.96 \pm 0.21$ & $-0.03 \pm 0.18$ & 0.88\\
$1-2$ & $F_X$	& 9~~~~~~ & 2021    &	$8.54 \pm 0.42$ & $-1.83 \pm 0.40$ & 0.00\\
$1-3.5$ & $F_X$	& 6~~~~~~ & 1008    &	$8.51 \pm 0.31$ & $-1.95 \pm 0.29$ & 0.00\\
$1-3.5$ & Q85\% $F_X$	& 10~~~~~~ & 3271 &	$7.44 \pm 0.26$ & $-0.18 \pm 0.23$ & 0.45\\
$3.5-5$ & Q85\% $F_X$	& 5~~~~~~ & 224 &	$9.39 \pm 0.42$ & $-2.69 \pm 0.38$ & 0.01\\
$3.5-7$ & Q85\% $F_X$	& 7~~~~~~ & 374 &	$9.29 \pm 0.58$ & $-2.61 \pm 0.61$ & 0.01\\
0.75,0.8,0.9 & $L_{bol}$	& \nodata & \nodata & -0.15,-0.17,-0.39 & -0.43,-0.35,-0.03 & \nodata\\
0.95,1,1.5 & $L_{bol}$	& \nodata & \nodata & -0.56,-0.77,0.14 & 0.20,0.47,0.52 & \nodata\\
2,3.5,5,7 & $L_{bol}$	& \nodata & \nodata & 1.29,2.11,2.68,3.07 & -0.03,0.02,0.08,0.22 & \nodata\\
0.75,0.8,0.9 & $R_{\star}$	& \nodata & \nodata & 0.26,0.25,0.23 & -0.26,-0.22,-0.20 & \nodata\\
0.95,1,1.5 & $R_{\star}$	& \nodata & \nodata & 0.22,0.20,0.57 & -0.18,-0.15,-0.32 & \nodata\\
2,3.5,5,7 & $R_{\star}$	& \nodata & \nodata & 0.36,0.27,0.35,0.36 & -0.15,0.03,0.07,0.16 & \nodata\\
\enddata
\tablecomments{Column 1: Mass stratum. Column 2: X-ray luminosity ($L_X$ in units erg~s$^{-1}$),  X-ray surface flux ($F_X$ in units erg~s$^{-1}$~cm$^{-2}$), median for lower mass stars and 85\% quantiles for higher mass stars (Appendix \S \ref{sec:temporal_highest_masses}). Bolometric luminosity ($L_{bol}$ in units $L_{\odot}$) and stellar radius ($R_{\star}$ in units $R_{\odot}$) are from PARSEC 1.2S evolutionary models. Columns 3-4: Numbers of open clusters and cluster members. Columns 5-7: Results from the linear regression fits for the relations $Q=a \times t_{Myr}^b$ within the 7--25~Myr age range, where Q is one of the above stellar quantities. The results include the intercept and slope with 68\% errors, and $p$-value for the hypothesis of zero slope.}
\end{deluxetable*}

Four findings emerge: 

{\bf 1.} For the X-ray detected MYStIX/SFiNCs stars in the low-mass range 0.75--1~M$_{\odot}$, within the first few Myr of evolution, their X-ray luminosities do not exhibit any statistically significant temporal changes; the median X-ray luminosities remain at a constant level of $\log(L_X) \sim 30.3$~erg~s$^{-1}$. The young stars within these mass and age ranges are still on Hayashi tracks; they are fully convective and keep gravitationally contracting. Since their X-ray luminosities are at a constant level, their surface fluxes rise drastically with time at the pace of the decline of their squared stellar radii. The associated astrophysical issue we consider in \S \ref{sec:dynamo_activity} is: {\it What causes the X-ray luminosity to remain nearly constant while the fully convective star keeps contracting?}

{\bf 2.} In contrast, the X-ray luminosities of the low-mass 0.75--1~M$_{\odot}$ members of the open clusters decline with time slightly during the 7--25~Myr period, but the effect is statistically significant. The majority of the stars in clusters younger than 15~Myr are still descending Hayashi tracks (e-PMS phase). But for older clusters, Henyey stars start significantly contributing or even dominating the stellar samples. The median X-ray luminosities decrease from $\log(L_X) \sim 29.8$~erg~s$^{-1}$ (at 7~Myr) to  $\log(L_X) \sim 29.5$~erg~s$^{-1}$ (at 25~Myr). The associated slope $b$ in the $L_X \propto t^b$ relation is around $b \sim -0.6$. This temporal trend is similar to the rate of decrease of stellar volume $V \propto R_{\star}^3$ based on the PARSEC 1.2S models (Figure~\ref{fig:lbol_r_vs_time}, Table~\ref{tab:temporal_evolution_table}).  The associated astrophysical issue is: {\it What causes the X-ray power to scale with stellar volume?}

{\bf 3.} Within the mass ranges 1--2~M$_{\odot}$ and 1--3.5~M$_{\odot}$, the open cluster stars are predominantly located on either Henyey (l-PMS phase) or ZAMS tracks. The stellar radii exhibit complex temporal behaviours --- some are rising, some are still decreasing, and some are roughly constant. The X-ray luminosity undergoes a rapid plunge. The slope of the $L_X \propto t^{b}$ relation is $b \simeq -1.8$. {\it What causes the X-ray power to plunge in stars undergoing most significant interior changes?} 

{\bf 4.} For the highest mass range available to us, $3.5 \lesssim M \lesssim 7$~M$_{\odot}$, the stars are entirely on the ZAMS.  X-ray luminosities are so low that the Kaplan-Meier estimator does not give a median value; we measure instead the temporal behavior of the 85\% quantile of the $L_X$ KM estimator.  These stars show an even more rapid decay rate than the 1--3.5~M$_{\odot}$ mass stratum. The decay slope for the 85\% quantile of $L_X$ changes from $b \sim -0.7$ in 1--3.5~M$_{\odot}$ stars to $b \simeq -[2-4]$ in more massive stars.  Appendix \ref{sec:temporal_highest_masses} details these changes. {\it Why does X-ray luminosity keep nosediving in more massive ZAMS stars?} 

The X-ray activity evolution results given above are based on the entire sample of 6,003 young stars located within the central regions of the ten open clusters (Table~\ref{tab:stellar_props_open_clusters}). These results are listed in Table~\ref{tab:temporal_evolution_table}. For the sample culled of 485 stars with inconsistent $G$-band vs. BP- and RP-band photometry, the inferred X-ray luminosity decay slopes are $b=(-0.7\pm0.2, -0.4\pm0.2, -2.0\pm0.4, -1.8\pm0.2)$ for the mass strata $M= (0.75-0.9), (0.75-1), (1-2), (1-3.5)$~M$_{\odot}$, respectively, and are statistically indistinguishable from the ones given in Table~\ref{tab:temporal_evolution_table}.   

\section{Comparison With Previous Literature} \label{sec:prev_literature}

Using approach A (i.e., relying on ages for individual stars; \S \ref{sec:intro_to_past_studies}), for the first $\sim 10$~Myr of PMS evolution \citet{PreibischFeigelson2005} and \citet{Gregory16} report a mild decay in the stellar X-ray luminosity with the slopes $b$ in the $L_X \propto t^{b}$ relation as $b \sim -0.3$ and $b \sim -0.5$, for the 0.4--1~M$_{\odot}$ and 1--2~M$_{\odot}$ mass ranges, respectively. \citet{Gregory16} further find a steeper decay slope for more massive stars, e.g., $b=-1.19 \pm 0.35$ in 2--3~M$_{\odot}$ stars. 

Our results differ from those in two ways: 1) we do not find any X-ray luminosity decreases within the first few Myr of PMS evolution; 2) but for the time interval after 7~Myr, we find significantly faster $L_X$ decays in $>1$~M$_{\odot}$ stars (Table \ref{tab:temporal_evolution_table}).

Our results are more consistent with the outcomes of the approach B in \citet{PreibischFeigelson2005}. In this case, the authors mention no indications for $L_X$ changes within the first few Myr of evolution and derive a steeper decay slope of $b=-0.75$ for 0.5--1.2~M$_{\odot}$ stars after 7~Myr.

Qualitatively, our results are consistent with the findings of \citet{Gregory16} that the X-ray luminosity decays faster in more massive PMS stars. 

\begin{figure*}[ht!]
\epsscale{1.15}
\plotone{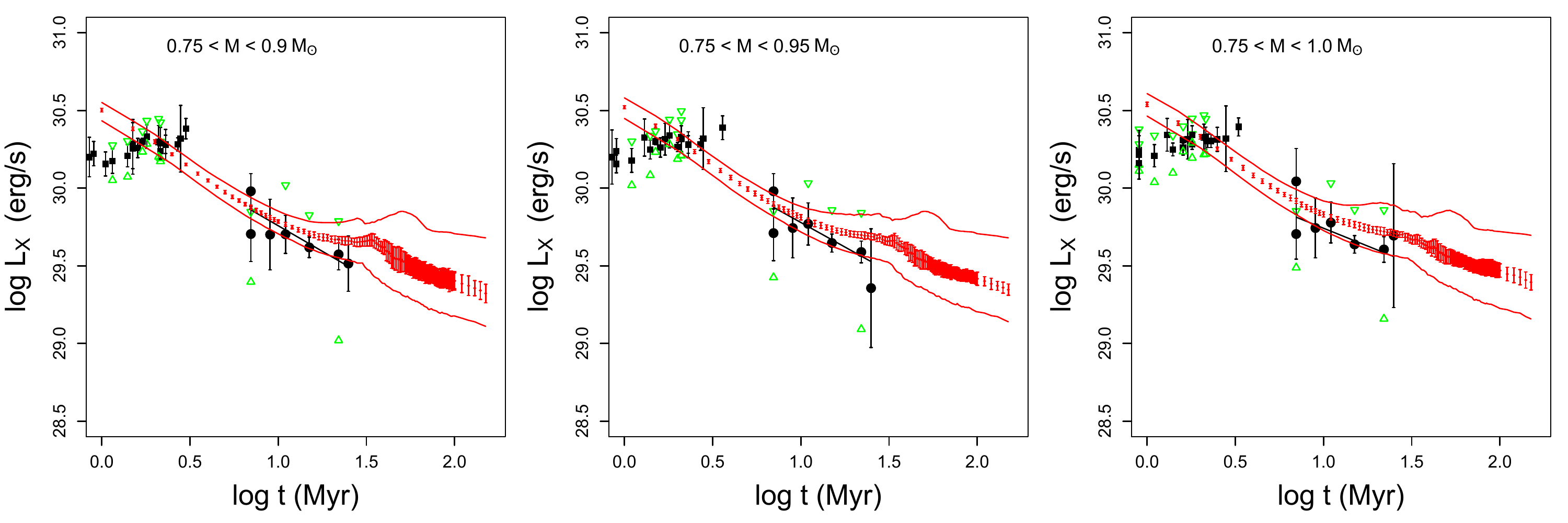}
\caption{Comparison with the modeling results of \citet{Johnstone2021}. Temporal evolution of the 25-50-75\% quartiles for the X-ray luminosity in 3 mass bins. The mass ranges are given in figure legends. Our empirical points from the current study, as medians with 68\% confidence intervals (black points with error bars) and 25-75\% quartiles for $N>20$ stellar samples (green triangles), are the same as in Figure \ref{fig:lx_vs_time}. For the modeled data of \citet{Johnstone2021}, median and its 68\% confidence intervals (red points with error bars) and 25-75\% quartiles (red lines). \label{fig:lx_comparison_with_johnstone}}
\end{figure*}

For e-PMS age ranges, \citet{Tu2015,Johnstone2021} calibrate their analytic rotation evolution approach against X-ray and rotation empirical data of only two similar-aged young ($\sim$2~Myr) star forming regions, Taurus cloud and Lagoon Nebula. For the l-PMS age range of interest here, 7--25~Myr, their calibration includes only one $\sim 13$~Myr old cluster, h~Per. The outcome from the model of \citet{Johnstone2021} and our empirical results are compared in Figure~\ref{fig:lx_comparison_with_johnstone}. For the modelled data of Johnstone et al., we calculate 25-50-75\% quartiles and 68\% confidence intervals on median $L_X$ based on their on-line electronic data-set entitled ``TrackGrid\_Distribution''. The following main inconsistencies between the model and empirical results are evident. First, the empirical distributions of X-ray luminosities have higher spreads (interquartile ranges) than the modeled distributions. For l-PMS stars such wide X-ray radiation spreads can be explained by possible dependencies of $L_X$ on rotation, with faster rotators having systematically high X-ray luminosities \citep[e.g.,][]{Tu2015}.  Second, for the first few Myr (at least up to 3~Myr), the model predicts a monotonic decrease in $L_X$, while the empirical data show nearly constant (within statistical uncertainties) X-ray emission. Third, for $<0.95$~M$_{\odot}$ stars in the $>10$~Myr age range, there is an indication for possible slightly steeper temporal decline of the empirical median $L_X$. Forth, closer to one-solar mass $>7$~Myr stars, the empirical data give systematically lower median X-ray luminosities.

Our future {\it Chandra} projects will be targeting more open clusters within the age ranges 5--7~Myr and 25--150~Myr for further improvements of the empirical trends and comparison with the models. \citet{Johnstone2021} are encouraged to use our data for calibration and refinement of their models.

\section{Discussion: Interior Dynamos and Surface Activity } \label{sec:dynamo_activity}

The evolution of PMS stars is strongly mass-dependent as they proceed from the e-PMS Hayashi track to the rapid crossing of the Hertzsprung-Russell diagram during the l-PMS phase (Figure~\ref{fig:lbol_r_vs_time}).  We therefore divide our discussion into mass strata considered earlier. 

{\bf The 0.75--1~M$_{\odot}$ stars} ~~ For this mass range, the X-ray detected stellar populations for nearby MYStIX/SFiNCs star forming regions are nearly complete in mass and all stars are still descending Hayashi HRD tracks. In contrast to most previous results \citep{PreibischFeigelson2005,Gregory16,Tu2015,Johnstone2021}, we find that the X-ray luminosity levels of these lower mass e-PMS stars remain nearly constant in time during the first few Myr of evolution (at least at $\sim (0.5-3)$~Myr). The surface fluxes rapidly increase with time simply as consequences of decreasing radii during the gravitational contraction phase of a fully convective PMS star. Our finding suggests that at this evolutionary stage, the X-ray levels do not respond to the temporal changes in the size of a star.

In addition, previous work has found no correlation between $L_X$ and rotation rate during the fully convective e-PMS phase \citep{Preibisch05, Alexander2012, Henderson2012}.  The independence of surface magnetic activity to rotation rates, the high levels of X-ray emission during the e-PMS phase with $\log L_X/L_{bol} \simeq -3$ to $-4$, and similar behaviors of older fully convective late-dM stars \citep{Magaudda2021}, lead to the conclusion that magnetic activity processes may be saturated.  We will discuss this issue for the ten older open clusters in our forthcoming companion paper concentrated on measuring ZTF and TESS stellar rotation periods.

But in low-mass e-PMS stars, it is not clear which stage is saturated: interior dynamos, surface filling factors,  or some process above the stellar surface such as centrifugal stripping of X-ray coronal structures \citep{Vilhu1984,Jardine1999, Pizzolato2003, Wright2011}. It is possible that a combination of some of these processes is responsible for the X-ray saturation. 

Qualitatively, the link between interior field generation and the observed X-ray emission is clear.  Fields generated in the interior dynamo, probably a distributed $\alpha^2$ dynamo process, erupt onto the stellar surface where reconnection commonly occurs as field loop footprints are twisted by turbulent cells.  \citet{Getman2021} and \citet{Getman2021b} find that super- and mega-flares are common during the e-PMS phase for all stellar masses.  The surface fields confine hot X-ray emitting plasma as the flare developed with the hotter plasma component originating in higher-altitude coronal loops and cooler plasma dominating in the lower-altitude corona. This is in line with the dynamic MHD simulations of \citet{Cohen2017}. The simulations suggest that surface magnetic fields in fully convective fast rotating stars may be concentrated in high-latitude areas leading to formation of large-scale dipolar coronal structures that due to their strong magnetic fields would be able to confine hot X-ray emitting plasma. This is alternative to the scenarios relying on the saturation of surface filling factors or dynamos.

A few lines of independent empirical evidence based on our COUP and MYStIX/SFiNCs data favor the scenario of \citet{Cohen2017}. First, the Orion COUP data show that the emission measure of the hot ($T \sim 20-60$~MK) plasma component dominates that of the cool ($T \sim 10$~MK) component by a factor of 3 in low mass e-PMS stars  \citep{Preibisch05,Getman10}.  Second, plasma modeling of COUP and MYStIX/SFiNCs super- and mega-flares provide strong evidence for the existence of giant X-ray hot flaring coronal structures extending $0.5-10$~R$_{\star}$ above the stellar surface \citep{Favata2005, Getman08b, Getman2021b}. In diskless stars, the heights of such structures can often exceed one corotation radius indicating that the structures may withstand the effects of centrifugal forces, thanks to the high strengths of their associated large-scale magnetic fields \citep{Getman08b}. Third, in the current study, we find that the X-ray luminosity is insensitive to the age and stellar size changes. This is in line with the idea that the higher altitudes of X-ray emitting structures imply less dependence on the properties of the surface far below the giant polar loops.

As the 0.75--1~M$_{\odot}$ stars progress into the l-PMS phase represented by our open cluster members, the X-ray luminosity starts with a drop sometime between 3 and 7~Myr.  While we have a gap in our age coverage during this interval, the drop is evident in  Figure~\ref{fig:lx_vs_time}. The X-ray luminosity continues to fall slowly during the following 7--25~Myr period. During this period, the PARSEC 1.2S interior models indicate that the bolometric luminosity continues to fall and reaches a minimum while the radius continues to decrease (Figure~\ref{fig:lbol_r_vs_time}). X-ray luminosity changes with time as the stellar volume does.

The explanation for this change in X-ray behavior is unclear. Perhaps, unlike the MYStIX/SFiNCs e-PMS regime, the convective dynamo finally becomes responsive to changes in stellar volume and produces weaker surface fields and consequently weaker X-ray flares. And/or perhaps the smaller stellar surface is unable to sustain the enormous polar loops responsible for the strongest X-ray flares. A simplistic model might be that the X-ray coronal scale height is proportional to the stellar radius so that  $L_X \propto V_{loop} \propto V_{star}$.  

{\bf The 1--2~M$_{\odot}$ stars} ~~
In this mass range, most of the open cluster members lie on the l-PMS Henyey track and or (for the 20--25~Myr cluster) have reached the ZAMS.  These are the most X-ray luminous stars in the clusters, as indicated by a sharp peak in the $L_X-M$ diagrams in Figure~\ref{fig:lx_vs_mass}.  Figure~\ref{fig:lx_vs_mass} (last panel), Figure~\ref{fig:lx_vs_time} (lower left panel), and Table~\ref{tab:temporal_evolution_table} show that the X-ray luminosity is now decreasing rapidly during the 7--25~Myr period roughly as $L_X \propto t^{-1.8}$. This is consistent with the decay for older stars on the main sequence: \citet{Maggio1987} and \citet{Gudel1997} obtain  $L_X \propto t^{-1.5}$ for G-type field stars and members of old open clusters with ages between $\gtrsim$70~Myr and 9~Gyr. \citet{Feigelson2004} report $L_X \propto t^{-2}$ for magnetically active M-, K-, and G-type Galactic stars. The magnetic activity of these main sequence stars is believed to be driven by the tachoclinal $\alpha\Omega$ dynamo. The great majority of older main sequence stars lie on the unsaturated locus of the X-ray activity $-$ rotation diagrams \citep{Maggio1987,Gudel1997}.

This similar decay law between our open cluster stars and main sequence stars in this mass range suggests that the $\alpha\Omega$ dynamo now dominates over the $\alpha^2$ dynamo during the l-PMS phase: a fundamental transition has occurred in the magnetic field generation with observable consequences in the X-ray band.  Like in the contemporary Sun, the bulk of our 1--2~M$_{\odot}$ l-PMS stars have developed radiative zones, but the outer parts are convective. Hints of changes in the large scale magnetic topologies of such stars, from simple (i.e., dipolar) axisymmetric to more complex (i.e., octopolar) non-axisymmetric, support this transition \citep{Gregory2014,Gregory16}. For NGC~869, \citet{Argiroffi2016} reach a similar conclusion based on their discovery of a handful of unsaturated 1--1.4~M$_{\odot}$ stars on the activity-rotation diagram.

It is interesting to note that for older (70~Myr~--~9~Gyr) solar-mass stars \citet{Gudel1997} report a strong negative correlation between the temperature of the hot X-ray emitting plasma component and stellar age, implying that the frequency of larger X-ray flares associated with hotter plasma is decreasing with age. Our 7--25~Myr old l-PMS stars (averaged across all masses) have {\it Chandra}-based median hot temperature component of $\sim 40$~MK (Table~\ref{tab:xspec_fits}) and fit nicely into the {\it ASCA}-based hot temperature versus age relation of \citet{Gudel1997}. And so do the 440~Myr and 210~Myr old solar-mass stars $\epsilon$~Eri and Kepler-63, whose {\it XMM-Newton}-based hot plasma temperatures range between 9 and 12~MK \citep{Coffaro2020,Coffaro2022}. Since the heights of X-ray flaring coronal structures on young stars, older active stars, and the Sun positively correlate with plasma temperature \citep[Figure~7 in][]{Getman2021b}, it is reasonable to propose that the X-ray coronas of our solar-mass l-PMS stars are more extended than those of older ($t>70$~Myr) solar-mass stars, but are not as large as the coronas of solar-mass e-PMS stars. Our future work on X-ray super-flares from l-PMS stars will provide more quantitative treatment of this issue.

\citet{Coffaro2020,Coffaro2022} also find that the X-ray cycle amplitudes of $\epsilon$~Eri and Kepler-63 are the smallest (compared to several older solar-mass stars with known X-ray cycles) and their surfaces may be profusely (60-100\%) covered by solar-type X-ray emitting magnetic structures (such as active region cores and flares). Careful analyses of mass-stratified X-ray spectral stacking, mass-stratified super-flare energies and frequencies, and rotation rates for l-PMS stars need to be performed in order to examine whether the solar-mass l-PMS stars exhibit similar X-ray surface saturation.

{\bf The higher-mass 2--7~M$_{\odot}$ stars}~~ Samples of MYStIX/SFiNCs e-PMS stars are likely incomplete above $2$~M$_{\odot}$ and involvement of Gaia data will not recover all young members due to often high and variable source extinctions and effects of high source crowding and background nebula emission \citep{Getman2019}. Nevertheless, the upper envelopes of apparent X-ray distributions for e-PMS MYStIX/SFiNCs stars seem still higher than those of l-PMS (Figure~\ref{fig:lx_km85_vs_time}).

The X-ray luminosity keeps decreasing as $L_X \propto t^{-1.8}$ when 1--2 and 2--3.5~M$_{\odot}$ l-PMS stars are combined. Many of the 2--3.5~M$_{\odot}$ stars reach ZAMS but remain partially convective. It is likely that as their 1--2~M$_{\odot}$ siblings they are now powered by the $\alpha\Omega$ dynamo.

$L_X$ falls precipitously during the l-PMS phase (Figure~\ref{fig:lx_km85_vs_time}). For these masses, most of the l-PMS stars are undetected by {\it Chandra} (Figure~\ref{fig:lx_vs_mass}, all panels) so it is difficult to quantify this dropoff. Fitting a power law to the upper  envelope of the $L_X$ distribution suggests that $b \simeq -2$ to $-4$ in the $L_{X,85\%} \propto t^b$ relationship  (Table~\ref{tab:temporal_evolution_table}).  The dropoff is more dramatic in the 3.5--7~M$_\odot$ interval than in the 1--3.5~M$_\odot$ (Appendix \ref{sec:temporal_highest_masses}). 

Vast majority of the 3.5--7~M$_{\odot}$ stars are fully radiative and already reached the main sequence; a few are approaching their post-main sequence regime (Figure~\ref{fig:lbol_r_vs_time}). Many of them are A- and late-B type stars. Our findings of diminishing X-ray emission levels are consistent with the well known fact that such stars lack both the convection zones driving magnetic activity in lower-mass stars and the strong radiation driven stellar winds with shocks responsible for X-ray emission in O- and early B-type stars \citep[e.g.,][]{Gudel2009, Stelzer09, Drake2014, Nunez2021}. Occasionally, the X-ray emission is detected from some of such systems and probably is not produced by the massive primary, but rather by lower mass unresolved secondaries in multiple systems \citep{Stelzer05, Stelzer09}.

\section{Discussion: Effects on the molecular environs on PMS stars}
\label{sec:disks_atmos}

\subsection{X-ray effects on protoplanetary disks} \label{sec:disks}

High fluence ionizing PMS X-ray radiation is predicted to penetrate deep into molecular environments and substantially increase the degree of ionization, stimulate magnetorotational instability and associated turbulence, create conditions necessary for launching a magnetocentrifugal disk wind and a collimated jet, induce non-equilibrium ion-molecular chemistry, and sputter grain surfaces \citep[e.g.,][]{Glassgold2000,Fromang02,Shang2002,Ilgner06,Gressel13,Cleeves2017,Dupuy18,Waggoner2019,Waggoner2022}. 

X-rays are expected to play an important role in photoevaporative flows and dispersal of protoplanetary disks \citep[reviewed by][]{Williams2011, Alexander2014, Ercolano2017}. Disk photoevaporation linked to high-energy X-ray and ultraviolet radiation from the central star is now directly detected in a number of systems \citep{Alexander2014,Picogna2019}.  X-rays seem most important in the final stages of disk dispersal \citep{Owen13}. There is some empirical evidence that X-rays heat disks and diminish accretion due to photoevaporation of disks \citep{Drake2009,Flaccomio2018,Flaischlen2021}.

Within the framework of the hydrodynamic escape model by \citet{Owen2012}, the X-ray driven gas removal rate in the disk around a $1$~M$_{\odot}$ young star is $\dot{M}_{phot} = 6 \times 10^{-9} \times (L_X/10^{30})^{1.14}$~M$_{\odot}$~yr$^{-1}$. Figure \ref{fig:lx_comparison_with_johnstone} shows that for star age of 1~Myr, $L_X$ is $10^{30.2}$~erg~s$^{-1}$ based on the MYStIX/SFiNCs data but $10^{30.6}$~erg~s$^{-1}$ based on the modeling of \citet{Johnstone2021}. There is thus a factor of 3 discrepancy  between the disk gas rates expected from our empirical MYStIX/SFiNCs data and modeling by \citet{Johnstone2021}.

The bifurcation of e-PMS stars on the $\log(L_X) - \log(M)$ plane around $\log(L_X) \sim 30.5$~erg~s$^{-1}$ (Figure \ref{fig:xray_vs_mass_mystixsfincs}) is partially due to the presence of numerous large X-ray flares. The occurrence rate of super ($E_X>10^{34}$~erg) and mega ($E_X > 10^{36.2}$~erg) flares in solar-mass e-PMS stars is $\sim 200$ and $\sim 2$ flares per star per year, respectively \citep{Getman2021}. Large flares from PMS stars may significantly impact the physical and chemical evolution of circumstellar disks. Possible observational signatures of such an impact have been detected from the disk around the young solar-mass star IM Lupi via time-dependent flux variations of H$^{13}$CO$^{+}$ (J=3-2) ion in the sub-mm band observed by ALMA \citep{Cleeves2017}. The thermo-chemical disk models suggest that the formation of this ion is highly sensitive to the ionization rate of the disk through the formation of H$_{3}^{+}$, which reacts with the abundant molecules CO and N$_2$ to produce H$^{13}$CO$^{+}$ and N$_2$H$^{+}$, respectively \citep{Rab2017,Waggoner2022}. Disk chemistry simulations involving a large X-ray flare with the duration from several hours to 2 days predict enhanced ion abundances for up to 20 days \citep{Cleeves2017}. However, to date no X-ray observations have been performed in the same epoch as disk ion abundance variability \citep{Cleeves2017}. Simultaneous X-ray and sub-mm observations are required to test these scenarios. 

Over the typical 5~Myr disk-lifetime \citep{Richert18}, the disk around a solar-mass star is predicted to be irradiated by 1 billion super- and mega-flares with X-ray energies $34 < \log(E_X) < 38$~erg \citep{Getman2021}. Such large flares are expected to be accompanied by fast and energetic coronal mass ejections propagating at a few thousands of km~s$^{-1}$ \citep{Alvarado-Gomez2022}. Such fast CMEs are expected to initiate strong shocks and drive high fluence hard spectra stellar energetic particles, mostly protons, penetrating the equatorial regions of the disk at over 100 AU from the star, and igniting significant ionization with efficient production of H$^{13}$CO$^{+}$ and N$_2$H$^{+}$ ions on a time scale of a few weeks \citep{Rab2017,Rab2020,Hu2022}. 

The present open cluster observations contribute little to the X-ray effects on disks as the disks are mostly gone by $5-7$ Myr \citep{Richert18}.

\subsection{Impact of X-ray and associated EUV emission on primordial planetary atmospheres}
\label{sec:planatmos}

As shown in Figure \ref{fig:lx_vs_time}, the median time-averaged X-ray luminosity for 1--2~M$_{\odot}$ stars and the 75\% $L_X$ quartile for 0.75--1~M$_{\odot}$ PMS stars can reach up to $3 \times 10^{30}$~erg~s$^{-1}$ in the first few Myr of evolution. Assuming that $L_{EUV} \sim 4 \times L_{X}$ in the $100-1200$~\AA\ band, the total X-ray and EUV (XUV) power for many solar-mass e-PMS stars may exceed $10^{31}$~erg~s$^{-1}$.  This is over 1000 times higher than the current Sun's time-averaged XUV luminosity of $1.5 \times 10^{28}$~erg~s$^{-1}$ \citep{Airapetian2020,Johnstone2020}. While this enhanced XUV flux declines during the l-PMS phase in a mass-dependent fashion (\S\ref{sec:xray_activity_slopes}) when  protoplanetary disks have mostly dissipated, the PMS XUV irradiance should have significant impact on atmospheric evolution of primary atmospheres of protoplanets and young exoplanets. 	

The stellar XUV radiation is absorbed by atmospheric gas producing excitation, dissociation, and ionization of atmospheric species. Photoionization generates photoelectrons with energies sufficient for the subsequent ionization and excitation of atomic and molecular hydrogen. These photoelectrons excite secondary electrons that lose their excess kinetic energy via Coulomb collisions with the ambient atmospheric particles, and contribute to the heating and expansion of exoplanetary upper atmospheres \citep{Glocer2012,Airapetian2017}. This in turn can drive several thermal, nonthermal, and chemical escape processes forming the ionosphere and thermosphere \citep{Airapetian2020,Gronoff2020}.  

The XUV driven heating rate depends critically on the chemical composition of the primordial planetary atmosphere  because different atmospheric species have different wavelength-dependent absorption cross sections. Hydrogen-rich primordial atmospheres of protoplanets around PMS stars efficiently absorb stellar radiation shortward of 1120~\AA~~via molecular and atomic hydrogen. XUV driven heating increases the temperature at the exobase, the atmospheric layer, where the particle mean free path is comparable to the pressure scale height. In this layer, energetic particles from the tail of Maxwellian distribution move with outgoing velocity exceeding the planet's escape velocity and thus escape into space. The escape rate is controlled by the Jean's escape parameter $\lambda_C$, represented by the ratio of gravitational energy to the mean particle's thermal energy.  For a pure hydrogen atmosphere, the hydrodynamic escape occurs at $\lambda_C < 3.5$ or at the exospheric temperatures over 4,200~K. 

However, the atmosphere should also contain rock vapor formed due to frequent impacts in the early phase of evolution of protoatmospheres \citep{Canup2004}.  This may raise the temperature in excess of 21,000~K required for the onset of hydrodynamic escape via adiabatic expansion \citep{Massol2016}.

Recent hydrodynamic and magnetohydrodynamic models provide satisfactory agreement with observational signatures of atmospheric escape from hot giant and mini-Neptune exoplanets \citep{Johnstone2020,GarciaMunoz2021,Shaikhislamov2020,Khodachenko2021,Zhang2022}. In particular, the measurement of Ly$\alpha$ absorption during transits of mini-Neptune HD~63433c orbiting at 0.145 AU from a young (440 Myr) solar-like G5 star leads to an estimate of the hydrodynamic escape of a hydrogen rich atmosphere at  $3 \times 10^{10}$~g~s$^{-1}$ \citep{Zhang2022}. The stellar XUV fluence is estimated to be $\sim 1000$ times of the Sun's XUV flux at 1~AU.  HD 63433c has a surface gravity of the Earth and is thus a reasonable representative case for a rocky exoplanet orbiting within $\sim 0.1$~AU.  Its inferred atmospheric escape rate is close to a theoretical calculation of $\sim 1.5 \times 10^{10}$~g~s$^{-1}$ from an Earth-like steam (H$_{2}$O-rich) atmosphere subject to similar level of irradiation \citep{Johnstone2020}.

According to \citet{Lammer2014}, a protoplanet with an Earth-mass rocky core can capture hydrogen envelopes between $\sim 5 \times 10^{22}$~g and $\sim 1.5 \times 10^{24}$~g.  The hydrogen rich atmosphere of an Earth mass planet exposed to the 1000$F_{X,Sun}$ flux at 1~AU from an e-PMS host star will escape at the rate of $3 \times 10^{10}$~g~s$^{-1}$, which suggests that the planet will lose its primary atmosphere within $0.1-2$~Myr, considerably shorter than the span of the e-PMS phase for stars with masses $\geq 2.5$~M$_\odot$ stars (Figure~\ref{fig:lbol_r_vs_time}). We thus find that e-PMS XUV radiation may lead to the loss of the entire primordial atmosphere of rocky planets orbiting within $\sim 1$~AU. 

The atmospheric loss discussed so far is driven by a continuous ``characteristic'' XUV flux from a superposition of weak flares. Mega-flares with X-ray energies of $>10^{36}$~erg are superposed on this characteristic emission and can be individually studied in {\it Chandra} observations.  They have occurrence rates of a few flares per year from solar-mass e-PMS stars, contributing an additional $10-20$\% to the characteristic X-ray fluence of e-PMS stars \citep{Getman2021}. Mega-flares produce more energetic X-rays that will penetrate more deeply into the planetary atmosphere. If the response of the planetary atmosphere escape to the additional ionization is much slower than the mega-flare timescale ($\sim 1$~day), then the total effect of the mega-flares will be modest. But if the response is rapid, then the effects can be substantial.  

Rapid atmospheric loss of a primary atmosphere will have other effects on the planetary evolution.  At typical distances of $\sim$1~AU atmospheric loss drives magma ocean crystallization, while close-in exoplanets will undergo strong tidal heating \citep{Kite2020}. At this stage, if an Earth-like planet is massive enough, its mantle convection can drive formation of crustal rock and water oceans with subsequent volcanic and tectonic outgassing forming a secondary atmosphere. The mantle redox state affects the volatile partitioning during mantle melting and its volatile speciation close to the surface \citep{Ortenzi2020}. Mantle oxygen fugacity also plays a significant role in specifying the atmospheric thickness. Earth-like planets with high oxygen fugacity can regenerate a high mean molecular weight  (N$_2$-CO$_{2}$-H$_{2}$O rich) atmosphere. 

Recent studies suggest that the fate of secondary atmospheres forming during $\sim$30 Myr after the loss of the primary atmosphere strongly depends on the stellar XUV flux and is not sensitive to CO$_2$ abundance \citep{JohnstoneLammer2021}.  For XUV fluxes of $\sim 30 \times$ solar, even in the case of 100\% CO$_2$ abundance the atmosphere will experience significant hydrodynamic escape. 

At stellar ages $\sim 10$~Myr, we find that l-PMS stars with masses 0.75--1~M$_{\odot}$ and 1--2~M$_{\odot}$ have typical median $\log(L_X)$ of 29.7~erg~s$^{-1}$ and 75\%-quartiles up to $\log(L_X) \sim 30.2$~erg~s$^{-1}$ (\S\ref{sec:xray_activity_slopes}). Assuming the X-ray-to-EUV conversion of $L_{EUV} = 4 \times L_X$, the XUV luminosities of many $\sim 10$~Myr solar-mass l-PMS stars would be in the range $(150-550) \times L_{XUV,{\odot}}$. At such high XUV fluxes, the photoionization driven energy deposition will drive hydrodynamic escape of secondary atmospheres at the rates $\sim 3 \times 10^9$~g~s$^{-1}$ from an unmagnetized planet and $\gtrsim 7$ times lower rates from a magnetized Earth-like planet.  This suggests a dense 100 bar CO$_2$-rich atmosphere will evaporate in $\sim$10 Myr (unmagnetized) and $\sim 100$ Myr (magnetized case) unless it is replenished by internal dynamics processes including volcanic, tectonic processes and tidally induced outgassing.  We thus conclude that l-PMS XUV emission may (partially) remove the early secondary atmosphere as well as the primordial atmosphere of an Earth mass planet unless the planet migrated to 1 AU from the outer regions of the disk at $\geq$ 5~AU \citep{Batygin2015} and its secondary atmosphere can be replenished during its evolution.

\section{Future Work} \label{sec:future_work}

The current paper is the first of a series of planned studies based on the {\it Chandra} study of magnetic activity in 7--25~Myr and older ZAMS star clusters.  

First, we will examine the relation of $L_X$ to stellar rotation in our l-PMS clusters. Rotation period analyses for numerous l-PMS stars are underway using lightcurves from the TESS satellite and Zwicky Transient Facility survey. Both $L_X$ and $P_{rot}$ are evolving rapidly during the l-PMS phase, and the distribution of stellar angular momenta splits into slow- and rapidly-rotating tracks \citep{Barnes03, Wright11, Vidotto2014}. \citet{Argiroffi2016} has examined X-ray and rotation in NGC 869 = h Per and found hints that by $\sim 13$~Myr some young stars have drifted towards the ``non-saturation'' locus on the activity-age plane, a manifestation of possible changes in internal dynamos. Our deeper X-ray sensitivity for h Per, and expansion to other l-PMS clusters of different ages, should give samples sufficiently rich to capture the effects of rapid changes in stellar structure on surface activity as stars approach the ZAMS. 

Second, we will extend the {\it Chandra} findings on super- and mega-flares in the e-PMS phase \citep{Getman2021,Getman2021b} to the ten clusters in the l-PMS phase.  These studies provide a homogeneous methodology to quantify the  discovery, frequencies, energetics, and flare loop geometries of powerful X-ray flares. Combining flare properties for the younger e-PMS MYStIX and SFiNCs cluster members to the older l-PMS open clusters discussed here will extend empirical relations on flare evolution and magnetic loop geometries.  Flares in the l-PMS phase are particularly important for understanding effects of X-rays on planetary atmospheres (\S\ref{sec:planatmos}). 

Third, we will extend our {\it Chandra} observations to older clusters in the 25--150~Myr age range. The $\alpha\Omega$ dynamo should now be the sole process for generating surface magnetic fields of many $>0.7$~M$_{\odot}$ PMS stars.  The decay of characteristic and flare XUV luminosities can be quantified for improved modeling of planetary atmospheres.  Fast- and slow-rotating stars may have different XUV evolutions.

\section{Conclusions} \label{sec:conclusions}

Using new {\it Chandra} observations of $\sim 6000$ stellar members of 10 open clusters with ages 7--25~Myr, we have improved characterization of magnetic activity for the late pre-main sequence (l-PMS) phase of stellar evolution.  In addition to quantifying the evolution of X-ray activity in young stars for a range of stellar masses (\S\S \ref{sec:xrays_vs_mass}-\ref{sec:xray_activity_slopes}), the results give insight into two astrophysical issues: the response of magnetic dynamo processes to rapid l-PMS changes in interior structure (\S\S \ref{sec:intro2}, \ref{sec:dynamo_activity}), and the effects of high-energy radiation on protoplanetary disks and primordial planetary atmospheres (\S \ref{sec:disks_atmos}).

Our expanded sample of stars in the 7--25~Myr age range benefits from a joint analysis of X-ray emitting and X-ray non-emitting cluster members from {\it Chandra} and {\it Gaia} surveys (\S\S \ref{sec:targets_chandra}-\ref{sec:Xray_luminosities}, Table 5).  These are then combined with previous studies of e-PMS MYStIX and SFiNCs star forming regions (\S\ref{sec:mystix_sfincs}, Table~\ref{tab:mystix_sfincs_table}) for a systematic study of mass-stratified magnetic activity measurement in the wider age range of 0.5--25~Myr that covers the e-PMS, l-PMS, and ZAMS phases (\S \ref{sec:xrays_vs_mass}, \ref{sec:xray_activity_slopes}).  The samples are analyzed in a homogeneous fashion with statistical methods correctly treating X-ray nondetections.  Due to wide scatter in X-ray luminosities, X-ray luminosity function quantiles, as well as medians, are provided when available (Figures~\ref{fig:xray_vs_mass_mystixsfincs}-\ref{fig:lx_vs_mass}, Appendix).    

The results improve the previous studies of mass-stratified  activity-age relations such as \citet{PreibischFeigelson2005} and \citet{Gregory16}. We do not find decreases in X-ray luminosity during the e-PMS phase and report steeper temporal decays for later l-PMS phases (\S \ref{sec:prev_literature}). Our empirical distributions of PMS X-ray luminosities have wider spreads, large $L_X-t$ slope differences for the e-PMS phase, and small systematic differences in the median $L_X$ for the l-PMS phase than those derived from semi-analytic rotation evolution models \citep{Johnstone2021}. 

In \S \ref{sec:dynamo_activity} we discuss the magnetic origin of the inferred activity-age trends. X-ray luminosity is constant during the e-PMS phase despite strong drops in stellar luminosity and radius.  This may indicate that extended X-ray emitting loops are insensitive to stellar size. The X-ray luminosity decays rapidly during the 7--25~Myr period: $L_X \propto t^{-0.5}$ for $<1$~M$_{\odot}$ stars, $L_X \propto t^{-1.8}$ for 1--3.5~M$_{\odot}$ stars, and $L_X \propto t^{-4}$ and steeper for intermediate-mass stars. This is attributed to decreasing efficiency of the $\alpha^2$ dynamo at $<1$~M$_{\odot}$ regime, switch to $\alpha\Omega$ dynamo at 1--3.5~M$_{\odot}$ regime as radiative cores grow, and the lack of any dynamo as the more massive stars become fully radiative. 

Our findings provide improved empirical inputs into calculations of the effects of high energy radiation on the young stellar environment, first the protoplanetary disk and then the atmospheres of young planets (\S \ref{sec:disks_atmos}). A few estimations are made. The disk around a solar-mass e-PMS star can be photoevaporated within a few Myr years; but our empirical gas rate estimates may be a few times lower than those based on the semi-analytic rotation evolution approach of \citet{Johnstone2021}. X-ray luminosity values inferred here for PMS stars imply a very rapid escape rate of a primordial hydrogen-rich atmosphere in an Earth-like planet on a 1~AU-orbit around a solar-mass e-PMS star of $\sim 3 \times 10^{10}$~g~s$^{-1}$, suggesting a complete loss of the atmosphere within 2~Myr. An early secondary CO$_2$-rich atmosphere might also be evaporated depending on the planet's magnetization.

\acknowledgments
We thank the anonymous referee for his time and many useful comments that improved this work. We thank L. Townsley (Penn State) for sharing with us the {\it Chandra} data products for the open clusters NGC 3293 and NGC 869. This project is supported by the {\it Chandra} grant GO9-20011X (K. Getman, Principal Investigator) and the {\it Chandra} ACIS Team contract SV474018 (G. Garmire \& L. Townsley, Principal Investigators), issued by the {\it Chandra} X-ray Center, which is operated by the Smithsonian Astrophysical Observatory for and on behalf of NASA under contract NAS8-03060. The {\it Chandra} Guaranteed Time Observations (GTO) data used here were selected by the ACIS Instrument Principal Investigator, Gordon P. Garmire, of the Huntingdon Institute for X-ray Astronomy, LLC, which is under contract to the Smithsonian Astrophysical Observatory; contract SV2-82024. The work of T.P. was supported by the Deutsche Forschungsgemeinschaft (DFG, German Research Foundation) under project numbers 325594231 and 362051796. V.S. Airaoetian acknowledges support from the NASA/GSFC Sellers Exoplanet Environments Collaboration (SEEC), which is funded by the NASA Planetary Science Division’s Internal Scientist Funding Model (ISFM). This work has made use of data from the European Space Agency (ESA) mission {\it Gaia} (\url{https://www.cosmos.esa.int/gaia}), processed by the {\it Gaia} Data Processing and Analysis Consortium (DPAC, \url{https://www.cosmos.esa.int/web/gaia/dpac/consortium}). Funding for the DPAC has been provided by national institutions, in particular the institutions participating in the {\it Gaia} Multilateral Agreement.  

\vspace{5mm}
\facilities{CXO, Gaia}

\software{ACIS Extract \citep{Broos10}, 
        R \citep{RCoreTeam20}, XSPEC \citep{Arnaud1996}}

\clearpage
\appendix

\section{Chandra Observations Of The 10 Open Clusters} \label{sec:appendix_chandra_log_table}
This section presents Table \ref{tab:log_chandra_observations} detailing the {\it Chandra} observations of the 10 open clusters.

\begin{deluxetable}{lrrcccr}
\tabletypesize{\small}
\tablecaption{Log of {\it Chandra}-ACIS-I Observations  \label{tab:log_chandra_observations}}
\tablewidth{0pt}
\tablehead{
\colhead{Cluster} & \colhead{ObsId} &
\colhead{Exposure} & \colhead{R.A.} & \colhead{Decl.} & \colhead{Start Time} & \colhead{PI}\\
\colhead{} & \colhead{} &
\colhead{(ksec)} & \colhead{$\alpha_{J2000}$} & \colhead{$\delta_{J2000}$} & \colhead{(UT)} & \colhead{}\\
\colhead{(1)} & \colhead{(2)} & \colhead{(3)} & \colhead{(4)} & \colhead{(5)} & \colhead{(6)} &
\colhead{(7)}
}
\startdata
NGC 1502 & 21138	& 66.6~~~~~ &	04:07:50.00 &	+62:19:54.00	& 2019-04-30T13:26:51 & G. Garmire\\
NGC 2169 & 21139  & 30.7~~~~~ &  06:08:24.00	& +13:57:54.00  & 2018-12-10T08:15:05 & G. Garmire\\
NGC 2169 & 21997  & 33.3~~~~~ &  06:08:24.00	& +13:57:54.00  & 2018-12-14T10:34:54 & G. Garmire\\
IC 2395  & 21137  & 23.8~~~~~ &  08:42:30.00	& -48:07:00.00  & 2020-08-19T05:06:54 & G. Garmire\\
IC 2395  & 24613  & 41.5~~~~~ &  08:42:30.00	& -48:07:00.00  & 2020-08-19T22:14:15 & G. Garmire\\
NGC 3293 & 16648	& 70.9~~~~~ &  10:35:50.00	& -58:14:00.00	& 2015-10-07T10:13:16 & T. Preibisch\\
NGC 869	 & 5407	  &	41.1~~~~~	& 02:19:02.20 &	+57:07:12.00  &	2004-12-02T06:28:44	& N. Evans\\	 	 	 
NGC 869  & 9912	  &	101.7~~~~~	& 02:19:02.20 &	+57:07:12.00  &	2009-11-11T07:20:00	& G. Micela\\	 	 	 
NGC 869  & 9913	  &	36.7~~~~~	& 02:19:02.20 &	+57:07:12.00  &	2009-10-16T05:56:43	& G. Micela\\	 	 	 
NGC 869  & 12021	&	51.4~~~~~	& 02:19:02.20 &	+57:07:12.00  &	2009-11-08T22:58:35	& G. Micela\\ 
NGC 884  &  21172	& 25.7~~~~~	& 02:22:00.40	& +57:07:40.00	& 2019-11-13T16:17:56	& K. Getman\\	 	 	 
NGC 884  &	21173	& 16.8~~~~~	& 02:22:00.40	& +57:07:40.00	&	2020-03-30T03:23:21	& K. Getman\\	 	 	 
NGC 884  &	23070	& 24.7~~~~~	& 02:22:00.40	& +57:07:40.00	&	2019-11-14T07:20:29	& K. Getman\\
NGC 884  &	23071 & 25.8~~~~~	& 02:22:00.40	& +57:07:40.00	&	2019-11-15T10:07:30	& K. Getman\\
NGC 884  &	23072 & 19.8~~~~~	& 02:22:00.40	& +57:07:40.00	& 2019-11-17T15:28:10	& K. Getman\\
NGC 884  &	23202	& 53.5~~~~~	& 02:22:00.40	& +57:07:40.00	&	2020-03-31T14:11:34	& K. Getman\\
NGC 884  &	23203	& 15.9~~~~~	& 02:22:00.40	& +57:07:40.00	& 2020-04-01T18:53:32	& K. Getman\\
NGC 4755 & 21169	& 45.5~~~~~  &	12:53:40.00	& -60:22:05.00	& 2019-12-02T03:52:00	& K. Getman\\	 	 	 
NGC 4755 & 22889	& 47.4~~~~~	&	12:53:40.00	& -60:22:05.00	&	2019-11-14T16:55:24	& K. Getman\\	 	 	 
NGC 4755 & 22890	& 49.4~~~~~	&	12:53:40.00	& -60:22:05.00	&	2019-11-25T03:52:48	& K. Getman\\
NGC 1960 & 21168	& 17.8~~~~~	& 05:36:18.00	& +34:08:24.00	& 2018-12-26T19:57:53 & K. Getman\\
NGC 1960 & 22031	& 21.8~~~~~	& 05:36:18.00	& +34:08:24.00	& 2018-12-27T11:21:49	& K. Getman\\	 
NGC 1960 & 22032	& 14.9~~~~~	& 05:36:18.00	& +34:08:24.00	& 2019-01-01T00:06:59	& K. Getman\\ 
NGC 1960 & 22033	& 11.9~~~~~	& 05:36:18.00	& +34:08:24.00	& 2019-01-06T23:51:05 & K. Getman\\
NGC 3766 & 21170	& 49.4~~~~~	& 11:36:14.00 &	-61:36:30.00	& 2019-11-05T18:31:43 & K. Getman\\
NGC 3766 & 21171	& 24.7~~~~~	& 11:36:14.00 &	-61:36:30.00	&	2019-12-10T10:16:50	& K. Getman\\
NGC 3766 & 22891	& 88.9~~~~~	& 11:36:14.00 &	-61:36:30.00	&	2019-11-16T07:38:03 & K. Getman\\
NGC 3766 & 23094	& 24.7~~~~~	& 11:36:14.00 &	-61:36:30.00	&	2019-12-11T04:41:11	& K. Getman\\
NGC 2232 & 22895	& 9.9~~~~~   &	06:27:56.15	& -05:01:53.30	& 2020-01-06T08:03:26	& G. Garmire\\ 
NGC 2232 & 22896	& 9.8~~~~~   &	06:27:55.95 &	-04:44:56.20	& 2020-01-06T11:10:33	& G. Garmire\\
NGC 2232 & 22897	& 9.8~~~~~   &	06:29:04.60	& -05:01:50.50	& 2020-01-06T14:05:44	& G. Garmire\\
NGC 2232 & 22898	& 9.9~~~~~	  & 06:26:47.56	& -05:01:49.20	& 2020-01-11T02:48:39	& G. Garmire\\
NGC 2232 & 22899	& 9.7~~~~~   &	06:27:55.77	& -04:27:46.80	& 2020-01-12T20:08:15	& G. Garmire\\
NGC 2232 & 22900  & 9.9~~~~~  	& 06:29:04.13	& -04:44:48.00	& 2020-01-17T01:11:08	& G. Garmire\\
NGC 2232 & 22901	& 9.8~~~~~	  & 06:26:47.73	& -04:44:47.08	& 2020-01-17T04:26:33	& G. Garmire\\
NGC 2232 & 22902	& 9.8~~~~~  	& 06:29:03.90	& -04:27:46.90	& 2020-01-17T07:21:49	& G. Garmire\\
NGC 2232 & 22903	& 9.9~~~~~   &	06:26:47.69	& -04:27:47.90	& 2020-01-13T21:21:24	& G. Garmire\\
\enddata
\tablecomments{Column 1: Cluster name. Columns 2-3: {\it Chandra}-ACIS-I observation identifier (ObsID) and net exposure time in kiloseconds. Columns 4-5:  Aimpoint of the {\it Chandra} observation in right ascension and declination.  Column 6: Start time of {\it Chandra} exposure. Column 7: Principal investigator of {\it Chandra} observation.}
\end{deluxetable}

\section{Evolution of Fractional X-ray Luminosity and X-ray Flux in PMS Stars} \label{sec:appendix_temporal_fx_lxlbol_pms}

\begin{figure*}[ht!]
\epsscale{1.15}
\plotone{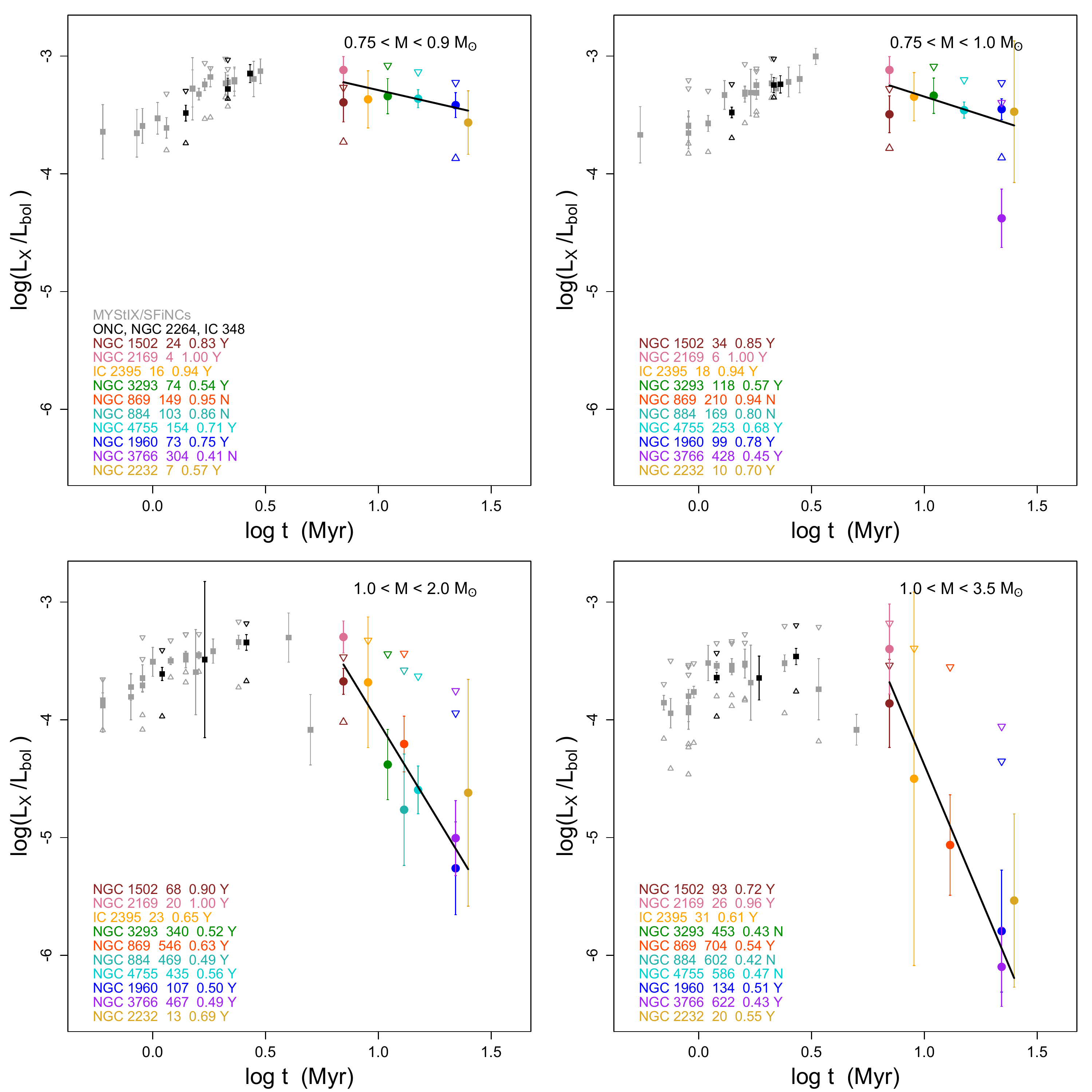}
\caption{Temporal evolution of fractional X-ray luminosity ($L_X/L_{bol}$) for four mass strata. See Figure~\ref{fig:lx_vs_time} for a detailed description. \label{fig:lxlbol_vs_time}}
\end{figure*}

\begin{figure*}[ht!]
\epsscale{1.15}
\plotone{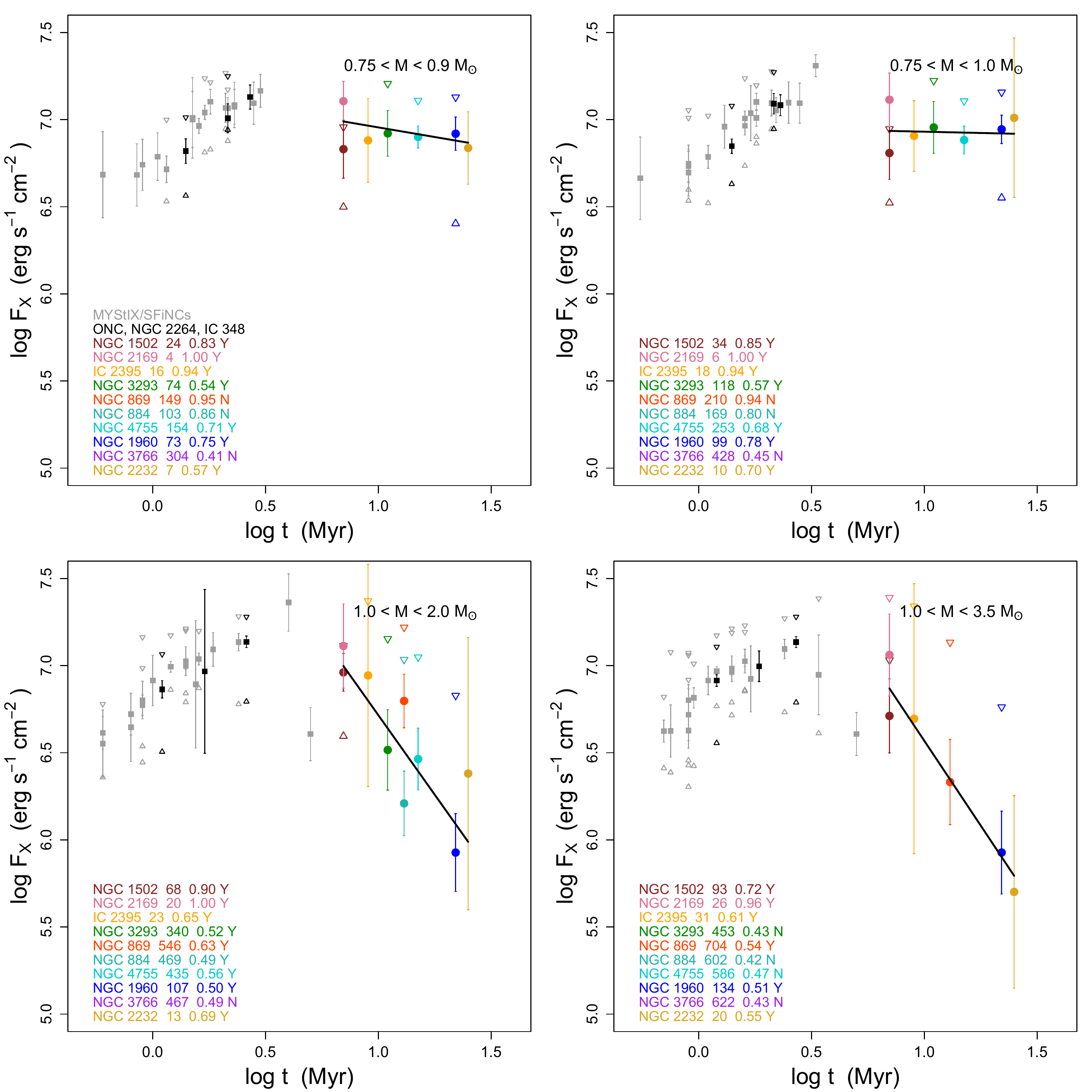}
\caption{Temporal evolution of X-ray surface flux for four mass strata. See Figure~\ref{fig:lx_vs_time} for a detailed description. \label{fig:fx_vs_time}}
\end{figure*}

In \S~\ref{sec:xray_activity_slopes} we report mass-stratified temporal evolution of PMS X-ray luminosity and in \S~\ref{sec:dynamo_activity} we link these findings to possible changes in underlying interior dynamos and X-ray coronal structures.  
In addition to the temporal evolution of X-ray luminosity itself (Figure~\ref{fig:lx_vs_time}), here Figures~\ref{fig:lxlbol_vs_time} and \ref{fig:fx_vs_time} show the evolution of X-ray luminosity normalized by the stellar bolometric luminosity and stellar surface. As mathematical expressions of normalized X-ray luminosities, their temporal changes can be explained by invoking the changes in the X-ray luminosity and stellar interior.

For younger e-PMS MYStIX/SFiNCs stars, X-ray luminosities remain nearly constant in time (Figure~\ref{fig:lx_vs_time}) but bolometric luminosities keep decreasing with time (for vast majority of the stars) as stars travel along Hayashi tracks, resulting in positive correlations between $L_X/L_{bol}$ and $t$ (Figure~\ref{fig:lxlbol_vs_time}). For older open cluster stars, within the mass range 0.75--1~M$_{\odot}$, the temporal rise and decay of the bolometric luminosities for lower- and higher-mass stars, respectively, counterbalance each other (Figure~\ref{fig:lbol_r_vs_time}) resulting in small integral temporal changes of $L_{bol}$ and similar decay slopes $b \sim -0.6$ in the $L_X \propto t^b$ and $L_X/L_{bol} \propto t^b$ relations (Table~\ref{tab:temporal_evolution_table}). Within the mass range 1--3.5~M$_{\odot}$, the bolometric luminosities of open cluster stars, which lie mainly on Henyey or ZAMS tracks, either keep rising with time or settle at a constant level, resulting on-average in positive temporal changes and thus steeper decay slopes in the $L_X/L_{bol} - t$ versus $L_X - t$ relations (Table~\ref{tab:temporal_evolution_table}).


Younger e-PMS MYStIX/SFiNCs stars are still gravitationally contracting, and since their X-ray luminosities remain constant with time their X-ray fluxes appear to increase with time (Figure~\ref{fig:fx_vs_time}). For older open cluster stars with masses 0.75--1~M$_{\odot}$, their surface areas keep decreasing with time as approximately $\propto t^{-0.4}$ leading to a mild or no decay of stellar flux (Table~\ref{tab:temporal_evolution_table}, Figure~\ref{fig:fx_vs_time}). Many 1--3.5~M$_{\odot}$ open cluster stars are on Henyey or ZAMS tracks; since they stop contracting, their $F_X - t$ and $L_X - t$ relations exhibit similar decay slopes of $b \sim -2$ (Table~\ref{tab:temporal_evolution_table}).

\section{Evolution of 85\% Quantiles for X-ray Luminosities of High-mass Stars} \label{sec:temporal_highest_masses}

\begin{figure*}[ht!]
\epsscale{1.15}
\plotone{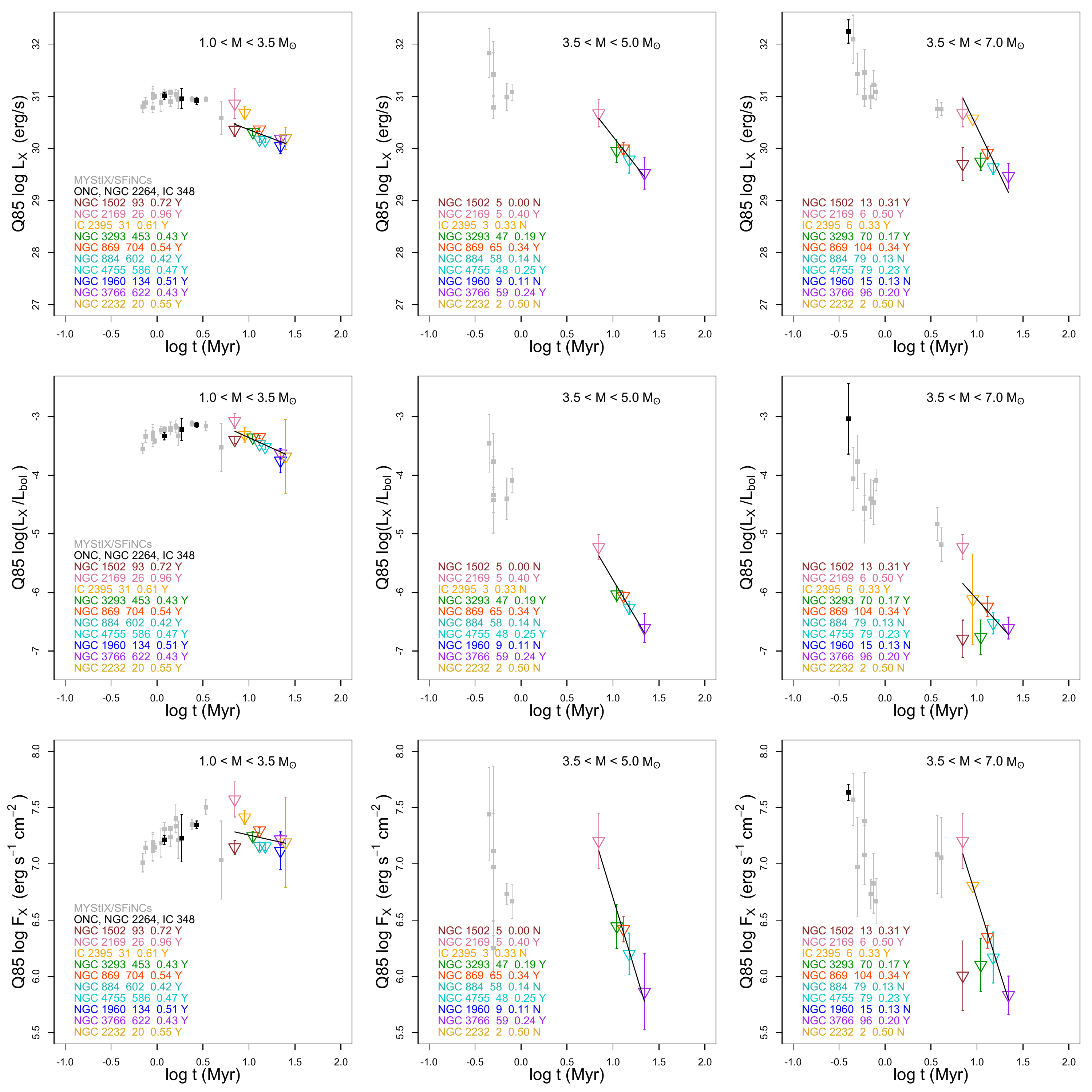}
\caption{Evolution of the 85\% quantiles of the Kaplan-Meier estimator of the X-ray luminosity (upper row), fractional X-ray luminosity (middle row), and X-ray surface flux (bottom row) in three mass strata. See Figure \ref{fig:lx_vs_time} for details. \label{fig:lx_km85_vs_time}}
\end{figure*}

For the higher-mass $\gtrsim 3.5$~M$_{\odot}$ ZAMS and post-ZAMS stars, the ratio of the non-X-ray to X-ray stars is so high that only 85\% quartiles are available for the KM estimators of $L_X$, $L_X/L_{bol}$, and $F_X$. Linear regression results are provided in Table~\ref{tab:temporal_evolution_table} and we present here in Figure \ref{fig:lx_km85_vs_time} plots showing the individual 7--25~Myr clusters.  The decay of the X-ray luminosity, fractional X-ray luminosity, and surface X-ray flux is much stronger for the higher-mass stars (3.5--7~M$_{\odot}$) than for the lower-mass stars (1--3.5~M$_{\odot}$).

\clearpage
\newpage

\bibliography{my_bibliography}{}

\begin{thebibliography}{}
\expandafter\ifx\csname natexlab\endcsname\relax\def\natexlab#1{#1}\fi
\providecommand{\url}[1]{\href{#1}{#1}}
\providecommand{\dodoi}[1]{doi:~\href{http://doi.org/#1}{\nolinkurl{#1}}}
\providecommand{\doeprint}[1]{\href{http://ascl.net/#1}{\nolinkurl{http://ascl.net/#1}}}
\providecommand{\doarXiv}[1]{\href{https://arxiv.org/abs/#1}{\nolinkurl{https://arxiv.org/abs/#1}}}

\bibitem[{{Airapetian} {et~al.}(2017){Airapetian}, {Glocer}, {Khazanov},
  {Loyd}, {France}, {Sojka}, {Danchi}, \& {Liemohn}}]{Airapetian2017}
{Airapetian}, V.~S., {Glocer}, A., {Khazanov}, G.~V., {et~al.} 2017, \apjl,
  836, L3, \dodoi{10.3847/2041-8213/836/1/L3}

\bibitem[{{Airapetian} {et~al.}(2020){Airapetian}, {Barnes}, {Cohen},
  {Collinson}, {Danchi}, {Dong}, {Del Genio}, {France}, {Garcia-Sage},
  {Glocer}, {Gopalswamy}, {Grenfell}, {Gronoff}, {G{\"u}del}, {Herbst},
  {Henning}, {Jackman}, {Jin}, {Johnstone}, {Kaltenegger}, {Kay}, {Kobayashi},
  {Kuang}, {Li}, {Lynch}, {L{\"u}ftinger}, {Luhmann}, {Maehara}, {Mlynczak},
  {Notsu}, {Osten}, {Ramirez}, {Rugheimer}, {Scheucher}, {Schlieder},
  {Shibata}, {Sousa-Silva}, {Stamenkovi{\'c}}, {Strangeway}, {Usmanov},
  {Vergados}, {Verkhoglyadova}, {Vidotto}, {Voytek}, {Way}, {Zank}, \&
  {Yamashiki}}]{Airapetian2020}
{Airapetian}, V.~S., {Barnes}, R., {Cohen}, O., {et~al.} 2020, International
  Journal of Astrobiology, 19, 136, \dodoi{10.1017/S1473550419000132}

\bibitem[{{Alexander} \& {Preibisch}(2012)}]{Alexander2012}
{Alexander}, F., \& {Preibisch}, T. 2012, \aap, 539, A64,
  \dodoi{10.1051/0004-6361/201118100}

\bibitem[{{Alexander} {et~al.}(2014){Alexander}, {Pascucci}, {Andrews},
  {Armitage}, \& {Cieza}}]{Alexander2014}
{Alexander}, R., {Pascucci}, I., {Andrews}, S., {Armitage}, P., \& {Cieza}, L.
  2014, in Protostars and Planets VI, ed. H.~{Beuther}, R.~S. {Klessen}, C.~P.
  {Dullemond}, \& T.~{Henning}, 475,
  \dodoi{10.2458/azu_uapress_9780816531240-ch021}

\bibitem[{{Alvarado-G{\'o}mez} {et~al.}(2022){Alvarado-G{\'o}mez}, {Cohen},
  {Drake}, {Fraschetti}, {Poppenh{\"a}ger}, {Garraffo}, {Chebly}, {Ilin},
  {Harbach}, \& {Kochukhov}}]{Alvarado-Gomez2022}
{Alvarado-G{\'o}mez}, J.~D., {Cohen}, O., {Drake}, J.~J., {et~al.} 2022, arXiv
  e-prints, arXiv:2202.07949.
\newblock \doarXiv{2202.07949}

\bibitem[{{Anders} \& {Grevesse}(1989)}]{AndersGrevesse1989}
{Anders}, E., \& {Grevesse}, N. 1989, \gca, 53, 197,
  \dodoi{10.1016/0016-7037(89)90286-X}

\bibitem[{{Anders} {et~al.}(2022){Anders}, {Khalatyan}, {Queiroz}, {Chiappini},
  {Ard{\`e}vol}, {Casamiquela}, {Figueras}, {Jim{\'e}nez-Arranz}, {Jordi},
  {Mongui{\'o}}, {Romero-G{\'o}mez}, {Altamirano}, {Antoja}, {Assaad},
  {Cantat-Gaudin}, {Castro-Ginard}, {Enke}, {Girardi}, {Guiglion}, {Khan},
  {Luri}, {Miglio}, {Minchev}, {Ramos}, {Santiago}, \&
  {Steinmetz}}]{Anders2022}
{Anders}, F., {Khalatyan}, A., {Queiroz}, A.~B.~A., {et~al.} 2022, \aap, 658,
  A91, \dodoi{10.1051/0004-6361/202142369}

\bibitem[{{Argiroffi} {et~al.}(2016){Argiroffi}, {Caramazza}, {Micela},
  {Sciortino}, {Moraux}, {Bouvier}, \& {Flaccomio}}]{Argiroffi2016}
{Argiroffi}, C., {Caramazza}, M., {Micela}, G., {et~al.} 2016, \aap, 589, A113,
  \dodoi{10.1051/0004-6361/201526539}

\bibitem[{{Argiroffi} {et~al.}(2017){Argiroffi}, {Drake}, {Bonito}, {Orlando},
  {Peres}, \& {Miceli}}]{Argiroffi2017}
{Argiroffi}, C., {Drake}, J.~J., {Bonito}, R., {et~al.} 2017, \aap, 607, A14,
  \dodoi{10.1051/0004-6361/201731342}

\bibitem[{{Arnaud}(1996)}]{Arnaud1996}
{Arnaud}, K.~A. 1996, in Astronomical Society of the Pacific Conference Series,
  Vol. 101, Astronomical Data Analysis Software and Systems V, ed. G.~H.
  {Jacoby} \& J.~{Barnes}, 17

\bibitem[{{Bailer-Jones} {et~al.}(2018){Bailer-Jones}, {Rybizki}, {Fouesneau},
  {Mantelet}, \& {Andrae}}]{Bailer-Jones2018}
{Bailer-Jones}, C.~A.~L., {Rybizki}, J., {Fouesneau}, M., {Mantelet}, G., \&
  {Andrae}, R. 2018, \aj, 156, 58, \dodoi{10.3847/1538-3881/aacb21}

\bibitem[{{Barnes}(2003)}]{Barnes03}
{Barnes}, S.~A. 2003, \apj, 586, 464, \dodoi{10.1086/367639}

\bibitem[{{Batygin} \& {Laughlin}(2015)}]{Batygin2015}
{Batygin}, K., \& {Laughlin}, G. 2015, Proceedings of the National Academy of
  Science, 112, 4214, \dodoi{10.1073/pnas.1423252112}

\bibitem[{{Beccari} {et~al.}(2017){Beccari}, {Petr-Gotzens}, {Boffin},
  {Romaniello}, {Fedele}, {Carraro}, {De Marchi}, {de Wit}, {Drew}, {Kalari},
  {Manara}, {Martin}, {Mieske}, {Panagia}, {Testi}, {Vink}, {Walsh}, \&
  {Wright}}]{Beccari2017}
{Beccari}, G., {Petr-Gotzens}, M.~G., {Boffin}, H.~M.~J., {et~al.} 2017, \aap,
  604, A22, \dodoi{10.1051/0004-6361/201730432}

\bibitem[{{Bouvier} {et~al.}(1993){Bouvier}, {Cabrit}, {Fernandez}, {Martin},
  \& {Matthews}}]{Bouvier93}
{Bouvier}, J., {Cabrit}, S., {Fernandez}, M., {Martin}, E.~L., \& {Matthews},
  J.~M. 1993, \aap, 272, 176

\bibitem[{{Bressan} {et~al.}(2012){Bressan}, {Marigo}, {Girardi}, {Salasnich},
  {Dal Cero}, {Rubele}, \& {Nanni}}]{Bressan12}
{Bressan}, A., {Marigo}, P., {Girardi}, L., {et~al.} 2012, \mnras, 427, 127,
  \dodoi{10.1111/j.1365-2966.2012.21948.x}

\bibitem[{{Broos} {et~al.}(2012){Broos}, {Townsley}, {Getman}, \&
  {Bauer}}]{Broos2012}
{Broos}, P., {Townsley}, L., {Getman}, K., \& {Bauer}, F. 2012, {AE: ACIS
  Extract}.
\newblock \doeprint{1203.001}

\bibitem[{{Broos} {et~al.}(2010){Broos}, {Townsley}, {Feigelson}, {Getman},
  {Bauer}, \& {Garmire}}]{Broos10}
{Broos}, P.~S., {Townsley}, L.~K., {Feigelson}, E.~D., {et~al.} 2010, \apj,
  714, 1582, \dodoi{10.1088/0004-637X/714/2/1582}

\bibitem[{{Broos} {et~al.}(2013){Broos}, {Getman}, {Povich}, {Feigelson},
  {Townsley}, {Naylor}, {Kuhn}, {King}, \& {Busk}}]{Broos13}
{Broos}, P.~S., {Getman}, K.~V., {Povich}, M.~S., {et~al.} 2013, \apjs, 209,
  32, \dodoi{10.1088/0067-0049/209/2/32}

\bibitem[{{Cantat-Gaudin} \& {Anders}(2020)}]{CantatGaudinAnders2020}
{Cantat-Gaudin}, T., \& {Anders}, F. 2020, \aap, 633, A99,
  \dodoi{10.1051/0004-6361/201936691}

\bibitem[{{Cantat-Gaudin} {et~al.}(2020){Cantat-Gaudin}, {Anders},
  {Castro-Ginard}, {Jordi}, {Romero-G{\'o}mez}, {Soubiran}, {Casamiquela},
  {Tarricq}, {Moitinho}, {Vallenari}, {Bragaglia}, {Krone-Martins}, \&
  {Kounkel}}]{Cantat-Gaudin2020}
{Cantat-Gaudin}, T., {Anders}, F., {Castro-Ginard}, A., {et~al.} 2020, \aap,
  640, A1, \dodoi{10.1051/0004-6361/202038192}

\bibitem[{{Canup}(2004)}]{Canup2004}
{Canup}, R.~M. 2004, \araa, 42, 441,
  \dodoi{10.1146/annurev.astro.41.082201.113457}

\bibitem[{{Chen} {et~al.}(2014){Chen}, {Girardi}, {Bressan}, {Marigo},
  {Barbieri}, \& {Kong}}]{Chen14}
{Chen}, Y., {Girardi}, L., {Bressan}, A., {et~al.} 2014, \mnras, 444, 2525,
  \dodoi{10.1093/mnras/stu1605}

\bibitem[{{Cleeves} {et~al.}(2017){Cleeves}, {Bergin}, {{\"O}berg}, {Andrews},
  {Wilner}, \& {Loomis}}]{Cleeves2017}
{Cleeves}, L.~I., {Bergin}, E.~A., {{\"O}berg}, K.~I., {et~al.} 2017, \apjl,
  843, L3, \dodoi{10.3847/2041-8213/aa76e2}

\bibitem[{Cleveland {et~al.}(1992)Cleveland, Grosse, \& Shyu}]{Cleveland1992}
Cleveland, W.~S., Grosse, E., \& Shyu, W.~M. 1992, in Statistical Models in S,
  ed. J.~Chambers \& T.~Hastie (Wadsworth \& Brooks/Cole).
\newblock
  \url{https://www.bibsonomy.org/bibtex/249dc786266d86d0cc54ed6ccb2fe9894/gron}

\bibitem[{{Coffaro} {et~al.}(2022){Coffaro}, {Stelzer}, \&
  {Orlando}}]{Coffaro2022}
{Coffaro}, M., {Stelzer}, B., \& {Orlando}, S. 2022, \aap, 661, A79,
  \dodoi{10.1051/0004-6361/202142298}

\bibitem[{{Coffaro} {et~al.}(2020){Coffaro}, {Stelzer}, {Orlando}, {Hall},
  {Metcalfe}, {Wolter}, {Mittag}, {Sanz-Forcada}, {Schneider}, \&
  {Ducci}}]{Coffaro2020}
{Coffaro}, M., {Stelzer}, B., {Orlando}, S., {et~al.} 2020, \aap, 636, A49,
  \dodoi{10.1051/0004-6361/201936479}

\bibitem[{{Cohen} {et~al.}(2017){Cohen}, {Yadav}, {Garraffo}, {Saar}, {Wolk},
  {Kashyap}, {Drake}, \& {Pillitteri}}]{Cohen2017}
{Cohen}, O., {Yadav}, R., {Garraffo}, C., {et~al.} 2017, \apj, 834, 14,
  \dodoi{10.3847/1538-4357/834/1/14}

\bibitem[{{Donati} {et~al.}(1997){Donati}, {Semel}, {Carter}, {Rees}, \&
  {Collier Cameron}}]{Donati97}
{Donati}, J.~F., {Semel}, M., {Carter}, B.~D., {Rees}, D.~E., \& {Collier
  Cameron}, A. 1997, \mnras, 291, 658, \dodoi{10.1093/mnras/291.4.658}

\bibitem[{{Donati} {et~al.}(2007){Donati}, {Jardine}, {Gregory}, {Petit},
  {Bouvier}, {Dougados}, {M{\'e}nard}, {Collier Cameron}, {Harries}, {Jeffers},
  \& {Paletou}}]{Donati07}
{Donati}, J.~F., {Jardine}, M.~M., {Gregory}, S.~G., {et~al.} 2007, \mnras,
  380, 1297, \dodoi{10.1111/j.1365-2966.2007.12194.x}

\bibitem[{{Drake} {et~al.}(2014){Drake}, {Braithwaite}, {Kashyap},
  {G{\"u}nther}, \& {Wright}}]{Drake2014}
{Drake}, J.~J., {Braithwaite}, J., {Kashyap}, V., {G{\"u}nther}, H.~M., \&
  {Wright}, N.~J. 2014, \apj, 786, 136, \dodoi{10.1088/0004-637X/786/2/136}

\bibitem[{{Drake} {et~al.}(2009){Drake}, {Ercolano}, {Flaccomio}, \&
  {Micela}}]{Drake2009}
{Drake}, J.~J., {Ercolano}, B., {Flaccomio}, E., \& {Micela}, G. 2009, \apjl,
  699, L35, \dodoi{10.1088/0004-637X/699/1/L35}

\bibitem[{{Dupuy} {et~al.}(2018){Dupuy}, {Bertin}, {F{\'e}raud}, {Hassenfratz},
  {Michaut}, {Putaud}, {Philippe}, {Jeseck}, {Angelucci}, {Cimino}, {Baglin},
  {Romanzin}, \& {Fillion}}]{Dupuy18}
{Dupuy}, R., {Bertin}, M., {F{\'e}raud}, G., {et~al.} 2018, Nature Astronomy,
  2, 796, \dodoi{10.1038/s41550-018-0532-y}

\bibitem[{{Durney} {et~al.}(1993){Durney}, {De Young}, \&
  {Roxburgh}}]{Durney1993}
{Durney}, B.~R., {De Young}, D.~S., \& {Roxburgh}, I.~W. 1993, \solphys, 145,
  207, \dodoi{10.1007/BF00690652}

\bibitem[{{Ercolano} \& {Pascucci}(2017)}]{Ercolano2017}
{Ercolano}, B., \& {Pascucci}, I. 2017, Royal Society Open Science, 4, 170114,
  \dodoi{10.1098/rsos.170114}

\bibitem[{{Evans} {et~al.}(2018){Evans}, {Riello}, {De Angeli}, {Carrasco},
  {Montegriffo}, {Fabricius}, {Jordi}, {Palaversa}, {Diener}, {Busso},
  {Cacciari}, {van Leeuwen}, {Burgess}, {Davidson}, {Harrison}, {Hodgkin},
  {Pancino}, {Richards}, {Altavilla}, {Balaguer-N{\'u}{\~n}ez}, {Barstow},
  {Bellazzini}, {Brown}, {Castellani}, {Cocozza}, {De Luise}, {Delgado},
  {Ducourant}, {Galleti}, {Gilmore}, {Giuffrida}, {Holl}, {Kewley}, {Koposov},
  {Marinoni}, {Marrese}, {Osborne}, {Piersimoni}, {Portell}, {Pulone},
  {Ragaini}, {Sanna}, {Terrett}, {Walton}, {Wevers}, \&
  {Wyrzykowski}}]{GaiaDR22018}
{Evans}, D.~W., {Riello}, M., {De Angeli}, F., {et~al.} 2018, \aap, 616, A4,
  \dodoi{10.1051/0004-6361/201832756}

\bibitem[{{Favata} {et~al.}(2005){Favata}, {Flaccomio}, {Reale}, {Micela},
  {Sciortino}, {Shang}, {Stassun}, \& {Feigelson}}]{Favata2005}
{Favata}, F., {Flaccomio}, E., {Reale}, F., {et~al.} 2005, \apjs, 160, 469,
  \dodoi{10.1086/432542}

\bibitem[{{Feigelson}(1996)}]{Feigelson96}
{Feigelson}, E.~D. 1996, \apj, 468, 306, \dodoi{10.1086/177691}

\bibitem[{{Feigelson}(2010)}]{Feigelson10}
---. 2010, Proceedings of the National Academy of Science, 107, 7153,
  \dodoi{10.1073/pnas.0913952107}

\bibitem[{{Feigelson} {et~al.}(2002){Feigelson}, {Broos}, {Gaffney}, {Garmire},
  {Hillenbrand}, {Pravdo}, {Townsley}, \& {Tsuboi}}]{Feigelson2002}
{Feigelson}, E.~D., {Broos}, P., {Gaffney}, James~A., I., {et~al.} 2002, \apj,
  574, 258, \dodoi{10.1086/340936}

\bibitem[{{Feigelson} \& {Montmerle}(1999)}]{Feigelson1999}
{Feigelson}, E.~D., \& {Montmerle}, T. 1999, \araa, 37, 363,
  \dodoi{10.1146/annurev.astro.37.1.363}

\bibitem[{{Feigelson} {et~al.}(2004){Feigelson}, {Hornschemeier}, {Micela},
  {Bauer}, {Alexander}, {Brandt}, {Favata}, {Sciortino}, \&
  {Garmire}}]{Feigelson2004}
{Feigelson}, E.~D., {Hornschemeier}, A.~E., {Micela}, G., {et~al.} 2004, \apj,
  611, 1107, \dodoi{10.1086/422248}

\bibitem[{{Feigelson} {et~al.}(2013){Feigelson}, {Townsley}, {Broos}, {Busk},
  {Getman}, {King}, {Kuhn}, {Naylor}, {Povich}, {Baddeley}, {Bate},
  {Indebetouw}, {Luhman}, {McCaughrean}, {Pittard}, {Pudritz}, {Sills}, {Song},
  \& {Wadsley}}]{Feigelson13}
{Feigelson}, E.~D., {Townsley}, L.~K., {Broos}, P.~S., {et~al.} 2013, \apjs,
  209, 26, \dodoi{10.1088/0067-0049/209/2/26}

\bibitem[{{Flaccomio} {et~al.}(2018){Flaccomio}, {Micela}, {Sciortino}, {Cody},
  {Guarcello}, {Morales-Calder{\`o}n}, {Rebull}, \& {Stauffer}}]{Flaccomio2018}
{Flaccomio}, E., {Micela}, G., {Sciortino}, S., {et~al.} 2018, \aap, 620, A55,
  \dodoi{10.1051/0004-6361/201833308}

\bibitem[{{Flaischlen} {et~al.}(2021){Flaischlen}, {Preibisch}, {Manara}, \&
  {Ercolano}}]{Flaischlen2021}
{Flaischlen}, S., {Preibisch}, T., {Manara}, C.~F., \& {Ercolano}, B. 2021,
  arXiv e-prints, arXiv:2103.03039.
\newblock \doarXiv{2103.03039}

\bibitem[{{Fromang} {et~al.}(2002){Fromang}, {Terquem}, \&
  {Balbus}}]{Fromang02}
{Fromang}, S., {Terquem}, C., \& {Balbus}, S.~A. 2002, \mnras, 329, 18,
  \dodoi{10.1046/j.1365-8711.2002.04940.x}

\bibitem[{{Gaia Collaboration} {et~al.}(2016){Gaia Collaboration}, {Prusti},
  {de Bruijne}, {Brown}, {Vallenari}, {Babusiaux}, {Bailer-Jones}, {Bastian},
  {Biermann}, {Evans}, {Eyer}, {Jansen}, {Jordi}, {Klioner}, {Lammers},
  {Lindegren}, {Luri}, {Mignard}, {Milligan}, {Panem}, {Poinsignon},
  {Pourbaix}, {Randich}, {Sarri}, {Sartoretti}, {Siddiqui}, {Soubiran},
  {Valette}, {van Leeuwen}, {Walton}, {Aerts}, {Arenou}, {Cropper}, {Drimmel},
  {H{\o}g}, {Katz}, {Lattanzi}, {O'Mullane}, {Grebel}, {Holland}, {Huc},
  {Passot}, {Bramante}, {Cacciari}, {Casta{\~n}eda}, {Chaoul}, {Cheek}, {De
  Angeli}, {Fabricius}, {Guerra}, {Hern{\'a}ndez}, {Jean-Antoine-Piccolo},
  {Masana}, {Messineo}, {Mowlavi}, {Nienartowicz}, {Ord{\'o}{\~n}ez-Blanco},
  {Panuzzo}, {Portell}, {Richards}, {Riello}, {Seabroke}, {Tanga},
  {Th{\'e}venin}, {Torra}, {Els}, {Gracia-Abril}, {Comoretto},
  {Garcia-Reinaldos}, {Lock}, {Mercier}, {Altmann}, {Andrae}, {Astraatmadja},
  {Bellas-Velidis}, {Benson}, {Berthier}, {Blomme}, {Busso}, {Carry},
  {Cellino}, {Clementini}, {Cowell}, {Creevey}, {Cuypers}, {Davidson}, {De
  Ridder}, {de Torres}, {Delchambre}, {Dell'Oro}, {Ducourant}, {Fr{\'e}mat},
  {Garc{\'\i}a-Torres}, {Gosset}, {Halbwachs}, {Hambly}, {Harrison}, {Hauser},
  {Hestroffer}, {Hodgkin}, {Huckle}, {Hutton}, {Jasniewicz}, {Jordan},
  {Kontizas}, {Korn}, {Lanzafame}, {Manteiga}, {Moitinho}, {Muinonen},
  {Osinde}, {Pancino}, {Pauwels}, {Petit}, {Recio-Blanco}, {Robin}, {Sarro},
  {Siopis}, {Smith}, {Smith}, {Sozzetti}, {Thuillot}, {van Reeven}, {Viala},
  {Abbas}, {Abreu Aramburu}, {Accart}, {Aguado}, {Allan}, {Allasia},
  {Altavilla}, {{\'A}lvarez}, {Alves}, {Anderson}, {Andrei}, {Anglada Varela},
  {Antiche}, {Antoja}, {Ant{\'o}n}, {Arcay}, {Atzei}, {Ayache}, {Bach},
  {Baker}, {Balaguer-N{\'u}{\~n}ez}, {Barache}, {Barata}, {Barbier}, {Barblan},
  {Baroni}, {Barrado y Navascu{\'e}s}, {Barros}, {Barstow}, {Becciani},
  {Bellazzini}, {Bellei}, {Bello Garc{\'\i}a}, {Belokurov}, {Bendjoya},
  {Berihuete}, {Bianchi}, {Bienaym{\'e}}, {Billebaud}, {Blagorodnova},
  {Blanco-Cuaresma}, {Boch}, {Bombrun}, {Borrachero}, {Bouquillon}, {Bourda},
  {Bouy}, {Bragaglia}, {Breddels}, {Brouillet}, {Br{\"u}semeister},
  {Bucciarelli}, {Budnik}, {Burgess}, {Burgon}, {Burlacu}, {Busonero}, {Buzzi},
  {Caffau}, {Cambras}, {Campbell}, {Cancelliere}, {Cantat-Gaudin}, {Carlucci},
  {Carrasco}, {Castellani}, {Charlot}, {Charnas}, {Charvet}, {Chassat},
  {Chiavassa}, {Clotet}, {Cocozza}, {Collins}, {Collins}, {Costigan}, {Crifo},
  {Cross}, {Crosta}, {Crowley}, {Dafonte}, {Damerdji}, {Dapergolas}, {David},
  {David}, {De Cat}, {de Felice}, {de Laverny}, {De Luise}, {De March}, {de
  Martino}, {de Souza}, {Debosscher}, {del Pozo}, {Delbo}, {Delgado},
  {Delgado}, {di Marco}, {Di Matteo}, {Diakite}, {Distefano}, {Dolding}, {Dos
  Anjos}, {Drazinos}, {Dur{\'a}n}, {Dzigan}, {Ecale}, {Edvardsson}, {Enke},
  {Erdmann}, {Escolar}, {Espina}, {Evans}, {Eynard Bontemps}, {Fabre},
  {Fabrizio}, {Faigler}, {Falc{\~a}o}, {Farr{\`a}s Casas}, {Faye}, {Federici},
  {Fedorets}, {Fern{\'a}ndez-Hern{\'a}ndez}, {Fernique}, {Fienga}, {Figueras},
  {Filippi}, {Findeisen}, {Fonti}, {Fouesneau}, {Fraile}, {Fraser}, {Fuchs},
  {Furnell}, {Gai}, {Galleti}, {Galluccio}, {Garabato}, {Garc{\'\i}a-Sedano},
  {Gar{\'e}}, {Garofalo}, {Garralda}, {Gavras}, {Gerssen}, {Geyer}, {Gilmore},
  {Girona}, {Giuffrida}, {Gomes}, {Gonz{\'a}lez-Marcos},
  {Gonz{\'a}lez-N{\'u}{\~n}ez}, {Gonz{\'a}lez-Vidal}, {Granvik}, {Guerrier},
  {Guillout}, {Guiraud}, {G{\'u}rpide}, {Guti{\'e}rrez-S{\'a}nchez}, {Guy},
  {Haigron}, {Hatzidimitriou}, {Haywood}, {Heiter}, {Helmi}, {Hobbs},
  {Hofmann}, {Holl}, {Holland}, {Hunt}, {Hypki}, {Icardi}, {Irwin}, {Jevardat
  de Fombelle}, {Jofr{\'e}}, {Jonker}, {Jorissen}, {Julbe}, {Karampelas},
  {Kochoska}, {Kohley}, {Kolenberg}, {Kontizas}, {Koposov}, {Kordopatis},
  {Koubsky}, {Kowalczyk}, {Krone-Martins}, {Kudryashova}, {Kull}, {Bachchan},
  {Lacoste-Seris}, {Lanza}, {Lavigne}, {Le Poncin-Lafitte}, {Lebreton},
  {Lebzelter}, {Leccia}, {Leclerc}, {Lecoeur-Taibi}, {Lemaitre}, {Lenhardt},
  {Leroux}, {Liao}, {Licata}, {Lindstr{\o}m}, {Lister}, {Livanou}, {Lobel},
  {L{\"o}ffler}, {L{\'o}pez}, {Lopez-Lozano}, {Lorenz}, {Loureiro},
  {MacDonald}, {Magalh{\~a}es Fernandes}, {Managau}, {Mann}, {Mantelet},
  {Marchal}, {Marchant}, {Marconi}, {Marie}, {Marinoni}, {Marrese},
  {Marschalk{\'o}}, {Marshall}, {Mart{\'\i}n-Fleitas}, {Martino}, {Mary},
  {Matijevi{\v{c}}}, {Mazeh}, {McMillan}, {Messina}, {Mestre}, {Michalik},
  {Millar}, {Miranda}, {Molina}, {Molinaro}, {Molinaro}, {Moln{\'a}r},
  {Moniez}, {Montegriffo}, {Monteiro}, {Mor}, {Mora}, {Morbidelli}, {Morel},
  {Morgenthaler}, {Morley}, {Morris}, {Mulone}, {Muraveva}, {Musella},
  {Narbonne}, {Nelemans}, {Nicastro}, {Noval}, {Ord{\'e}novic},
  {Ordieres-Mer{\'e}}, {Osborne}, {Pagani}, {Pagano}, {Pailler}, {Palacin},
  {Palaversa}, {Parsons}, {Paulsen}, {Pecoraro}, {Pedrosa}, {Pentik{\"a}inen},
  {Pereira}, {Pichon}, {Piersimoni}, {Pineau}, {Plachy}, {Plum}, {Poujoulet},
  {Pr{\v{s}}a}, {Pulone}, {Ragaini}, {Rago}, {Rambaux}, {Ramos-Lerate},
  {Ranalli}, {Rauw}, {Read}, {Regibo}, {Renk}, {Reyl{\'e}}, {Ribeiro},
  {Rimoldini}, {Ripepi}, {Riva}, {Rixon}, {Roelens}, {Romero-G{\'o}mez},
  {Rowell}, {Royer}, {Rudolph}, {Ruiz-Dern}, {Sadowski}, {Sagrist{\`a}
  Sell{\'e}s}, {Sahlmann}, {Salgado}, {Salguero}, {Sarasso}, {Savietto},
  {Schnorhk}, {Schultheis}, {Sciacca}, {Segol}, {Segovia}, {Segransan},
  {Serpell}, {Shih}, {Smareglia}, {Smart}, {Smith}, {Solano}, {Solitro},
  {Sordo}, {Soria Nieto}, {Souchay}, {Spagna}, {Spoto}, {Stampa}, {Steele},
  {Steidelm{\"u}ller}, {Stephenson}, {Stoev}, {Suess}, {S{\"u}veges}, {Surdej},
  {Szabados}, {Szegedi-Elek}, {Tapiador}, {Taris}, {Tauran}, {Taylor},
  {Teixeira}, {Terrett}, {Tingley}, {Trager}, {Turon}, {Ulla}, {Utrilla},
  {Valentini}, {van Elteren}, {Van Hemelryck}, {van Leeuwen}, {Varadi},
  {Vecchiato}, {Veljanoski}, {Via}, {Vicente}, {Vogt}, {Voss}, {Votruba},
  {Voutsinas}, {Walmsley}, {Weiler}, {Weingrill}, {Werner}, {Wevers},
  {Whitehead}, {Wyrzykowski}, {Yoldas}, {{\v{Z}}erjal}, {Zucker}, {Zurbach},
  {Zwitter}, {Alecu}, {Allen}, {Allende Prieto}, {Amorim},
  {Anglada-Escud{\'e}}, {Arsenijevic}, {Azaz}, {Balm}, {Beck}, {Bernstein},
  {Bigot}, {Bijaoui}, {Blasco}, {Bonfigli}, {Bono}, {Boudreault}, {Bressan},
  {Brown}, {Brunet}, {Bunclark}, {Buonanno}, {Butkevich}, {Carret}, {Carrion},
  {Chemin}, {Ch{\'e}reau}, {Corcione}, {Darmigny}, {de Boer}, {de Teodoro}, {de
  Zeeuw}, {Delle Luche}, {Domingues}, {Dubath}, {Fodor}, {Fr{\'e}zouls},
  {Fries}, {Fustes}, {Fyfe}, {Gallardo}, {Gallegos}, {Gardiol}, {Gebran},
  {Gomboc}, {G{\'o}mez}, {Grux}, {Gueguen}, {Heyrovsky}, {Hoar}, {Iannicola},
  {Isasi Parache}, {Janotto}, {Joliet}, {Jonckheere}, {Keil}, {Kim},
  {Klagyivik}, {Klar}, {Knude}, {Kochukhov}, {Kolka}, {Kos}, {Kutka}, {Lainey},
  {LeBouquin}, {Liu}, {Loreggia}, {Makarov}, {Marseille}, {Martayan},
  {Martinez-Rubi}, {Massart}, {Meynadier}, {Mignot}, {Munari}, {Nguyen},
  {Nordlander}, {Ocvirk}, {O'Flaherty}, {Olias Sanz}, {Ortiz}, {Osorio},
  {Oszkiewicz}, {Ouzounis}, {Palmer}, {Park}, {Pasquato}, {Peltzer}, {Peralta},
  {P{\'e}turaud}, {Pieniluoma}, {Pigozzi}, {Poels}, {Prat}, {Prod'homme},
  {Raison}, {Rebordao}, {Risquez}, {Rocca-Volmerange}, {Rosen}, {Ruiz-Fuertes},
  {Russo}, {Sembay}, {Serraller Vizcaino}, {Short}, {Siebert}, {Silva},
  {Sinachopoulos}, {Slezak}, {Soffel}, {Sosnowska}, {Strai{\v{z}}ys}, {ter
  Linden}, {Terrell}, {Theil}, {Tiede}, {Troisi}, {Tsalmantza}, {Tur},
  {Vaccari}, {Vachier}, {Valles}, {Van Hamme}, {Veltz}, {Virtanen}, {Wallut},
  {Wichmann}, {Wilkinson}, {Ziaeepour}, \& {Zschocke}}]{GaiaMission2016}
{Gaia Collaboration}, {Prusti}, T., {de Bruijne}, J.~H.~J., {et~al.} 2016,
  \aap, 595, A1, \dodoi{10.1051/0004-6361/201629272}

\bibitem[{{Gaia Collaboration} {et~al.}(2021){Gaia Collaboration}, {Brown},
  {Vallenari}, {Prusti}, {de Bruijne}, {Babusiaux}, {Biermann}, {Creevey},
  {Evans}, {Eyer}, {Hutton}, {Jansen}, {Jordi}, {Klioner}, {Lammers},
  {Lindegren}, {Luri}, {Mignard}, {Panem}, {Pourbaix}, {Randich}, {Sartoretti},
  {Soubiran}, {Walton}, {Arenou}, {Bailer-Jones}, {Bastian}, {Cropper},
  {Drimmel}, {Katz}, {Lattanzi}, {van Leeuwen}, {Bakker}, {Cacciari},
  {Casta{\~n}eda}, {De Angeli}, {Ducourant}, {Fabricius}, {Fouesneau},
  {Fr{\'e}mat}, {Guerra}, {Guerrier}, {Guiraud}, {Jean-Antoine Piccolo},
  {Masana}, {Messineo}, {Mowlavi}, {Nicolas}, {Nienartowicz}, {Pailler},
  {Panuzzo}, {Riclet}, {Roux}, {Seabroke}, {Sordo}, {Tanga}, {Th{\'e}venin},
  {Gracia-Abril}, {Portell}, {Teyssier}, {Altmann}, {Andrae}, {Bellas-Velidis},
  {Benson}, {Berthier}, {Blomme}, {Brugaletta}, {Burgess}, {Busso}, {Carry},
  {Cellino}, {Cheek}, {Clementini}, {Damerdji}, {Davidson}, {Delchambre},
  {Dell'Oro}, {Fern{\'a}ndez-Hern{\'a}ndez}, {Galluccio}, {Garc{\'\i}a-Lario},
  {Garcia-Reinaldos}, {Gonz{\'a}lez-N{\'u}{\~n}ez}, {Gosset}, {Haigron},
  {Halbwachs}, {Hambly}, {Harrison}, {Hatzidimitriou}, {Heiter},
  {Hern{\'a}ndez}, {Hestroffer}, {Hodgkin}, {Holl}, {Jan{\ss}en}, {Jevardat de
  Fombelle}, {Jordan}, {Krone-Martins}, {Lanzafame}, {L{\"o}ffler}, {Lorca},
  {Manteiga}, {Marchal}, {Marrese}, {Moitinho}, {Mora}, {Muinonen}, {Osborne},
  {Pancino}, {Pauwels}, {Petit}, {Recio-Blanco}, {Richards}, {Riello},
  {Rimoldini}, {Robin}, {Roegiers}, {Rybizki}, {Sarro}, {Siopis}, {Smith},
  {Sozzetti}, {Ulla}, {Utrilla}, {van Leeuwen}, {van Reeven}, {Abbas}, {Abreu
  Aramburu}, {Accart}, {Aerts}, {Aguado}, {Ajaj}, {Altavilla}, {{\'A}lvarez},
  {{\'A}lvarez Cid-Fuentes}, {Alves}, {Anderson}, {Anglada Varela}, {Antoja},
  {Audard}, {Baines}, {Baker}, {Balaguer-N{\'u}{\~n}ez}, {Balbinot}, {Balog},
  {Barache}, {Barbato}, {Barros}, {Barstow}, {Bartolom{\'e}}, {Bassilana},
  {Bauchet}, {Baudesson-Stella}, {Becciani}, {Bellazzini}, {Bernet}, {Bertone},
  {Bianchi}, {Blanco-Cuaresma}, {Boch}, {Bombrun}, {Bossini}, {Bouquillon},
  {Bragaglia}, {Bramante}, {Breedt}, {Bressan}, {Brouillet}, {Bucciarelli},
  {Burlacu}, {Busonero}, {Butkevich}, {Buzzi}, {Caffau}, {Cancelliere},
  {C{\'a}novas}, {Cantat-Gaudin}, {Carballo}, {Carlucci}, {Carnerero},
  {Carrasco}, {Casamiquela}, {Castellani}, {Castro-Ginard}, {Castro Sampol},
  {Chaoul}, {Charlot}, {Chemin}, {Chiavassa}, {Cioni}, {Comoretto}, {Cooper},
  {Cornez}, {Cowell}, {Crifo}, {Crosta}, {Crowley}, {Dafonte}, {Dapergolas},
  {David}, {David}, {de Laverny}, {De Luise}, {De March}, {De Ridder}, {de
  Souza}, {de Teodoro}, {de Torres}, {del Peloso}, {del Pozo}, {Delbo},
  {Delgado}, {Delgado}, {Delisle}, {Di Matteo}, {Diakite}, {Diener},
  {Distefano}, {Dolding}, {Eappachen}, {Edvardsson}, {Enke}, {Esquej}, {Fabre},
  {Fabrizio}, {Faigler}, {Fedorets}, {Fernique}, {Fienga}, {Figueras},
  {Fouron}, {Fragkoudi}, {Fraile}, {Franke}, {Gai}, {Garabato},
  {Garcia-Gutierrez}, {Garc{\'\i}a-Torres}, {Garofalo}, {Gavras}, {Gerlach},
  {Geyer}, {Giacobbe}, {Gilmore}, {Girona}, {Giuffrida}, {Gomel}, {Gomez},
  {Gonzalez-Santamaria}, {Gonz{\'a}lez-Vidal}, {Granvik},
  {Guti{\'e}rrez-S{\'a}nchez}, {Guy}, {Hauser}, {Haywood}, {Helmi}, {Hidalgo},
  {Hilger}, {H{\l}adczuk}, {Hobbs}, {Holland}, {Huckle}, {Jasniewicz},
  {Jonker}, {Juaristi Campillo}, {Julbe}, {Karbevska}, {Kervella}, {Khanna},
  {Kochoska}, {Kontizas}, {Kordopatis}, {Korn}, {Kostrzewa-Rutkowska},
  {Kruszy{\'n}ska}, {Lambert}, {Lanza}, {Lasne}, {Le Campion}, {Le Fustec},
  {Lebreton}, {Lebzelter}, {Leccia}, {Leclerc}, {Lecoeur-Taibi}, {Liao},
  {Licata}, {Lindstr{\o}m}, {Lister}, {Livanou}, {Lobel}, {Madrero Pardo},
  {Managau}, {Mann}, {Marchant}, {Marconi}, {Marcos Santos}, {Marinoni},
  {Marocco}, {Marshall}, {Martin Polo}, {Mart{\'\i}n-Fleitas}, {Masip},
  {Massari}, {Mastrobuono-Battisti}, {Mazeh}, {McMillan}, {Messina},
  {Michalik}, {Millar}, {Mints}, {Molina}, {Molinaro}, {Moln{\'a}r},
  {Montegriffo}, {Mor}, {Morbidelli}, {Morel}, {Morris}, {Mulone}, {Munoz},
  {Muraveva}, {Murphy}, {Musella}, {Noval}, {Ord{\'e}novic}, {Orr{\`u}},
  {Osinde}, {Pagani}, {Pagano}, {Palaversa}, {Palicio}, {Panahi}, {Pawlak},
  {Pe{\~n}alosa Esteller}, {Penttil{\"a}}, {Piersimoni}, {Pineau}, {Plachy},
  {Plum}, {Poggio}, {Poretti}, {Poujoulet}, {Pr{\v{s}}a}, {Pulone}, {Racero},
  {Ragaini}, {Rainer}, {Raiteri}, {Rambaux}, {Ramos}, {Ramos-Lerate}, {Re
  Fiorentin}, {Regibo}, {Reyl{\'e}}, {Ripepi}, {Riva}, {Rixon}, {Robichon},
  {Robin}, {Roelens}, {Rohrbasser}, {Romero-G{\'o}mez}, {Rowell}, {Royer},
  {Rybicki}, {Sadowski}, {Sagrist{\`a} Sell{\'e}s}, {Sahlmann}, {Salgado},
  {Salguero}, {Samaras}, {Sanchez Gimenez}, {Sanna}, {Santove{\~n}a},
  {Sarasso}, {Schultheis}, {Sciacca}, {Segol}, {Segovia}, {S{\'e}gransan},
  {Semeux}, {Shahaf}, {Siddiqui}, {Siebert}, {Siltala}, {Slezak}, {Smart},
  {Solano}, {Solitro}, {Souami}, {Souchay}, {Spagna}, {Spoto}, {Steele},
  {Steidelm{\"u}ller}, {Stephenson}, {S{\"u}veges}, {Szabados}, {Szegedi-Elek},
  {Taris}, {Tauran}, {Taylor}, {Teixeira}, {Thuillot}, {Tonello}, {Torra},
  {Torra}, {Turon}, {Unger}, {Vaillant}, {van Dillen}, {Vanel}, {Vecchiato},
  {Viala}, {Vicente}, {Voutsinas}, {Weiler}, {Wevers}, {Wyrzykowski}, {Yoldas},
  {Yvard}, {Zhao}, {Zorec}, {Zucker}, {Zurbach}, \& {Zwitter}}]{GaiaEDR32021}
{Gaia Collaboration}, {Brown}, A.~G.~A., {Vallenari}, A., {et~al.} 2021, \aap,
  649, A1, \dodoi{10.1051/0004-6361/202039657}

\bibitem[{{Gallet} \& {Bouvier}(2015)}]{Gallet2015}
{Gallet}, F., \& {Bouvier}, J. 2015, \aap, 577, A98,
  \dodoi{10.1051/0004-6361/201525660}

\bibitem[{{Garc{\'\i}a Mu{\~n}oz} {et~al.}(2021){Garc{\'\i}a Mu{\~n}oz},
  {Fossati}, {Youngblood}, {Nettelmann}, {Gandolfi}, {Cabrera}, \&
  {Rauer}}]{GarciaMunoz2021}
{Garc{\'\i}a Mu{\~n}oz}, A., {Fossati}, L., {Youngblood}, A., {et~al.} 2021,
  \apjl, 907, L36, \dodoi{10.3847/2041-8213/abd9b8}

\bibitem[{{Garmire} {et~al.}(2003){Garmire}, {Bautz}, {Ford}, {Nousek}, \&
  {Ricker}}]{Garmire03}
{Garmire}, G.~P., {Bautz}, M.~W., {Ford}, P.~G., {Nousek}, J.~A., \& {Ricker},
  George~R., J. 2003, in Society of Photo-Optical Instrumentation Engineers
  (SPIE) Conference Series, Vol. 4851, X-Ray and Gamma-Ray Telescopes and
  Instruments for Astronomy., ed. J.~E. {Truemper} \& H.~D. {Tananbaum},
  28--44, \dodoi{10.1117/12.461599}

\bibitem[{{Gehrels}(1986)}]{Gehrels86}
{Gehrels}, N. 1986, \apj, 303, 336, \dodoi{10.1086/164079}

\bibitem[{{Getman} {et~al.}(2017){Getman}, {Broos}, {Kuhn}, {Feigelson},
  {Richert}, {Ota}, {Bate}, \& {Garmire}}]{Getman17}
{Getman}, K.~V., {Broos}, P.~S., {Kuhn}, M.~A., {et~al.} 2017, \apjs, 229, 28,
  \dodoi{10.3847/1538-4365/229/2/28}

\bibitem[{{Getman} \& {Feigelson}(2021)}]{Getman2021}
{Getman}, K.~V., \& {Feigelson}, E.~D. 2021, \apj, 916, 32,
  \dodoi{10.3847/1538-4357/ac00be}

\bibitem[{{Getman} {et~al.}(2010){Getman}, {Feigelson}, {Broos}, {Townsley}, \&
  {Garmire}}]{Getman10}
{Getman}, K.~V., {Feigelson}, E.~D., {Broos}, P.~S., {Townsley}, L.~K., \&
  {Garmire}, G.~P. 2010, \apj, 708, 1760, \dodoi{10.1088/0004-637X/708/2/1760}

\bibitem[{{Getman} {et~al.}(2021){Getman}, {Feigelson}, \&
  {Garmire}}]{Getman2021b}
{Getman}, K.~V., {Feigelson}, E.~D., \& {Garmire}, G.~P. 2021, \apj, 920, 154,
  \dodoi{10.3847/1538-4357/ac1746}

\bibitem[{{Getman} {et~al.}(2014{\natexlab{a}}){Getman}, {Feigelson}, \&
  {Kuhn}}]{Getman2014b}
{Getman}, K.~V., {Feigelson}, E.~D., \& {Kuhn}, M.~A. 2014{\natexlab{a}}, \apj,
  787, 109, \dodoi{10.1088/0004-637X/787/2/109}

\bibitem[{{Getman} {et~al.}(2018{\natexlab{a}}){Getman}, {Feigelson}, {Kuhn},
  {Bate}, {Broos}, \& {Garmire}}]{Getman2018a}
{Getman}, K.~V., {Feigelson}, E.~D., {Kuhn}, M.~A., {et~al.}
  2018{\natexlab{a}}, \mnras, 476, 1213, \dodoi{10.1093/mnras/sty302}

\bibitem[{{Getman} {et~al.}(2019){Getman}, {Feigelson}, {Kuhn}, \&
  {Garmire}}]{Getman2019}
{Getman}, K.~V., {Feigelson}, E.~D., {Kuhn}, M.~A., \& {Garmire}, G.~P. 2019,
  \mnras, 487, 2977, \dodoi{10.1093/mnras/stz1457}

\bibitem[{{Getman} {et~al.}(2008){Getman}, {Feigelson}, {Micela}, {Jardine},
  {Gregory}, \& {Garmire}}]{Getman08b}
{Getman}, K.~V., {Feigelson}, E.~D., {Micela}, G., {et~al.} 2008, \apj, 688,
  437, \dodoi{10.1086/592034}

\bibitem[{{Getman} {et~al.}(2018{\natexlab{b}}){Getman}, {Kuhn}, {Feigelson},
  {Broos}, {Bate}, \& {Garmire}}]{Getman18b}
{Getman}, K.~V., {Kuhn}, M.~A., {Feigelson}, E.~D., {et~al.}
  2018{\natexlab{b}}, \mnras, 477, 298, \dodoi{10.1093/mnras/sty473}

\bibitem[{{Getman} {et~al.}(2005){Getman}, {Flaccomio}, {Broos}, {Grosso},
  {Tsujimoto}, {Townsley}, {Garmire}, {Kastner}, {Li}, {Harnden}, {Wolk},
  {Murray}, {Lada}, {Muench}, {McCaughrean}, {Meeus}, {Damiani}, {Micela},
  {Sciortino}, {Bally}, {Hillenbrand }, {Herbst}, {Preibisch}, \&
  {Feigelson}}]{Getman05}
{Getman}, K.~V., {Flaccomio}, E., {Broos}, P.~S., {et~al.} 2005, \apjs, 160,
  319, \dodoi{10.1086/432092}

\bibitem[{{Getman} {et~al.}(2014{\natexlab{b}}){Getman}, {Feigelson}, {Kuhn},
  {Broos}, {Townsley}, {Naylor}, {Povich}, {Luhman}, \& {Garmire}}]{Getman14a}
{Getman}, K.~V., {Feigelson}, E.~D., {Kuhn}, M.~A., {et~al.}
  2014{\natexlab{b}}, \apj, 787, 108, \dodoi{10.1088/0004-637X/787/2/108}

\bibitem[{{Glassgold} {et~al.}(2000){Glassgold}, {Feigelson}, \&
  {Montmerle}}]{Glassgold2000}
{Glassgold}, A.~E., {Feigelson}, E.~D., \& {Montmerle}, T. 2000, in Protostars
  and Planets IV, ed. V.~{Mannings}, A.~P. {Boss}, \& S.~S. {Russell}, 429

\bibitem[{{Glocer} {et~al.}(2012){Glocer}, {Kitamura}, {Toth}, \&
  {Gombosi}}]{Glocer2012}
{Glocer}, A., {Kitamura}, N., {Toth}, G., \& {Gombosi}, T. 2012, Journal of
  Geophysical Research (Space Physics), 117, A04318,
  \dodoi{10.1029/2011JA017136}

\bibitem[{{Gregory} {et~al.}(2016){Gregory}, {Adams}, \& {Davies}}]{Gregory16}
{Gregory}, S.~G., {Adams}, F.~C., \& {Davies}, C.~L. 2016, \mnras, 457, 3836,
  \dodoi{10.1093/mnras/stw259}

\bibitem[{{Gregory} {et~al.}(2014){Gregory}, {Donati}, {Morin}, {Hussain},
  {Mayne}, {Hillenbrand}, \& {Jardine}}]{Gregory2014}
{Gregory}, S.~G., {Donati}, J.~F., {Morin}, J., {et~al.} 2014, in Magnetic
  Fields throughout Stellar Evolution, ed. P.~{Petit}, M.~{Jardine}, \& H.~C.
  {Spruit}, Vol. 302, 40--43, \dodoi{10.1017/S1743921314001677}

\bibitem[{{Gregory} {et~al.}(2010){Gregory}, {Jardine}, {Gray}, \&
  {Donati}}]{Gregory2010}
{Gregory}, S.~G., {Jardine}, M., {Gray}, C.~G., \& {Donati}, J.~F. 2010,
  Reports on Progress in Physics, 73, 126901,
  \dodoi{10.1088/0034-4885/73/12/126901}

\bibitem[{{Gressel} {et~al.}(2013){Gressel}, {Nelson}, {Turner}, \&
  {Ziegler}}]{Gressel13}
{Gressel}, O., {Nelson}, R.~P., {Turner}, N.~J., \& {Ziegler}, U. 2013, \apj,
  779, 59, \dodoi{10.1088/0004-637X/779/1/59}

\bibitem[{{Gronoff} {et~al.}(2020){Gronoff}, {Arras}, {Baraka}, {Bell},
  {Cessateur}, {Cohen}, {Curry}, {Drake}, {Elrod}, {Erwin}, {Garcia-Sage},
  {Garraffo}, {Glocer}, {Heavens}, {Lovato}, {Maggiolo}, {Parkinson}, {Simon
  Wedlund}, {Weimer}, \& {Moore}}]{Gronoff2020}
{Gronoff}, G., {Arras}, P., {Baraka}, S., {et~al.} 2020, Journal of Geophysical
  Research (Space Physics), 125, e27639, \dodoi{10.1029/2019JA027639}

\bibitem[{{G{\"u}del}(2004)}]{Gudel2004}
{G{\"u}del}, M. 2004, \aapr, 12, 71, \dodoi{10.1007/s00159-004-0023-2}

\bibitem[{{G{\"u}del} {et~al.}(1997){G{\"u}del}, {Guinan}, \&
  {Skinner}}]{Gudel1997}
{G{\"u}del}, M., {Guinan}, E.~F., \& {Skinner}, S.~L. 1997, \apj, 483, 947,
  \dodoi{10.1086/304264}

\bibitem[{{G{\"u}del} \& {Naz{\'e}}(2009)}]{Gudel2009}
{G{\"u}del}, M., \& {Naz{\'e}}, Y. 2009, \aapr, 17, 309,
  \dodoi{10.1007/s00159-009-0022-4}

\bibitem[{{G{\"u}del} {et~al.}(2007){G{\"u}del}, {Briggs}, {Arzner}, {Audard},
  {Bouvier}, {Feigelson}, {Franciosini}, {Glauser}, {Grosso}, {Micela},
  {Monin}, {Montmerle}, {Padgett}, {Palla}, {Pillitteri}, {Rebull}, {Scelsi},
  {Silva}, {Skinner}, {Stelzer}, \& {Telleschi}}]{Gudel2007b}
{G{\"u}del}, M., {Briggs}, K.~R., {Arzner}, K., {et~al.} 2007, \aap, 468, 353,
  \dodoi{10.1051/0004-6361:20065724}

\bibitem[{{Henderson} \& {Stassun}(2012)}]{Henderson2012}
{Henderson}, C.~B., \& {Stassun}, K.~G. 2012, \apj, 747, 51,
  \dodoi{10.1088/0004-637X/747/1/51}

\bibitem[{{Hu} {et~al.}(2022){Hu}, {Airapetian}, {Li}, {Zank}, \&
  {Jin}}]{Hu2022}
{Hu}, J., {Airapetian}, V.~S., {Li}, G., {Zank}, G., \& {Jin}, M. 2022, Science
  Advances in press.

\bibitem[{{Iben}(1965)}]{Iben65}
{Iben}, Icko, J. 1965, \apj, 141, 993, \dodoi{10.1086/148193}

\bibitem[{{Ilgner} \& {Nelson}(2006)}]{Ilgner06}
{Ilgner}, M., \& {Nelson}, R.~P. 2006, \aap, 455, 731,
  \dodoi{10.1051/0004-6361:20065308}

\bibitem[{{Jardine} \& {Unruh}(1999)}]{Jardine1999}
{Jardine}, M., \& {Unruh}, Y.~C. 1999, \aap, 346, 883

\bibitem[{{Johnstone}(2020)}]{Johnstone2020}
{Johnstone}, C.~P. 2020, \apj, 890, 79, \dodoi{10.3847/1538-4357/ab6224}

\bibitem[{{Johnstone} {et~al.}(2021{\natexlab{a}}){Johnstone}, {Bartel}, \&
  {G{\"u}del}}]{Johnstone2021}
{Johnstone}, C.~P., {Bartel}, M., \& {G{\"u}del}, M. 2021{\natexlab{a}}, \aap,
  649, A96, \dodoi{10.1051/0004-6361/202038407}

\bibitem[{{Johnstone} {et~al.}(2021{\natexlab{b}}){Johnstone}, {Lammer},
  {Kislyakova}, {Scherf}, \& {G{\"u}del}}]{JohnstoneLammer2021}
{Johnstone}, C.~P., {Lammer}, H., {Kislyakova}, K.~G., {Scherf}, M., \&
  {G{\"u}del}, M. 2021{\natexlab{b}}, Earth and Planetary Science Letters, 576,
  117197, \dodoi{10.1016/j.epsl.2021.117197}

\bibitem[{Kaplan \& Meier(1958)}]{KaplanMeier58}
Kaplan, E.~L., \& Meier, P. 1958, Journal of the American Statistical
  Association, 53, 457, \dodoi{10.1080/01621459.1958.10501452}

\bibitem[{{Khodachenko} {et~al.}(2021){Khodachenko}, {Shaikhislamov}, {Lammer},
  {Miroshnichenko}, {Rumenskikh}, {Berezutsky}, \& {Fossati}}]{Khodachenko2021}
{Khodachenko}, M.~L., {Shaikhislamov}, I.~F., {Lammer}, H., {et~al.} 2021,
  \mnras, 507, 3626, \dodoi{10.1093/mnras/stab2366}

\bibitem[{{Kite} \& {Barnett}(2020)}]{Kite2020}
{Kite}, E.~S., \& {Barnett}, M.~N. 2020, Proceedings of the National Academy of
  Science, 117, 18264.
\newblock \doarXiv{2006.02589}

\bibitem[{{Kuhn} {et~al.}(2013){Kuhn}, {Getman}, {Broos}, {Townsley}, \&
  {Feigelson}}]{Kuhn2013a}
{Kuhn}, M.~A., {Getman}, K.~V., {Broos}, P.~S., {Townsley}, L.~K., \&
  {Feigelson}, E.~D. 2013, \apjs, 209, 27, \dodoi{10.1088/0067-0049/209/2/27}

\bibitem[{{Kuhn} {et~al.}(2015){Kuhn}, {Getman}, \& {Feigelson}}]{Kuhn15a}
{Kuhn}, M.~A., {Getman}, K.~V., \& {Feigelson}, E.~D. 2015, \apj, 802, 60,
  \dodoi{10.1088/0004-637X/802/1/60}

\bibitem[{{Kuhn} {et~al.}(2019){Kuhn}, {Hillenbrand}, {Sills}, {Feigelson}, \&
  {Getman}}]{Kuhn19}
{Kuhn}, M.~A., {Hillenbrand}, L.~A., {Sills}, A., {Feigelson}, E.~D., \&
  {Getman}, K.~V. 2019, \apj, 870, 32, \dodoi{10.3847/1538-4357/aaef8c}

\bibitem[{{Lammer} {et~al.}(2003){Lammer}, {Selsis}, {Ribas}, {Guinan},
  {Bauer}, \& {Weiss}}]{Lammer2003}
{Lammer}, H., {Selsis}, F., {Ribas}, I., {et~al.} 2003, \apjl, 598, L121,
  \dodoi{10.1086/380815}

\bibitem[{{Lammer} {et~al.}(2014){Lammer}, {St{\"o}kl}, {Erkaev}, {Dorfi},
  {Odert}, {G{\"u}del}, {Kulikov}, {Kislyakova}, \& {Leitzinger}}]{Lammer2014}
{Lammer}, H., {St{\"o}kl}, A., {Erkaev}, N.~V., {et~al.} 2014, \mnras, 439,
  3225, \dodoi{10.1093/mnras/stu085}

\bibitem[{Loader(1999)}]{Loader99}
Loader, C. 1999, Local Regression and Likelihood (Springer-Verlag New York),
  \dodoi{10.1007/b98858}

\bibitem[{{Loader}(2020)}]{Loader20}
{Loader}, C. 2020, locfit: Local Regression, Likelihood and Density Estimation,
   codes.
\newblock \url{https://cran.r-project.org/web/packages/locfit/index.html}

\bibitem[{{Luhman}(2018)}]{Luhman2018}
{Luhman}, K.~L. 2018, \aj, 156, 271, \dodoi{10.3847/1538-3881/aae831}

\bibitem[{{Luhman} \& {Esplin}(2020)}]{Luhman2020}
{Luhman}, K.~L., \& {Esplin}, T.~L. 2020, \aj, 160, 44,
  \dodoi{10.3847/1538-3881/ab9599}

\bibitem[{{Magaudda} {et~al.}(2021){Magaudda}, {Stelzer}, {Raetz}, \&
  {A}}]{Magaudda2021}
{Magaudda}, E., {Stelzer}, B., {Raetz}, S., \& {A}, K. 2021, arXiv e-prints,
  arXiv:2106.14548.
\newblock \doarXiv{2106.14548}

\bibitem[{{Maggio} {et~al.}(1987){Maggio}, {Sciortino}, {Vaiana}, {Majer},
  {Bookbinder}, {Golub}, {Harnden}, \& {Rosner}}]{Maggio1987}
{Maggio}, A., {Sciortino}, S., {Vaiana}, G.~S., {et~al.} 1987, \apj, 315, 687,
  \dodoi{10.1086/165170}

\bibitem[{{Maschberger}(2013)}]{Maschberger2013}
{Maschberger}, T. 2013, \mnras, 429, 1725, \dodoi{10.1093/mnras/sts479}

\bibitem[{{Massol} {et~al.}(2016){Massol}, {Hamano}, {Tian}, {Ikoma}, {Abe},
  {Chassefi{\`e}re}, {Davaille}, {Genda}, {G{\"u}del}, {Hori}, {Leblanc},
  {Marcq}, {Sarda}, {Shematovich}, {St{\"o}kl}, \& {Lammer}}]{Massol2016}
{Massol}, H., {Hamano}, K., {Tian}, F., {et~al.} 2016, \ssr, 205, 153,
  \dodoi{10.1007/s11214-016-0280-1}

\bibitem[{Ng \& Maechler(2007)}]{Ng2007}
Ng, P., \& Maechler, M. 2007, Statistical Modelling, 7, 315.
\newblock \url{http://smj.sagepub.com/content/7/4/315.abstract}

\bibitem[{Ng \& Maechler(2020)}]{Ng2020}
Ng, P.~T., \& Maechler, M. 2020, COBS -- Constrained B-splines (Sparse matrix
  based).
\newblock \url{https://CRAN.R-project.org/package=cobs}

\bibitem[{{Nu{\~n}ez} {et~al.}(2021){Nu{\~n}ez}, {Povich}, {Binder},
  {Townsley}, \& {Broos}}]{Nunez2021}
{Nu{\~n}ez}, E.~H., {Povich}, M.~S., {Binder}, B.~A., {Townsley}, L.~K., \&
  {Broos}, P.~S. 2021, \aj, 162, 153, \dodoi{10.3847/1538-3881/ac0af8}

\bibitem[{{Ortenzi} {et~al.}(2020){Ortenzi}, {Noack}, {Sohl}, {Guimond},
  {Grenfell}, {Dorn}, {Schmidt}, {Vulpius}, {Katyal}, {Kitzmann}, \&
  {Rauer}}]{Ortenzi2020}
{Ortenzi}, G., {Noack}, L., {Sohl}, F., {et~al.} 2020, Scientific Reports, 10,
  10907, \dodoi{10.1038/s41598-020-67751-7}

\bibitem[{{Owen}(2019)}]{Owen19}
{Owen}, J.~E. 2019, Annual Review of Earth and Planetary Sciences, 47, 67,
  \dodoi{10.1146/annurev-earth-053018-060246}

\bibitem[{{Owen} {et~al.}(2012){Owen}, {Clarke}, \& {Ercolano}}]{Owen2012}
{Owen}, J.~E., {Clarke}, C.~J., \& {Ercolano}, B. 2012, \mnras, 422, 1880,
  \dodoi{10.1111/j.1365-2966.2011.20337.x}

\bibitem[{{Owen} {et~al.}(2013){Owen}, {Hudoba de Badyn}, {Clarke}, \&
  {Robins}}]{Owen13}
{Owen}, J.~E., {Hudoba de Badyn}, M., {Clarke}, C.~J., \& {Robins}, L. 2013,
  \mnras, 436, 1430, \dodoi{10.1093/mnras/stt1663}

\bibitem[{{Picogna} {et~al.}(2019){Picogna}, {Ercolano}, {Owen}, \&
  {Weber}}]{Picogna2019}
{Picogna}, G., {Ercolano}, B., {Owen}, J.~E., \& {Weber}, M.~L. 2019, \mnras,
  487, 691, \dodoi{10.1093/mnras/stz1166}

\bibitem[{{Pizzolato} {et~al.}(2003){Pizzolato}, {Maggio}, {Micela},
  {Sciortino}, \& {Ventura}}]{Pizzolato2003}
{Pizzolato}, N., {Maggio}, A., {Micela}, G., {Sciortino}, S., \& {Ventura}, P.
  2003, \aap, 397, 147, \dodoi{10.1051/0004-6361:20021560}

\bibitem[{{Preibisch}(2012)}]{Preibisch2012}
{Preibisch}, T. 2012, Research in Astronomy and Astrophysics, 12, 1,
  \dodoi{10.1088/1674-4527/12/1/001}

\bibitem[{{Preibisch} \& {Feigelson}(2005)}]{PreibischFeigelson2005}
{Preibisch}, T., \& {Feigelson}, E.~D. 2005, \apjs, 160, 390,
  \dodoi{10.1086/432094}

\bibitem[{{Preibisch} {et~al.}(2017){Preibisch}, {Flaischlen}, {Gaczkowski},
  {Townsley}, \& {Broos}}]{Preibisch2017}
{Preibisch}, T., {Flaischlen}, S., {Gaczkowski}, B., {Townsley}, L., \&
  {Broos}, P. 2017, \aap, 605, A85, \dodoi{10.1051/0004-6361/201730874}

\bibitem[{{Preibisch} {et~al.}(2005){Preibisch}, {Kim}, {Favata}, {Feigelson},
  {Flaccomio}, {Getman}, {Micela}, {Sciortino}, {Stassun}, {Stelzer}, \&
  {Zinnecker}}]{Preibisch05}
{Preibisch}, T., {Kim}, Y.-C., {Favata}, F., {et~al.} 2005, \apjs, 160, 401,
  \dodoi{10.1086/432891}

\bibitem[{{R Core Team}(2020)}]{RCoreTeam20}
{R Core Team}. 2020, R: A language and environment for statistical computing,
  R Foundation for Statistical Computing, Vienna, Austria.
\newblock \url{https://www.R-project.org}

\bibitem[{{Rab} {et~al.}(2017){Rab}, {G{\"u}del}, {Padovani}, {Kamp}, {Thi},
  {Woitke}, \& {Aresu}}]{Rab2017}
{Rab}, C., {G{\"u}del}, M., {Padovani}, M., {et~al.} 2017, \aap, 603, A96,
  \dodoi{10.1051/0004-6361/201630241}

\bibitem[{{Rab} {et~al.}(2020){Rab}, {Padovani}, {G{\"u}del}, {Kamp}, {Thi}, \&
  {Woitke}}]{Rab2020}
{Rab}, C., {Padovani}, M., {G{\"u}del}, M., {et~al.} 2020, in Origins: From the
  Protosun to the First Steps of Life, ed. B.~G. {Elmegreen}, L.~V. {T{\'o}th},
  \& M.~{G{\"u}del}, Vol. 345, 310--311, \dodoi{10.1017/S174392131900156X}

\bibitem[{{Reggiani} {et~al.}(2011){Reggiani}, {Robberto}, {Da Rio}, {Meyer},
  {Soderblom}, \& {Ricci}}]{Reggiani2011}
{Reggiani}, M., {Robberto}, M., {Da Rio}, N., {et~al.} 2011, \aap, 534, A83,
  \dodoi{10.1051/0004-6361/201116946}

\bibitem[{{Richert} {et~al.}(2018){Richert}, {Getman}, {Feigelson}, {Kuhn},
  {Broos}, {Povich}, {Bate}, \& {Garmire}}]{Richert18}
{Richert}, A.~J.~W., {Getman}, K.~V., {Feigelson}, E.~D., {et~al.} 2018,
  \mnras, 477, 5191, \dodoi{10.1093/mnras/sty949}

\bibitem[{{Riello} {et~al.}(2021){Riello}, {De Angeli}, {Evans}, {Montegriffo},
  {Carrasco}, {Busso}, {Palaversa}, {Burgess}, {Diener}, {Davidson}, {Rowell},
  {Fabricius}, {Jordi}, {Bellazzini}, {Pancino}, {Harrison}, {Cacciari}, {van
  Leeuwen}, {Hambly}, {Hodgkin}, {Osborne}, {Altavilla}, {Barstow}, {Brown},
  {Castellani}, {Cowell}, {De Luise}, {Gilmore}, {Giuffrida}, {Hidalgo},
  {Holland}, {Marinoni}, {Pagani}, {Piersimoni}, {Pulone}, {Ragaini}, {Rainer},
  {Richards}, {Sanna}, {Walton}, {Weiler}, \& {Yoldas}}]{Riello2021}
{Riello}, M., {De Angeli}, F., {Evans}, D.~W., {et~al.} 2021, \aap, 649, A3,
  \dodoi{10.1051/0004-6361/202039587}

\bibitem[{{Sanz-Forcada} {et~al.}(2003){Sanz-Forcada}, {Brickhouse}, \&
  {Dupree}}]{Sanz-Forcada2003}
{Sanz-Forcada}, J., {Brickhouse}, N.~S., \& {Dupree}, A.~K. 2003, \apjs, 145,
  147, \dodoi{10.1086/345815}

\bibitem[{{Shaikhislamov} {et~al.}(2020){Shaikhislamov}, {Fossati},
  {Khodachenko}, {Lammer}, {Garc{\'\i}a Mu{\~n}oz}, {Youngblood}, {Dwivedi}, \&
  {Rumenskikh}}]{Shaikhislamov2020}
{Shaikhislamov}, I.~F., {Fossati}, L., {Khodachenko}, M.~L., {et~al.} 2020,
  \aap, 639, A109, \dodoi{10.1051/0004-6361/202038363}

\bibitem[{{Shang} {et~al.}(2002){Shang}, {Glassgold}, {Shu}, \&
  {Lizano}}]{Shang2002}
{Shang}, H., {Glassgold}, A.~E., {Shu}, F.~H., \& {Lizano}, S. 2002, \apj, 564,
  853, \dodoi{10.1086/324197}

\bibitem[{Sheather(2009)}]{Sheather09}
Sheather, S. 2009, A Modern Approach to Regression with R (Springer New York),
  \dodoi{10.1007/978-0-387-09608-7}

\bibitem[{{Siess} {et~al.}(2000){Siess}, {Dufour}, \& {Forestini}}]{Siess00}
{Siess}, L., {Dufour}, E., \& {Forestini}, M. 2000, \aap, 358, 593.
\newblock \doarXiv{astro-ph/0003477}

\bibitem[{{Skumanich}(1972{\natexlab{a}})}]{Skumanich72}
{Skumanich}, A. 1972{\natexlab{a}}, \apj, 171, 565, \dodoi{10.1086/151310}

\bibitem[{{Skumanich}(1972{\natexlab{b}})}]{Skumanich1972}
---. 1972{\natexlab{b}}, \apj, 171, 565, \dodoi{10.1086/151310}

\bibitem[{{Smith} {et~al.}(2001){Smith}, {Brickhouse}, {Liedahl}, \&
  {Raymond}}]{Smith2001}
{Smith}, R.~K., {Brickhouse}, N.~S., {Liedahl}, D.~A., \& {Raymond}, J.~C.
  2001, \apjl, 556, L91, \dodoi{10.1086/322992}

\bibitem[{{Stelzer}(2017)}]{Stelzer2017}
{Stelzer}, B. 2017, Astronomische Nachrichten, 338, 195,
  \dodoi{10.1002/asna.201713330}

\bibitem[{{Stelzer} {et~al.}(2005){Stelzer}, {Flaccomio}, {Montmerle},
  {Micela}, {Sciortino}, {Favata}, {Preibisch}, \& {Feigelson}}]{Stelzer05}
{Stelzer}, B., {Flaccomio}, E., {Montmerle}, T., {et~al.} 2005, \apjs, 160,
  557, \dodoi{10.1086/432375}

\bibitem[{{Stelzer} {et~al.}(2009){Stelzer}, {Robrade}, {Schmitt}, \&
  {Bouvier}}]{Stelzer09}
{Stelzer}, B., {Robrade}, J., {Schmitt}, J.~H.~M.~M., \& {Bouvier}, J. 2009,
  \aap, 493, 1109, \dodoi{10.1051/0004-6361:200810540}

\bibitem[{{Telleschi} {et~al.}(2007){Telleschi}, {G{\"u}del}, {Briggs},
  {Audard}, \& {Palla}}]{Telleschi07}
{Telleschi}, A., {G{\"u}del}, M., {Briggs}, K.~R., {Audard}, M., \& {Palla}, F.
  2007, \aap, 468, 425, \dodoi{10.1051/0004-6361:20066565}

\bibitem[{Therneau(2020)}]{Therneau20}
Therneau, T.~M. 2020, A Package for Survival Analysis in R.
\newblock \url{https://CRAN.R-project.org/package=survival}

\bibitem[{{Townsley} {et~al.}(2019){Townsley}, {Broos}, {Garmire}, \&
  {Povich}}]{Townsley2019}
{Townsley}, L.~K., {Broos}, P.~S., {Garmire}, G.~P., \& {Povich}, M.~S. 2019,
  \apjs, 244, 28, \dodoi{10.3847/1538-4365/ab345b}

\bibitem[{{Tu} {et~al.}(2015){Tu}, {Johnstone}, {G{\"u}del}, \&
  {Lammer}}]{Tu2015}
{Tu}, L., {Johnstone}, C.~P., {G{\"u}del}, M., \& {Lammer}, H. 2015, \aap, 577,
  L3, \dodoi{10.1051/0004-6361/201526146}

\bibitem[{{Vidotto} {et~al.}(2014){Vidotto}, {Gregory}, {Jardine}, {Donati},
  {Petit}, {Morin}, {Folsom}, {Bouvier}, {Cameron}, {Hussain}, {Marsden},
  {Waite}, {Fares}, {Jeffers}, \& {do Nascimento}}]{Vidotto2014}
{Vidotto}, A.~A., {Gregory}, S.~G., {Jardine}, M., {et~al.} 2014, \mnras, 441,
  2361, \dodoi{10.1093/mnras/stu728}

\bibitem[{{Vilhu}(1984)}]{Vilhu1984}
{Vilhu}, O. 1984, \aap, 133, 117

\bibitem[{{Waggoner} \& {Cleeves}(2019)}]{Waggoner2019}
{Waggoner}, A.~R., \& {Cleeves}, L.~I. 2019, \apj, 883, 197,
  \dodoi{10.3847/1538-4357/ab3d38}

\bibitem[{{Waggoner} \& {Cleeves}(2022)}]{Waggoner2022}
---. 2022, arXiv e-prints, arXiv:2202.06962.
\newblock \doarXiv{2202.06962}

\bibitem[{{Warnecke} \& {K{\"a}pyl{\"a}}(2020)}]{Warnecke2020}
{Warnecke}, J., \& {K{\"a}pyl{\"a}}, M.~J. 2020, \aap, 642, A66,
  \dodoi{10.1051/0004-6361/201936922}

\bibitem[{{Williams} \& {Cieza}(2011)}]{Williams2011}
{Williams}, J.~P., \& {Cieza}, L.~A. 2011, \araa, 49, 67,
  \dodoi{10.1146/annurev-astro-081710-102548}

\bibitem[{{Wilms} {et~al.}(2000){Wilms}, {Allen}, \& {McCray}}]{Wilms2000}
{Wilms}, J., {Allen}, A., \& {McCray}, R. 2000, \apj, 542, 914,
  \dodoi{10.1086/317016}

\bibitem[{{Wright} {et~al.}(2011{\natexlab{a}}){Wright}, {Drake}, {Mamajek}, \&
  {Henry}}]{Wright2011}
{Wright}, N.~J., {Drake}, J.~J., {Mamajek}, E.~E., \& {Henry}, G.~W.
  2011{\natexlab{a}}, \apj, 743, 48, \dodoi{10.1088/0004-637X/743/1/48}

\bibitem[{{Wright} {et~al.}(2011{\natexlab{b}}){Wright}, {Drake}, {Mamajek}, \&
  {Henry}}]{Wright11}
---. 2011{\natexlab{b}}, \apj, 743, 48, \dodoi{10.1088/0004-637X/743/1/48}

\bibitem[{{Yadav} {et~al.}(2015){Yadav}, {Christensen}, {Morin}, {Gastine},
  {Reiners}, {Poppenhaeger}, \& {Wolk}}]{Yadav2015}
{Yadav}, R.~K., {Christensen}, U.~R., {Morin}, J., {et~al.} 2015, \apjl, 813,
  L31, \dodoi{10.1088/2041-8205/813/2/L31}

\bibitem[{{Zhang} {et~al.}(2022){Zhang}, {Knutson}, {Wang}, {Dai}, {dos
  Santos}, {Fossati}, {Henry}, {Ehrenreich}, {Alibert}, {Hoyer}, {Wilson}, \&
  {Bonfanti}}]{Zhang2022}
{Zhang}, M., {Knutson}, H.~A., {Wang}, L., {et~al.} 2022, \aj, 163, 68,
  \dodoi{10.3847/1538-3881/ac3f3b}

\bibitem[{{Zhu} {et~al.}(2017){Zhu}, {Tian}, {Li}, \& {Zhang}}]{Zhu2017}
{Zhu}, H., {Tian}, W., {Li}, A., \& {Zhang}, M. 2017, \mnras, 471, 3494,
  \dodoi{10.1093/mnras/stx1580}

\end{thebibliography}
\bibliographystyle{aasjournal}
\end{document}